\documentclass[aps,prb,twocolumn,amsmath,amssymb,eqsecnum,letterpaper,showpacs,floatfix,superscriptaddress]{revtex4-1}
\usepackage{graphicx}
\usepackage{hyperref}
\usepackage{longtable}
\usepackage{nicefrac}

\allowdisplaybreaks[0]

\begin{document}
\title{Resistance of helical edges formed in a semiconductor heterostructure
}
\author{Jukka I. V\"{a}yrynen}
\affiliation{Department of Physics, Yale University, New Haven, CT 06520, USA}
\author{Moshe Goldstein}
\affiliation{Raymond and Beverly Sackler School of Physics and Astronomy, Tel Aviv University, Tel Aviv 69978, Israel}
\author{Yuval Gefen}
\affiliation{Department of Condensed Matter Physics, Weizmann Institute of Science, Rehovot, Israel}
\author{Leonid I. Glazman}
\affiliation{Department of Physics, Yale University, New Haven, CT 06520, USA}
\date{August 31, 2014}
\begin{abstract}
Time-reversal symmetry prohibits elastic backscattering of electrons propagating within a helical edge of a two-dimensional topological insulator. However, small band gaps in these systems make them sensitive to doping disorder, which may lead to the formation of electron and hole puddles. Such a puddle -- a quantum dot -- tunnel-coupled to the edge may significantly enhance the inelastic backscattering rate, due to the long dwelling time of an electron in the dot. The added resistance is especially strong for dots carrying an odd number of electrons, due to the Kondo effect. For the same reason, the temperature dependence of the added resistance becomes rather weak.  We present a detailed theory of the quantum dot effect on the helical edge resistance. It allows us to make specific predictions for possible future experiments with artificially prepared dots in topological insulators. It also provides a qualitative explanation of the resistance fluctuations observed in short HgTe quantum wells. In addition to the single-dot theory, we develop a statistical description of the helical edge resistivity introduced by random charge puddles in a long heterostructure carrying helical edge states. The presence of charge puddles in long samples may explain the observed coexistence of a high sample resistance with the propagation of electrons along the sample edges.
\end{abstract}
\maketitle

\section{Introduction}
A two-dimensional topological insulator supports gapless helical boundary modes. \cite{kane05a,kane05b,bernevig06} Elastic backscattering is forbidden for two states counter-propagating along a ``helical edge'' of a time-reversal symmetric topological insulator.
In the absence of inelastic scattering mechanisms, that would lead to ideal conductance $G_0=e^2/(2\pi\hbar)$ of an edge, as long as the temperature is much smaller than the gap. The joint effect of electron-electron interaction and impurities leads to a temperature-dependent suppression of the conductance, $G=G_0-\Delta G(T)$. The function $\Delta G(T)$ has a power-law low-temperature asymptote; if the impurities are structureless, the only natural scale for the $T$-dependence is provided by the bulk gap $E_g$. In models~\cite{xu06,wu06,budich12} conserving one of the spin projections, $\Delta G\propto (T/E_g)^6$. Lifting that constraint~\cite{schmidt12} makes the $T$-dependence somewhat weaker, $\Delta G\propto (T/E_g)^4$ (hereinafter we dispense with Luttinger liquid effects~\cite{lezmy12,kainaris14} because of the relatively high dielectric constant, $\kappa \approx 13$, in HgTe quantum wells~\cite{qi_review}). At estimated $E_g\sim 10$ meV in HgTe/CdTe heterostructures \cite{qi_review} even the slowest of the two asymptotes provides a strong temperature dependence of $\Delta G$, which is apparently incompatible with observations \cite{konig07,gusev13} (similar results have been obtained  \cite{du13,spanton14} in experiments on InAs/GaSb quantum wells \cite{liu08}).

\begin{figure}
\includegraphics[width=0.7\columnwidth]{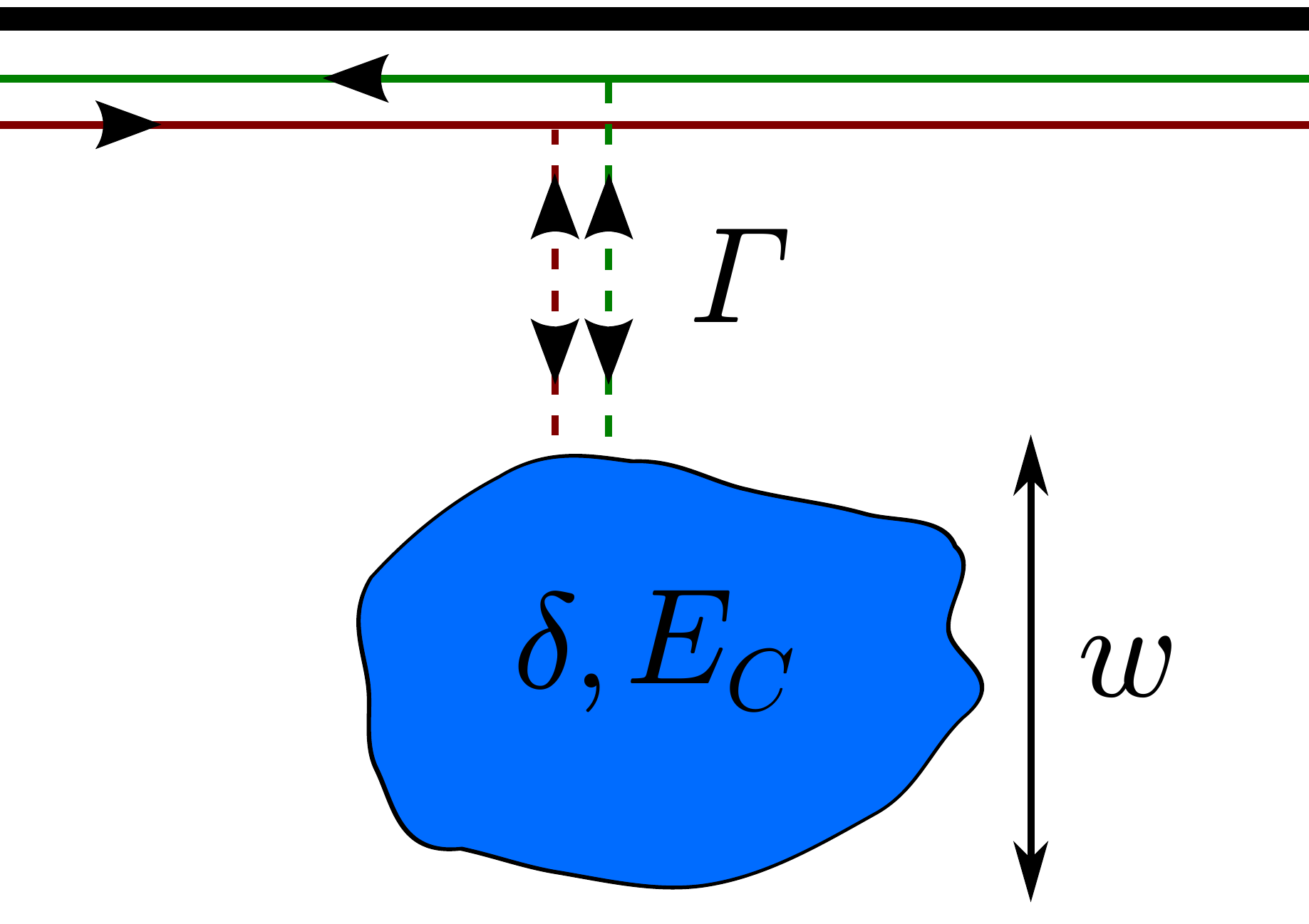}\caption{
(Color online)
A quantum dot of linear size $w$ (bottom) tunnel-coupled to a helical edge (top). The mean level spacing of the puddle is denoted by $\delta$ and its charging energy by $E_C$. The typical tunneling-induced level width is $\Gamma$.}
\label{fig:puddle1}
\end{figure}

In Ref. \onlinecite{vayrynen13} some of us suggested that puddles of electron liquid induced in the topological insulator by doping of a gated heterostructure may enhance backscattering.
One may think of a puddle affecting the edge conductance as a quantum dot side-coupled to the helical edge by a tunnel junction, see Fig. \ref{fig:puddle1}. The crucial difference of the ``new'' quantum dot physics compared to the ``conventional'' one~\cite{Kouwenhoven97,glazman05}
is that elastic scattering processes do not lead to {\it back}scattering, and therefore do not contribute to $\Delta G$. 
The confinement of charge carriers to a puddle -- or a quantum dot -- produces two new small energy scales: spacing  between the levels of spatial quantization $\delta$, and the level width $\Gamma$ associated with the dot-edge tunneling. Backscattering becomes sensitive to the position of the chemical potential with respect to the broadened single-electron levels. That makes $\Delta G(T)$ also dependent on the gate voltage $V_g$ which controls the chemical potential. If it is tuned to coincide with one of the levels, then the characteristic energy scale determining the low-energy asymptote of $\Delta G$ becomes $\sim\Gamma$, rather than $E_g$. In addition to boosting the coefficient in the $\Delta G(T)$ power-law asymptote, the scale $\Gamma$ also defines the range of temperatures above which $\Delta G(T)$ substantially deviates from the asymptote. In fact, the sign of the derivative $d\Delta G/dT$ can change at $T\sim\Gamma$. If the chemical potential is tuned away from a discrete level, then the temperature dependence of $\Delta G$ strongly depends on the electron number parity in the ground state of an isolated puddle. The earlier work \cite{vayrynen13}  concentrated exclusively on the even-parity states, where the characteristic energy separating the low- and ``high''-temperature regimes is $\sim\delta$. That consideration ignored the possibility of puddles which, if isolated, would contain an odd number of electrons and thus carry spin. The spin-carrying states become ubiquitous if the puddle's charging energy $E_C$ exceeds $\delta$. The presence of a spin, in turn, leads to a Kondo effect and to the appearance of a new energy scale $T_K$, which may be exponentially small in the parameter $\delta/\Gamma$. The appearance of the Kondo temperature scale $T_K$ strongly affects the temperature dependence of $\Delta G$ since it limits the validity of the power-law asymptote to  extremely low temperatures. For this reason the puddle-induced resistivity of a long edge shows remarkably weak temperature dependence. This behavior was not captured by the calculation of Ref. \onlinecite{vayrynen13}, which only considered  puddles with an even electron number.

\begin{figure*}[t]
\begin{tabular}{cc}
\includegraphics[width=0.9\columnwidth]{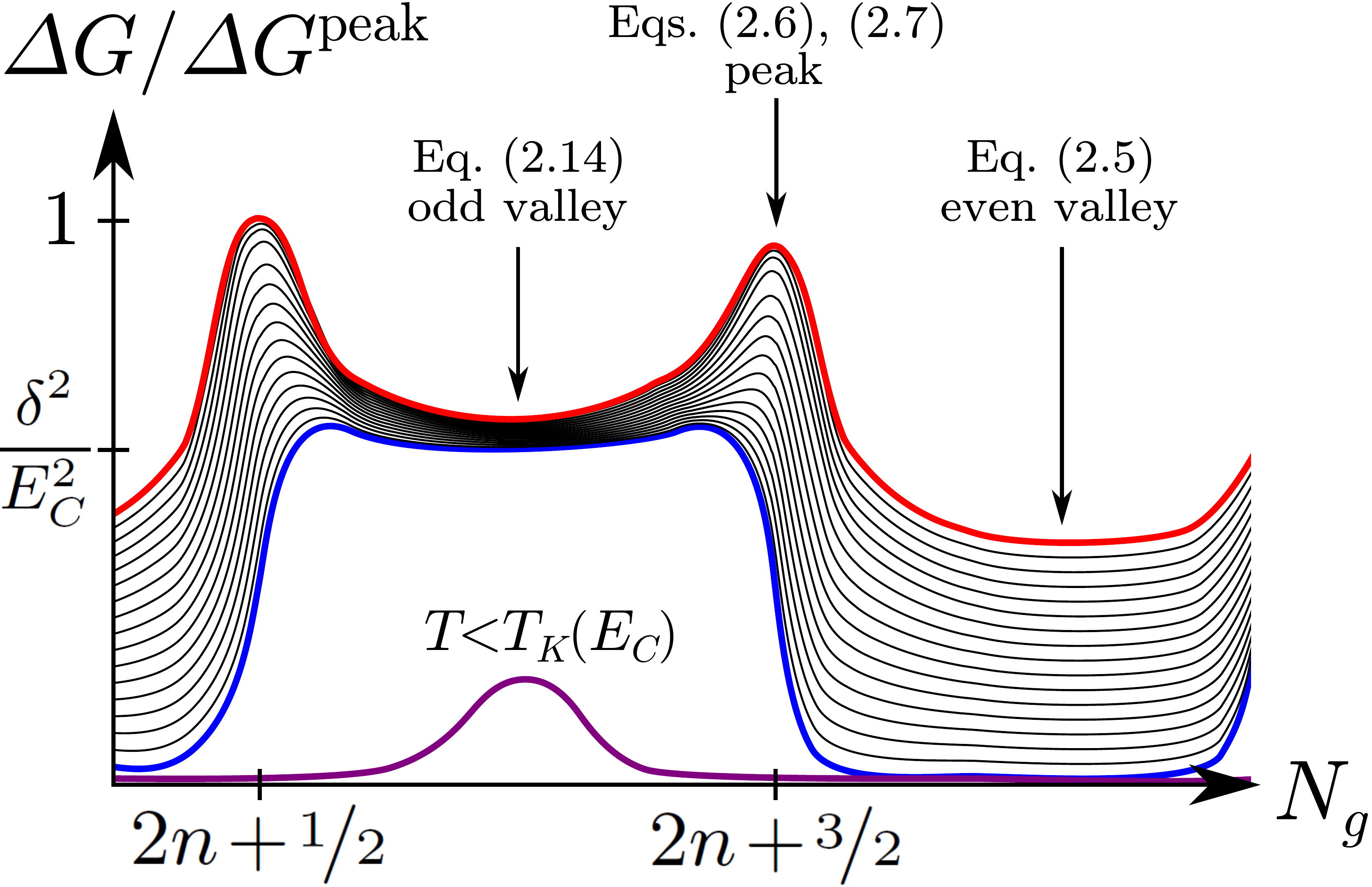} & \includegraphics[width=0.9\columnwidth]{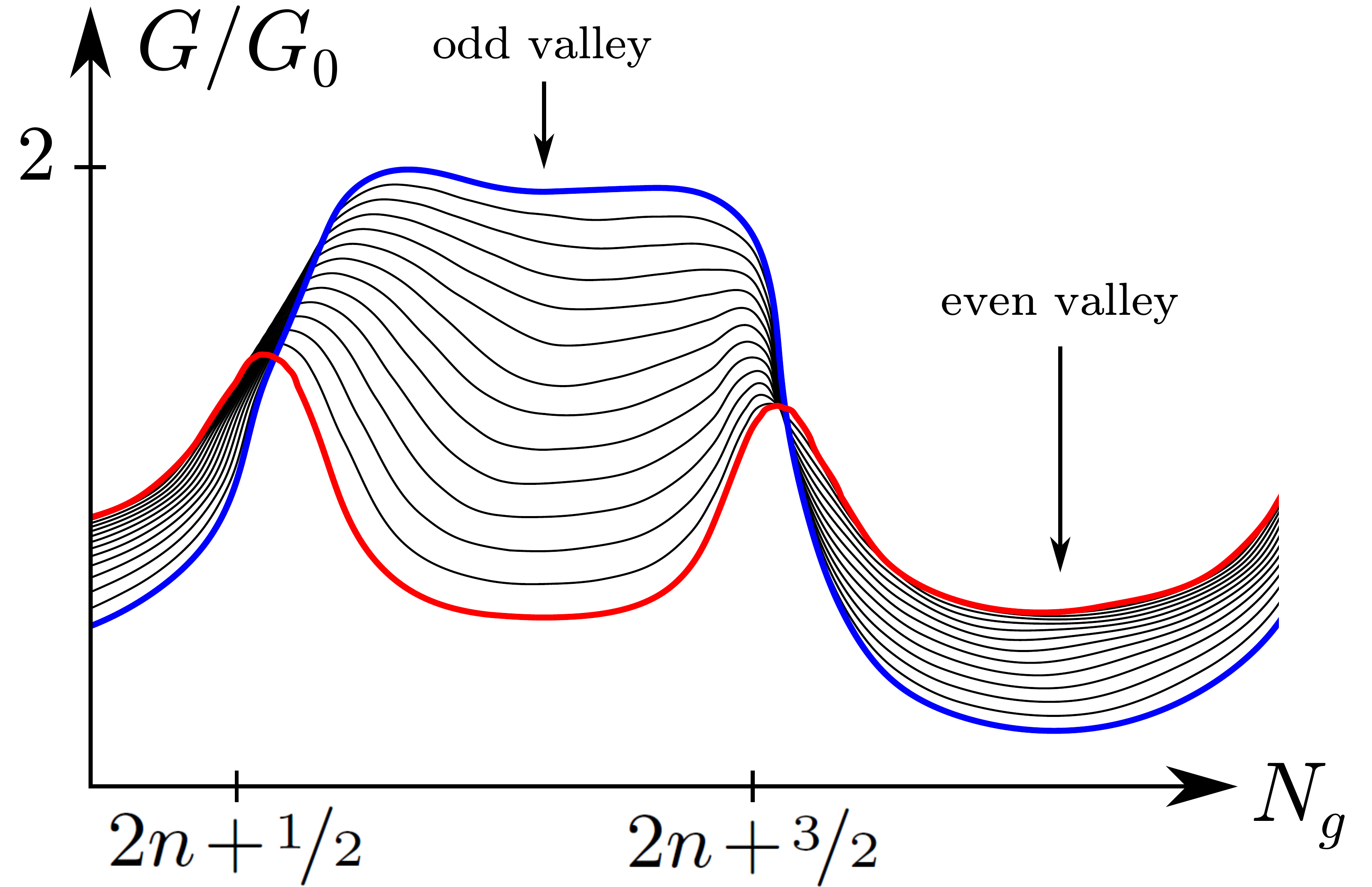}\tabularnewline
(a) & (b)\tabularnewline
\end{tabular}\caption{
(Color online)
(a) Helical edge conductance correction $\Delta G$ caused by a Coulomb blockaded quantum dot is plotted as a function of dimensionless gate voltage $N_{g}=C_{g}V_{g}/e$ and temperature $T$. (Here $C_g$ is the dot-gate capacitance.) Half-integer values of $N_g$ correspond to points of charge degeneracy. For convenience, we measure $\Delta G$ in units of its peak value $\Delta G^{\text{peak}}=G_0 \Gamma^{2}/g^{2}\delta^{2}$ at ``high'' temperature, $\delta > T\gg \Gamma$, see Eq.~(\ref{eq:QualitPeakHighTWithEC}).
 Temperature decreases from dot level spacing $T\sim\delta \gg \Gamma$ (red, topmost curve) down to a very low value $T\sim\delta e^{-\pi\delta/\Gamma}$ (blue, second lowest curve), corresponding to odd-valley Kondo temperature at distance $\delta/E_C$ from a charge degeneracy point.
 Ground state degeneracy in the odd valleys leads to a weak temperature dependence ${\Delta G\propto\ln^{2}(T/T_{K})}$ resulting in the conspicuous plateau seen in the graph. 
  The plateau persists down to temperature $\delta e^{-\pi\delta/\Gamma}$ below which it starts to narrow from the sides (not shown in the figure).
  Once temperature is further lowered below $T\sim \delta e^{-\pi E_{C}/2\Gamma}$ (Kondo temperature in the bottom of the valley),  the odd valley contribution starts to decrease (purple, lowest curve) as $\Delta G \propto T^4$, see Eq.~(\ref{eq:QualitDeltaGoddLowT}).
  In the even valley, the power-law $\Delta G \propto T^4$ holds throughout the wide interval from level spacing to zero temperature. 
  The same law applies to the peaks at temperatures below the level width, $T\ll\Gamma$.
  However, at temperatures above $\Gamma$ the peak $\Delta G(T)$ saturates to a constant value 
   because of off-resonant scattering, see Eq.~(\ref{eq:QualitPeakHighTWithEC}) and the discussion leading to it. 
 (b) For comparison: two-terminal conductance $G(N_g, T)$ of a conventional quantum dot in the regime of the Kondo effect. 
 Notice that $dG/dT<0$ in the odd valley because of Kondo effect,  while in the even one $dG/dT>0$. 
 This results in trace crossings in the graph, 
 contrary to Fig. \ref{FigKondoComparison}a.
 Elastic tunneling makes $G$ finite even in the limit $T\to 0$ (blue curve). 
 This is also in contrast with $\Delta G (T)$ in Fig. \ref{FigKondoComparison}a, which is monotonically decreasing at any $N_{g}$, and eventually vanishes at $T\to 0$ (purple, lowest curve) due to the inelastic nature of scattering.}
\label{FigKondoComparison}
\end{figure*}

The goal of this paper is to develop a comprehensive theory of electron transport along the helical edge channel coupled to quantum dots formed in the ``bulk'' of the two-dimensional topological insulator. We concentrate on the linear-in-bias regime, and investigate $\Delta G (T,V_g)$ for a helical edge in the presence of a single controllable dot. Some of the signature differences of the found $\Delta G (T,V_g)$ from the two-terminal conductance across a conventional quantum dot $G (T,V_g)$ are summarized in Fig.~\ref{FigKondoComparison}. We also aim at predicting the behavior of $\Delta G(T)$ averaged over many random charge puddles (modeled as quantum dots) along an edge of a macroscopic sample.  

As we already mentioned, the peculiarity introduced by the helical nature of the edge is the absence of elastic backscattering processes. That renders inapplicable the conventional elements of the quantum dot transport theory,\cite{glazman05} such as elastic tunneling in the sequential regime, elastic co-tunneling, 
 and elastic scattering off a dot in Kondo regime,
 and makes us to address anew the conduction mechanisms in a broad temperature interval, from $T\ll T_K$, to $T\gtrsim E_C$. 
Using our results we can account for the all the qualitative features seen in the experiments, namely the imperfect conductance of short samples at low temperatures and its fluctuations as function of the gate voltage~\cite{konig07,roth09} as well as the resistive behavior of long samples.~\cite{gusev13} We also provide detailed predictions for future experiments. Let us also note that when time reversal symmetry is broken (e.g., by applying a magnetic field) elastic backscattering caused by sources other then puddles may become significant; we therefore do not discuss this case.

The paper is organized as follows. In Section \ref{sec:Qualitative} we explain qualitatively the main results. Sections \ref{sec:BSinAbsenceOfEC} and  \ref{sec:CoulombBlockade} deal with the backscattering of the helical edge electrons off a single quantum dot with small or large charging energy $E_{C}$, respectively. In Section \ref{sec:LongEdge} we estimate the puddle parameters assuming the puddles of charge originate from dopant-induced potential fluctuations. In the same section we calculate the low-temperature resistivity $\varrho(T)$ of a long edge due to puddles. In Section \ref{sec:Analysis} we discuss how our theory connects with existing experimental data. Finally, conclusions are drawn in Section \ref{sec:Conclusions}.

Throughout this paper we use units such that $\hbar=k_B=1$.

\section{Qualitative discussion of the main results\label{sec:Qualitative}}

In this section we give qualitative description of our main results, leaving detailed discussion to the following sections. We start with estimates of the conductance correction $\Delta G(T)$ coming from a single dot. Being a source of inelastic scattering, it yields $\Delta G\propto T^4$ at low temperature; we relate the proportionality coefficient in that dependence to the parameters of the dot and establish the temperature range for this limiting behavior. The temperature dependence of $\Delta G$ outside that range turns out to be much slower, as we demonstrate next. Finally, we use the single-dot results to estimate the effect of random charge puddles on the edge resistivity of a long sample.

\subsubsection{Electron backscattering off a single quantum dot.\label{sub:QualitSingle}}

At sufficiently low energy scales, electron scattering may be considered in the long-wavelength limit, {\sl i.e.}, the dot behaves as a point-like scatterer. The latter is described by an effective Hamiltonian~\cite{schmidt12,lezmy12}
\begin{equation}
H_{\lambda}=\lambda\left.\{[\psi_{L}^{\dagger}\partial_{x}\psi_{L}^{\dagger}\psi_{L}\psi_{R}
-\psi_{R}^{\dagger}\psi_{L}^{\dagger}\psi_{R}\partial_{x}\psi_{R}]
+\text{h.c.}\}\right|_{x=x_0}\,,
\label{eq:HPhenomenological}
\end{equation}
which respects the time-reversal symmetry; $x_0$ here is the (coarse-grained) coordinate of the scatterer. The spatial derivative in Eq.~(\ref{eq:HPhenomenological}) accounts for the Pauli principle ($[\psi_{L}(x)]^2=[\psi_{R}(x)]^2=0$). The long-wavelength limit of the Hamiltonian is applicable in some energy band $|\varepsilon|\lesssim D$. The bandwidth $D$ is determined by the structure of the impurity, which also determines the value of  $\lambda$. Upon 
transformation of Eq.~(\ref{eq:HPhenomenological}) to the momentum ($k$) representation, the spatial derivative yields a factor $\propto k$ in the interaction amplitude. That in turn leads to an extra factor $T^2$, in the scattering cross-section compared to the ``standard'' Fermi-liquid result,\cite{nozieresBook} and yields $\Delta G\sim \lambda^2 \rho^4 T^4$, as can be checked with the help of the Fermi Golden Rule. (Here $\rho$ is the density of states of the edge.)

To estimate $\lambda$, we match the results for some scattering cross-section, evaluated in two ways: microscopically, and with the help of Eq.~(\ref{eq:HPhenomenological}). For instance, we consider scattering of a left- and a right-mover into two left-movers, $d^\dagger_{k_1L}d^\dagger_{k_2R}|G\rangle\,\to\,d^\dagger_{k_3L}d^\dagger_{k_4L}|G\rangle$. Here $|G\rangle$ is the direct product of ground states for edge and dot, $\psi_{\gamma}(x)=\sum_{k}d_{k\gamma}e^{ikx}$, and the indices $\gamma=L,R$ 
denote the propagation direction along the edge. Each state in the dot is Kramers degenerate; for each Kramers doublet we choose as a basis the states whose spin projections \emph{at the point of contact} match those of the R/L edge modes at the Fermi level. We denote these dot states by R/L too. Hence, tunneling conserves the R/L index.

\begin{figure}
\includegraphics[width=\columnwidth]{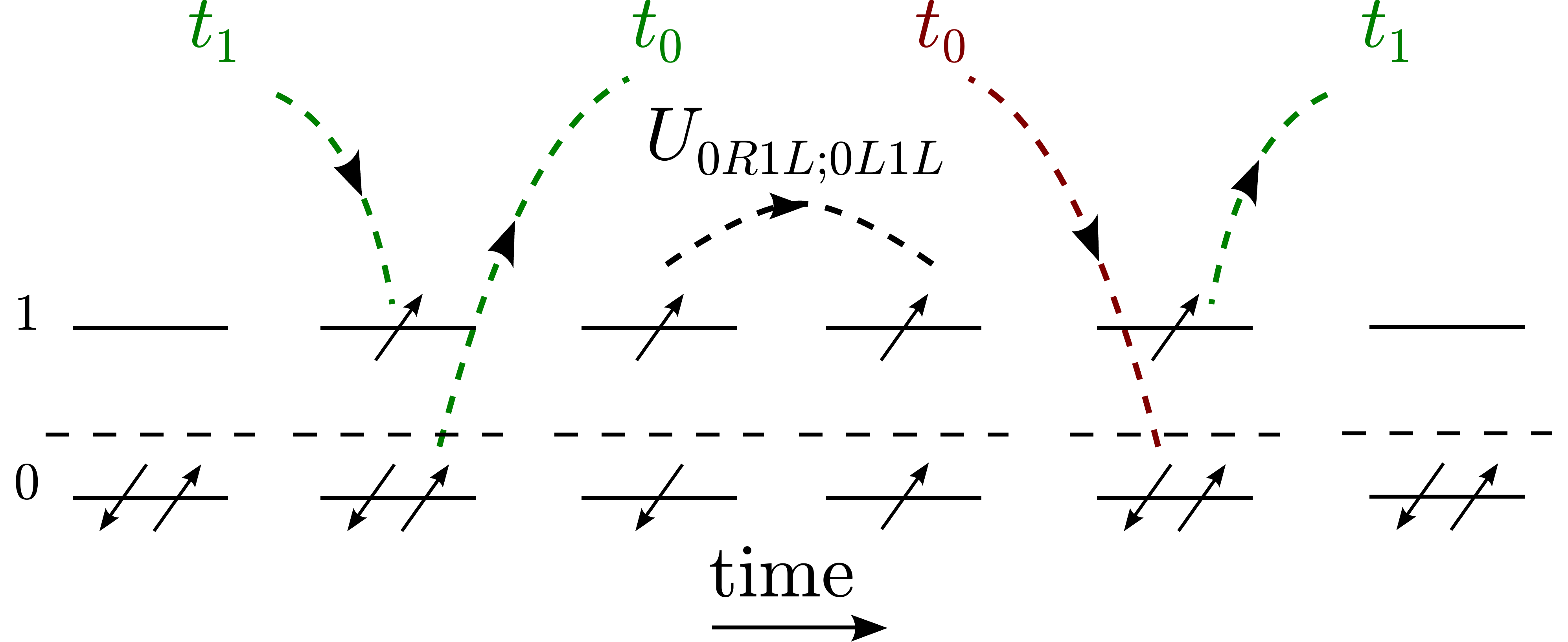}\caption{
(Color online)
An example of a virtual process contributing to the amplitude (\ref{eq:AmplitudeEvenVal}). The initial, intermediate, and final states of the dot are shown (from left to right). The dashed line marks the Fermi level. Here the upwards and downwards arrows correspond to the labels $L$ and $R$, respectively.
}
\label{Fig:EvenProcess}
\end{figure} 

The scattering amplitude in question depends strongly on the parity of the electron number on the dot. We start with an even electron number and sketch the perturbative-in-tunneling evaluation of the amplitude. Moreover, we consider here a ``toy dot'' with only two orbital states.  At even electron number, the state $\varepsilon_0$ is doubly-occupied, and the state $\varepsilon_1$ is empty (energies are measured from the Fermi level of the edge, $\varepsilon_{0}<0$). A typical contribution to the scattering amplitude goes through a sequence of intermediate states (see Fig. \ref{Fig:EvenProcess}):
$d^\dagger_{k_1L}d^\dagger_{k_2R}|G\rangle\,\to\,
c^\dagger_{1L}d^\dagger_{k_2R}|G\rangle\,\to\,
c^\dagger_{1L}d^\dagger_{k_2R}d^\dagger_{k_3L}c_{0L}|G\rangle\,\to\,
c^\dagger_{1L}d^\dagger_{k_2R}d^\dagger_{k_3L}c_{0R}|G\rangle\,\to\,
c^\dagger_{1L}d^\dagger_{k_3L}|G\rangle\,\to\,
d^\dagger_{k_4L}d^\dagger_{k_3L}|G\rangle$ (operators $c^\dagger$ create electrons in the dot). Here the first two transitions are facilitated, respectively, by tunneling of an electron into level $1$ and tunneling of an electron out of level $0$; the third transition is due to the matrix element of interaction within the dot $U_{0L1L;0R1L}=U$, allowed by the time-reversal symmetry (for brevity we denote all the backscattering matrix elements by $U$ in the following). The contribution to the scattering amplitude reads:
\begin{align}
 & A_{k_{1}k_{2}k_{3}k_{4}}=  \label{eq:AmplitudeEvenVal} \\
 & \frac{t_{1}t_{0}U t_{0}t_{1}}{(\varepsilon_{1}+E_+-E_{k_{4}})(\varepsilon_1-\varepsilon_0-E_{k_{1}}+E_{k_{3}})^{2}(\varepsilon_{1}+E_+-E_{k_{1}})}\,.
 \nonumber
\end{align}
Here $t_0, t_1$ are the matrix elements of tunneling Hamiltonian connecting the dot states $0,1$ with the edge, energies $E_{k_i}$ of the edge states are measured from the Fermi level, and $E_+$ is the proxy for the charging energy (coming from interaction of the additional electron on level $1$ with the two electrons on level $0$). The strength of momentum dispersion of the amplitude  is determined by the ratio of energies $E_{k_i}$ in the denominators of Eq.~(\ref{eq:AmplitudeEvenVal}) to the momentum-independent terms in the same denominator. The comparison indicates that the dispersion is negligible, as long as the energies of the involved electrons of the edge reside deep within the bandwidth $D={\min(\delta,\varepsilon_{1}+E_+)}$, where $\delta = \varepsilon_1-\varepsilon_0$ is the dot level spacing. In the linear response regime, that sets condition on temperature, $T\ll D$, at which one may use the effective low-energy theory.

There is a process otherwise identical to the one depicted in Fig. \ref{Fig:EvenProcess}, but with
the two momenta $k_{3}$ and $k_{4}$ interchanged. Adding these two
and using $|E_{k_i}|\ll D$ leads to 
\begin{equation}
H_{\text{eff}}\sim\sum_{k_{1}\dots k_{4}}\, A_{k_{1}k_{2}[k_{3}k_{4}]}d_{k_{4}L}^{\dagger}d_{k_{3}L}^{\dagger}d_{k_{2}R}d_{k_{1}L}\,,
\label{eq:heff}
\end{equation}
where the combined amplitude is 
\begin{equation}
A_{k_{1}k_{2}[k_{3}k_{4}]}=
 \frac{t_{0}^{2}t_{1}^{2}U [\delta+2(\varepsilon_{1}+E_{+})]}{\delta^{3}(\varepsilon_{1}+E_{+})^{3}} \cdot (E_{k_{4}}-E_{k_{3}})
 \,\label{eq:AmplitudeSymmetrized}
\end{equation}
(in deriving it, we accounted for the energy conservation, $E_{k_{1}}+E_{k_{2}}=E_{k_{3}}+E_{k_{4}}$).
The factor ${E_{k_{4}}-E_{k_{3}}}={u(k_{4}-k_{3})}$ in the amplitude
corresponds to $\partial_{x}$ in Eq.~(\ref{eq:HPhenomenological})
and nullifies the amplitude when $E_{k_{4}}=E_{k_{3}}$, in agreement
with the Pauli principle ($u$ is the Fermi velocity of an electron in the edge state). In the following we take the tunneling amplitudes
to be of the same order of magnitude, $t\sim t_{0},\, t_{1}$, and
define a level width $\Gamma = 2\pi t^{2}\rho$, where $\rho$ is the
edge density of states, $\rho={\sum_{k}\delta(E_{k}-E)}$.

The denominator $\varepsilon_1+E_+$ in the amplitude depends on the Fermi level position. Far away from the points of charge degeneracy, $\varepsilon_1+E_+\sim E_C\gg\delta$ in the presence of large charging energy, or $\varepsilon_1+E_+\sim\delta$ if charging is negligible. Using these estimates to simplify the amplitude (\ref{eq:AmplitudeSymmetrized}), and then applying the Fermi Golden Rule to the Hamiltonian (\ref{eq:heff}), we find
\begin{equation}
\Delta G\sim G_0\frac{U^{2}}{\delta^{2}}\frac{\Gamma^{4}}{\delta^{4}}\frac{T^{4}}{E_{C}^{4}}
\sim \frac{G_0}{g^{2}}\frac{\Gamma^{4}}{\delta^{4}}\frac{T^{4}}{E_{C}^{4}}
\,,\quad T\ll\delta\,,
\label{eq:QualitDeltaGevenLowT}
\end{equation}
in the presence of large charging energy [Eq.~(\ref{eq:TypicalDeltaG1ValleyLowTWithECCmb}), Subsection \ref{sub:Even-valley-conductance}]. The result for negligible charging can be obtained from Eq.~(\ref{eq:QualitDeltaGevenLowT}) by setting $E_C \to \delta$, see Eq.~(\ref{eq:TypicalDeltaG1ValleyLowTNoEC})  in Subsection \ref{sub:LowTwithoutEC}.
(Hereinafter in the final estimates of $\Delta G$ we use the typical value of the matrix elements $U^2$ for screened Coulomb interaction  in the symplectic ensemble [strong spin-orbit coupling]; $g$ is the dimensionless conductance of the dot defined as the ratio $E_T/\delta$ of its Thouless energy to level spacing.\cite{aleiner02}) Note that far from the charge degeneracy points $D={\min(\delta,\varepsilon_{1}+E_+)}\sim\delta$ regardless of the ratio $E_C/\delta$.

In the case where Fermi level is near a resonance, ${\varepsilon_{1}+E_{+}}\sim\Gamma$,
the bandwidth $D$ gets reduced to $\Gamma$ and the above process (\ref{eq:AmplitudeSymmetrized})
gives 
\begin{equation}
\Delta G\sim G_0\frac{U^{2}}{\delta^{2}}\frac{T^{4}}{\Gamma^{2}\delta^{2}}
\sim \frac{G_0}{g^{2}}\frac{T^{4}}{\Gamma^{2}\delta^{2}}
\,,\quad T\ll\Gamma\,.\label{eq:QualitPeakSubleadingNoEC}
\end{equation}
The resonances with respect to $E_{k_1}$ and $E_{k_4}$ in Eq.~(\ref{eq:AmplitudeEvenVal}) become available at  temperature $T\gg \varepsilon_{1}+E_{+}$. When in addition $T \ll \delta$, level 1 remains the only resonant level. The resonant denominators in Eq.~(\ref{eq:AmplitudeEvenVal}) are regularized by introducing an imaginary part $i\Gamma_1$ -- this is the level width arising from 
tunneling between the resonant level and the edge.
The correction to the conductance $\Delta G$ can be written in terms of a scattering cross section $\sigma (E)$ of an incoming particle of energy $E$: $\Delta G \propto T^{-1}\int' dE \sigma(E)$. (We denote here $\int' dE = \int_{|E|<T} dE$.) The cross section is related to scattering amplitude $A$ by $\sigma(E)=\int' dE_3 dE_4 \rho^4| A(E,E_3,E_4)|^2$, where $E_{3,4}$ are the energies of the outgoing particles, while the energy of the second incoming particle, $E_2$, is fixed by conservation of energy. The dominant contribution to $\sigma(E)$ comes from processes where one of the outgoing particles is in resonance, while the other is non-resonant and has energy $\sim T$ (\emph{e.g.}, $|E|,\,|E_3|\lesssim\Gamma$; $|E_2|,\,|E_4|\lesssim T$). The amplitude for such a process is estimated from Eq.~(\ref{eq:AmplitudeSymmetrized}) to be $\rho^2 A\sim \Gamma U/\delta^2 (E + i\Gamma)$, while its phase space volume is $\int' dE_3 dE_4 \sim \Gamma T$. We then find
\begin{equation}
\Delta G\sim G_0\frac{U^{2}}{\delta^{2}}\frac{\Gamma^{2}}{\delta^{2}}
\sim \frac{G_0}{g^{2}}\frac{\Gamma^{2}}{\delta^{2}}
\,,\quad T\gg\Gamma\,.\label{eq:QualitPeakHighTWithEC}
\end{equation}
A detailed discussion with more refined formulas 
for $\Delta G$ near the Coulomb blockade peak is given in Subsections \ref{sub:Cond-near-the-Pk-Even} and \ref{sub:Peak-conductance.}, Eqs.~(\ref{eq:DeltaG1NearPkLowT})--(\ref{eq:TypicalDeltaGEvenPkHighTWithEC}). 
 For the limit of weak charging interaction, see Eqs.~(\ref{eq:DeltaG1PeakLowTNoEC})--(\ref{eq:TypicalDeltaG1PeakLowTNoEC}).

We assumed above that the ground state of an isolated dot has an even number of electrons. While this is always the case for small charging energy, a strong charging interaction makes it possible to tune gate voltage so
that the dot ground state has an odd number of electrons. Such a ground
state is doubly degenerate and can be thought of as two states of
a spin-1/2 particle. One can then derive an effective Hamiltonian
\begin{equation}
H_{\rm ex}=J_{ij}S_{i}s_{j}(x_0)\,,\quad
\mathbf{s}(x)=\frac{1}{2}\!\!\!\sum_{\alpha,\beta=L,R}\psi_{\alpha}^{\dagger}(x)\boldsymbol{\sigma}_{\alpha\beta}
\psi_{\beta}(x)\,,
\label{eq:Kondo1}
\end{equation}
valid in the energy band $|\varepsilon|\lesssim\delta$ and describing exchange interaction between the dot spin $\mathbf{S}$ and the edge spin density $\mathbf{s}(x_0)$ at the point contact. The exchange coupling $J_{ij}$ can be split into isotropic and anisotropic
parts, $J_{ij}=J_{0}\delta_{ij}+\delta J_{ij}$. The isotropic part may be related to the properties of the single-occupied state, similar to how it is done for the Anderson impurity problem.\cite{anderson66} Notably, for a point-contact the spin-orbit interaction does not lead to exchange anisotropy in the single-level approximation.~\footnote{this limitation is lifted if the tunneling region is extended, see Appendix \ref{Adx:Tunneling-through-an}} 
  To find the anisotropic part $\delta J_{ij}$ (which reflects the presence of spin-orbit interaction), one has to account for the multi-level structure of the dot and the intra-dot interaction between the electrons.

Let us first consider only the isotropic part $J_{0}$. Connecting to our
microscopic Hamiltonian, $J_{0}$ can be written\cite{anderson66} as the amplitude of an edge electron tunneling into and out of the singly-occupied level $0$, 
\begin{equation}
J_{0}=2t_{0}^{2}(\frac{1}{E_{-}-\varepsilon_{0}}+\frac{1}{E_{+}+\varepsilon_{0}})\,.
\label{eq:QualitJ0}
\end{equation}
Here $E_{-}$ is the cost in charging energy to empty level $0$. The
isotropic part $J_{0}$ leads to the familiar definition of Kondo
temperature $T_{K}\sim \delta e^{-\frac{1}{|J_{0}\rho|}}$.
At $T\gg T_K$ one may generalize the ``poor man's scaling'' ideas \cite{anderson70} to find $\Delta G(T)$. The result is a weak (logarithmic) temperature dependence in the temperature interval $T_K\ll T\ll \delta$. The physics of Kondo effect in a helical edge is somewhat different from the conventional one in quantum dots,\cite{glazman05} and we sketch it next.

Unlike in the conventional quantum dot case,\cite{glazman05} the isotropic part of $J_{ij}$ does not contribute to backscattering. It can be shown with the following argument. The dc backscattering current
can be expressed in terms of time derivative of the numbers of right- and left-movers (of the edge) in a steady state, $\Delta I=(e/2)\langle\frac{d}{dt}(\hat N_R-\hat N_{L})\rangle$. 
None of the edge (or dot) spin-projections are in general conserved, but we can nevertheless define a hypothetical spin by using the orthogonal Kramers states. We take the $z$-projection of this edge net ``spin'' to be proportional to the above difference, $\hat S_z^{\rm edge}=(1/2)(\hat N_L-\hat N_R)$. 
If the exchange interaction between the dot and helical edge would be isotropic, then $z$-component of the total spin is conserved, $\frac{d}{dt}\hat S^{\rm tot}_z=\frac{d}{dt}(\hat S_{z}+\hat S_z^{\rm edge})=0$; as the result, $\Delta I\propto\langle\frac{d}{dt}\hat S_{z}\rangle$, where $\hat S_z$ is the $z$-component of the dot spin. The latter one is bounded, $||\hat S_z||=1/2$; hence, in steady state $\langle\frac{d}{dt}\hat S_{z}\rangle= \left.\langle[\hat S_{z}(t)-\hat S_z(0)]/t\rangle\right|_{t\to\infty}=0$ and~\cite{tanaka11} $\Delta I=0$. We note that $\langle \hat S_z\rangle\neq 0$ in a steady state at bias $V\neq 0$. Finding $\langle \hat S_z\rangle$ may be reduced to a problem of equilibrium statistical mechanics. Using the general principle \cite{LLvol5} of constructing Gibbs distribution $\hat\rho$, we form a linear combination of the Hamiltonian and integral of motion $\hat S_z^{\rm tot}$,
\begin{align}
\hat\rho\propto  \exp\left\{ \right. &-\frac{1}{T}\left[\sum_{k}E_{k}(d_{kL}^{\dagger}d_{kL}+d_{kR}^{\dagger}d_{kR})+H_{{\rm ex}}\right. \nonumber\\
 & \left.\left.-(eV/2)(\hat N_{L}-\hat N_{R}+2\hat S_{z})\right]\right\}
 \label{eq:Gibbs}
\end{align}
The factor $eV/2$ ensures that the chemical potentials of the left and right movers differ by $eV$ due to the applied bias.~\footnote{using the equilibrium distribution $\hat\rho$ is compatible with having different chemical potentials of the two fermion species due to the presence of the integral of motion $\hat S_z^{\rm tot}$} We may neglect $H_{\rm ex}$ in Eq.~(\ref{eq:Gibbs}) as long as $\rho J_0 \ll 1$ and $T\gg T_K$. After that, finding average spin polarizations becomes trivial, $\langle \hat S_z\rangle=eV/(4T)$ at $eV\ll T$; the dot spin polarization is in equilibrium with the itinerant electron polarization $S_{E_F}$ at the Fermi energy. 
(Polarization here is defined in the Kramers-pair sense.) 
Inclusion of a small anisotropic part $\delta J_{ij}$ into the exchange coupling breaks the integral of motion, and, in analogy with the Bloch equations \cite{bloch46} [see Eq.~(\ref{eq:BlochEqWithoutB})] we expect  $\frac{d}{dt}\langle\hat S^{\rm tot}_z\rangle=(-1/\tau_s)(a\langle\hat S_z\rangle+bS_{E_F})$. Here $1/\tau_s\sim (\rho\delta J)^2 T$ is the analogue of the Korringa relaxation rate \cite{korringa50} which in our case accounts only for the interaction violating the $\hat S_z^{\rm total}$ conservation; $a$ and $b$ are some dimensionless constants depending on the specific form of the tensor $\delta J_{ij}$. At the same time, small $1/\tau_s$ should leave the steady-state values $\langle \hat S_z\rangle$ and $S_{E_F}$ close to the equilibrium values at $1/\tau_s=0$. As the result, we find (see also Ref. \onlinecite{tanaka11}) [Eq.~(\ref{eq:DeltaGOddValleyAboveTKNonRGd})]
\begin{equation}
\Delta G=\frac{\Delta I}{V}\sim G_0(\rho\delta J)^2\left[1 - 2\rho J_0 \ln(\delta / T)\right] \,.
\label{eq:Kondo2}
\end{equation}
The logarithmic correction in the brackets is the harbinger of the Kondo effect, which develops at $T\sim T_K$. 

\begin{figure}
\includegraphics[width=0.9\columnwidth]{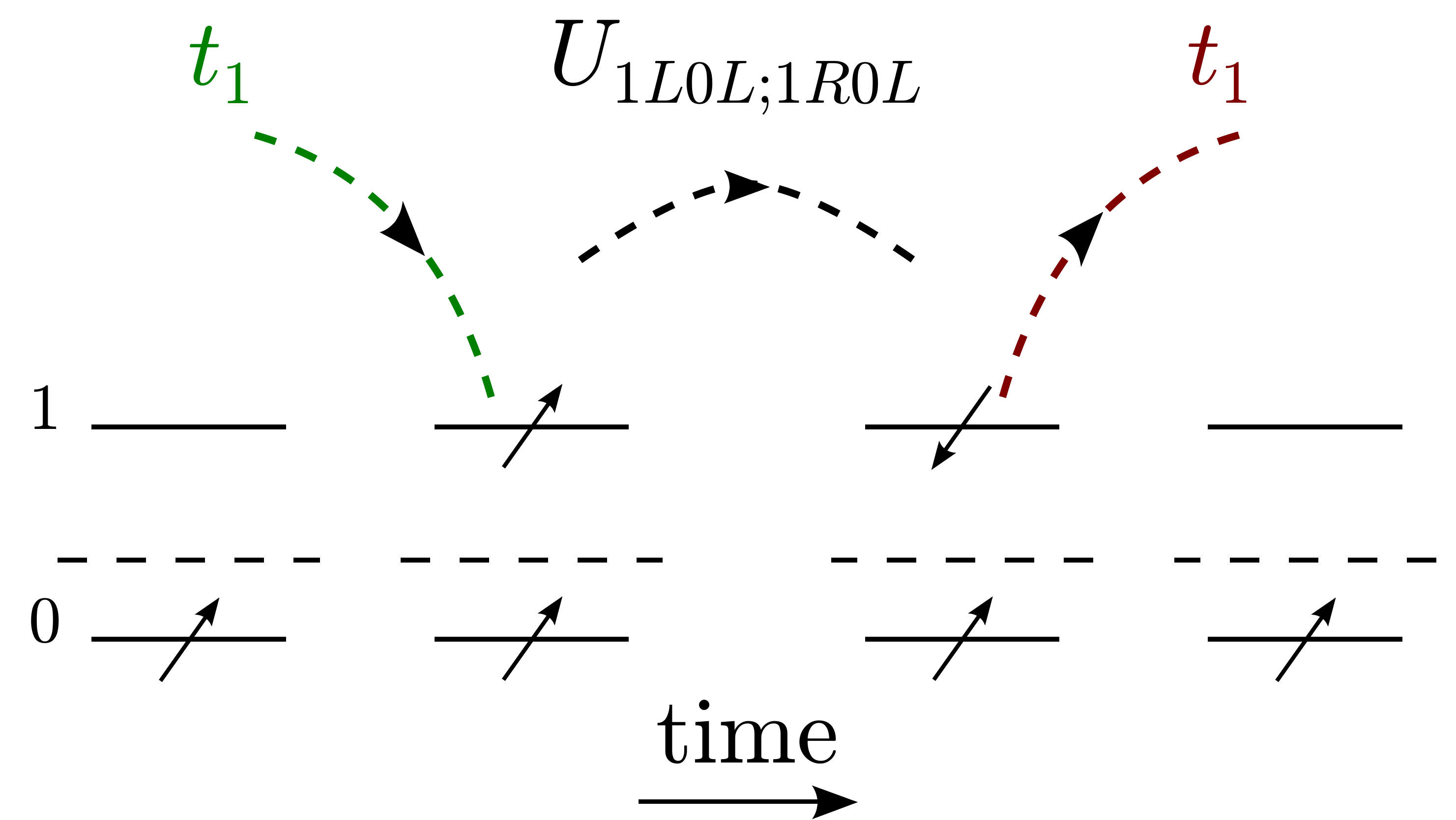}\caption{
(Color online)
An example of a virtual process contributing to the anisotropic part of
exchange couplings $J_{ij}$. The amplitude for this process is given in Eq.~(\ref{eq:AmplitudeOddValley}). The figure depicts the initial, intermediate,
and final states of the dot. The oddly 
occupied dot starts from a ``spin up'' 
 Kramers state (left). A left mover from the edge
tunnels into an empty dot level 1 (dashed line marks the Fermi level).
It flips its spin by interacting, via $U$, with the electron on level 0, and then returns back to the edge, leaving the dot state unchanged (right).}
\label{Fig:OddProcess}
\end{figure}

The anisotropic contribution to the exchange is generated by processes that involve, in addition to electron tunneling into and out of the dot, an interaction-driven transition between the levels within the dot. An example of such a process has an amplitude (see Fig. \ref{Fig:OddProcess})
\begin{equation}
A=\frac{t_{1}^{2}U}{(\delta + \varepsilon_{0}+E_{+})^{2}}\,, \label{eq:AmplitudeOddValley}
\end{equation}
where level 1 is empty in the ground state. 
The energy $\varepsilon_0+E_+$ can be directly controlled with gate voltage and ranges from $E_{C}$ to $0$ when moving from the middle of the valley to the charge degeneracy point 
 (Note however that the exchange Hamiltonian (\ref{eq:Kondo1}) is \emph{not} valid 
 for $|\varepsilon_0+E_+|<\Gamma$, see Subsection \ref{sub:validityOfPerturbation}.)
 For brevity of notation, hereinafter we absorb $\varepsilon_0$ in $E_+$, so that 
  $0\leq E_+\leq E_C$.
 Later on, in Subsection \ref{sub:Derivation-of-the}, we will be considering a generic large dot with many levels -- this leads to summation over empty levels, yielding an extra factor ${(\delta +E_{+})}/\delta$ in the amplitude $A$. To be consistent with the later general results, we introduce this factor 
  into the amplitude. 
  The resulting estimate for the anisotropic part of the (bare) exchange is $\delta J_{ij}\sim t^{2}U/(\delta +E_{+})\delta$. Using this estimate in Eq.~(\ref{eq:Kondo2}) leads to
\begin{equation}
\Delta G \sim \frac{G_0}{g^2}\frac{\Gamma^{2}}{(\delta + E_{+})^{2}} \left[1 - \frac{2\Gamma}{\pi E_{+}} \ln\frac{\delta}{T}\right]\,,\quad T_{K}\ll T\lesssim\delta\,.\label{eq:QualitDeltaGoddNonRGd}
\end{equation}
[We assumed here for simplicity that $E_{+} \leq E_{C}$ in Eq.~(\ref{eq:QualitJ0}).]
This result is perturbative in $J_{ij}$, thus requiring $T\gg T_{K}$.
As one lowers the temperature $T\to T_{K}$, the exchange couplings $J_{ij}$ get renormalized. 
 Summing the leading log-series leads to a suppression~\cite{anderson70} of $\delta J_{ij}$ by a logarithmic factor [Eq.~(\ref{eq:TypicalDeltaGOddValley2})],
\begin{equation}
\Delta G
\sim \frac{G_0}{g^2}\frac{\Gamma^{2}}{(\delta + E_{+})^{2}}\frac{\ln^{2}\frac{T}{T_{K}}}{\ln^{2}\frac{\delta}{T_{K}}}\,,\quad T_{K}\lesssim T\ll \delta\,.\label{eq:QualitDeltaGoddRgd}
\end{equation}

Below the Kondo temperature we can use the low-energy Hamiltonian (\ref{eq:HPhenomenological}) to get $\Delta G\sim\lambda^{2}\rho^{2}T^{4}$. By matching this with Eq.~(\ref{eq:QualitDeltaGoddRgd})
in the limit $T\to T_{K}$ we can read off $\lambda^{2}\rho^{2}T_{K}^{4}\sim\Gamma^{2}/(\delta + E_{+})^{2}g^{2}$.
The new low-energy bandwidth is $T_{K}$; for $T$ within it, we get  [Eq.~(\ref{eq:TypicalDeltaGOddValley2})]
\begin{equation}
\Delta G\sim\frac{G_0}{g^2}\frac{\Gamma^{2}}{(\delta + E_{+})^{2}}\frac{T^{4}}{T_{K}^{4}}\,,\quad T\ll T_{K} \label{eq:QualitDeltaGoddLowT}\,.
\end{equation}
Far from the points of charge degeneracy, $J_0$ is smallest, and consequently $T_{K}\sim \delta\exp(-\pi E_{C}/2\Gamma)$ is exponentially small. That makes the temperature dependence of $\Delta G$ weak, see Eqs.~(\ref{eq:QualitDeltaGoddNonRGd}) and (\ref{eq:QualitDeltaGoddRgd}), in a broad temperature interval $T\lesssim\delta$.

 Close to the charge degeneracy points, $E_{+}\to\Gamma$, the exchange constants substantially increase leading to $T_K\to\Gamma$ in Eqs.~(\ref{eq:QualitDeltaGoddNonRGd}), (\ref{eq:QualitDeltaGoddLowT}), which match then the estimates of the peak values of $\Delta G$ given in Eqs.~(\ref{eq:QualitPeakHighTWithEC}), (\ref{eq:QualitPeakSubleadingNoEC}). 

The key features of the temperature and gate voltage dependence of $\Delta G$ are illustrated in Fig.~\ref{FigKondoComparison}a. (For a contrast, we present in Fig.~\ref{FigKondoComparison}b the conductance {\sl vs.} gate voltage and $T$-dependence for a ``conventional'' quantum dot device.\cite{vanderWiel00}) At low temperatures $T\ll \delta$, the conductance correction displays strong dependence on gate voltage, seen as peaks [Eqs.~(\ref{eq:QualitPeakSubleadingNoEC}),
(\ref{eq:QualitPeakHighTWithEC})] and valleys [Eqs.~(\ref{eq:QualitDeltaGevenLowT}), (\ref{eq:QualitDeltaGoddRgd}), (\ref{eq:QualitDeltaGoddLowT})]. Furthermore, the valleys with an odd number of electrons  differ significantly from those with an even number  [cf. Eq.~(\ref{eq:QualitDeltaGevenLowT}) vs. Eq.~(\ref{eq:QualitDeltaGoddRgd})].

Upon increasing temperature, the difference between peaks and valleys is seen to decrease, 
 while the average $\Delta G$ increases as a function of temperature.  
 Also the distinction between odd and even valleys disappears once temperature is increased, $T\gtrsim \delta$, since spin-carrying particle-hole pairs become thermally populated regardless of dot particle number parity.
 
\begin{figure}[t]
\includegraphics[width=0.7\columnwidth]{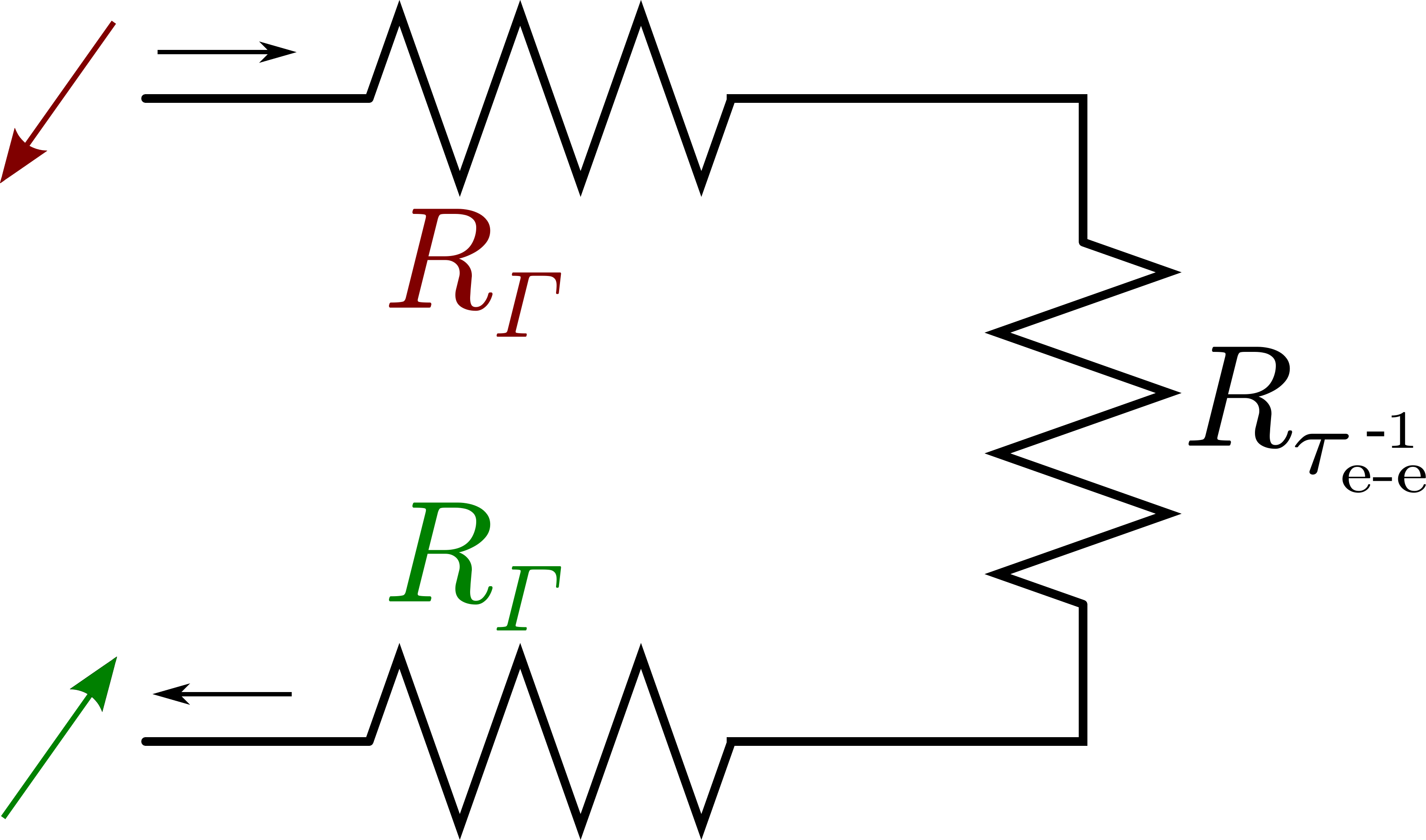}\caption{
(Color online)
At high temperatures the correction to the conductance is given by $\Delta G=(2R_{\Gamma}+R_{\tau^{-1}_{\text{e-e}}})^{-1}$, corresponding to three resistors in series: two resistors correspond to the weak tunneling between the edge (on the left) and the dot, while the third one describes the inelastic process in the dot, needed to convert right movers into left movers, and \emph{vice versa}.}
\label{Fig:3resistors}
\end{figure}

 Upon further increase of temperature, $T\gtrsim E_C$, also the particle number of the dot is allowed to fluctuate, washing away peaks and valleys, making $\Delta G$ weakly dependent on gate voltage. 
 At these high temperatures the virtual processes 
  described  above [Eqs.~(\ref{eq:QualitDeltaGevenLowT})--(\ref{eq:QualitDeltaGoddLowT})] give way to sequential tunneling. 
The conductance correction $\Delta G$ can then be thought of as the conductance of a quantum dot coupled to spin-polarized ``leads'' (the left- and right-moving edge channels), see Subsection \ref{sub:Kinetic-equation}. 
 Then $\Delta G$ corresponds to three resistors in series: two identical resistors 
corresponding to weak tunneling between the edge (or ``leads'') and the dot and a third one corresponding to the necessary spin-flip process 
inside the dot, see Fig. \ref{Fig:3resistors}. The latter is characterized by the intra-dot scattering rate $\tau_{e-e}^{-1}\sim T^{2}\delta/E_{T}^{2}$, and exceeds the rate of tunneling $\Gamma$ at temperatures $T\gtrsim E_{T}\sqrt{\Gamma/\delta}$.
 Above this temperature nearly every electron
that finds its way into the dot 
 loses memory of its origin and has equal probabilities to tunnel back as a left or a right mover. The bottleneck for backscattering then lies in tunneling in and out of the dot. Then $\Delta G$ saturates to the constant value [Eq.~(\ref{eq:KineticConductanceFinal})] given by tunneling rate times the dot density of states,
\begin{equation}
\Delta G= G_0\frac{\Gamma}{2\delta}\,,\quad\text{max}\left( E_{C},E_{T}\sqrt{\frac{\Gamma}{\delta}}\right) \lesssim T\,.
\label{eq:QualitDeltaGHighTSaturated}
\end{equation}
Taking the limit $\Gamma \to \delta$, corresponding to a dot 
 lying on the edge, we find $\Delta G=G_{0}/2$ in agreement with the theoretical model used in Ref. \onlinecite{roth09}. There the authors showed that a completely phase-randomizing dot on the edge has the same effect on edge conductance as an additional lead.
\begin{figure}[h]
\includegraphics[width=1.0\columnwidth]{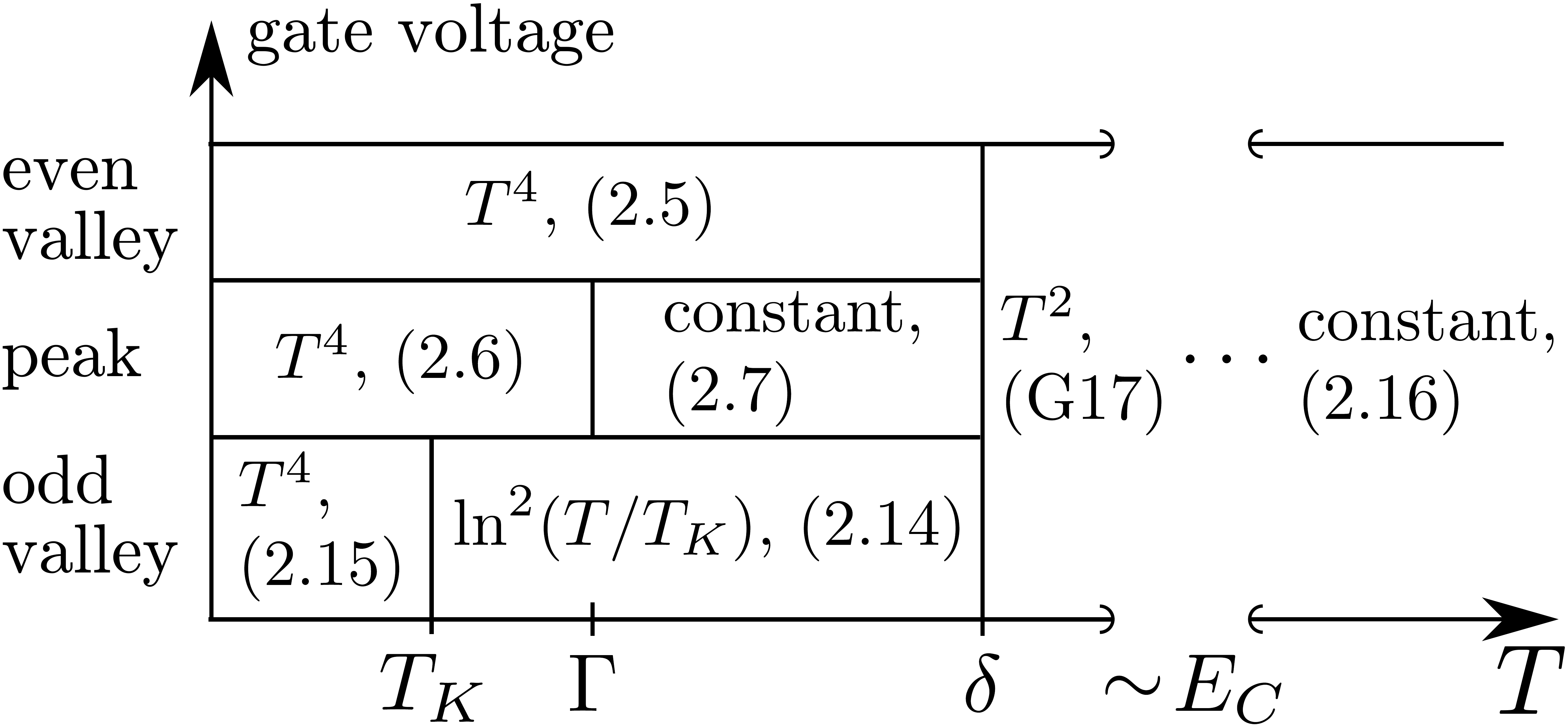}\caption{
The temperature and gate-voltage dependence of the single-dot correction $\Delta G$ to the helical edge conductance.
At very low temperatures, $T\ll T_K$, the low-temperature asymptote of $\Delta G$ for all gate voltages is a  power-law with a universal exponent, $\Delta G \propto T^4$. 
The  Kondo temperature $T_K$ depends on the gate voltage,  attaining its highest value $\sim \Gamma$ upon approaching the Coulomb blockade peaks. 
Above $T_K$, the correction $\Delta G$ at the peaks and in the odd valleys becomes  weakly-dependent on $T$, as also illustrated in Fig.~\ref{FigKondoComparison}a; for even valleys, the $\Delta G\propto T^4$ law  remains intact. 
Upon increasing the temperature above the level spacing $\delta$, the distinction between even and odd valleys disappears because of spinful thermal excitations in the dot; at the same time, thermal fluctuations of the dot electron number remain suppressed as long as $T \ll E_C$. In this limit $\Delta G \propto T^2$ -- a standard consequence  of interaction in the Fermi-liquid theory, see Eq.~(\ref{eq:KineticConductanceFinalWithCot}). 
Finally, at temperatures $T\gg E_C$, direct tunneling of electrons into and out of the dot becomes possible, resulting in a temperature- and gate-voltage-independent asymptote of $\Delta G$, see Eq.~(\ref{eq:QualitDeltaGHighTSaturated}) and the explanatory Fig.~\ref{Fig:3resistors}. The cross-over leading to Eq.~(\ref{eq:QualitDeltaGHighTSaturated}) depends on specific dot parameters, see Fig.~\ref{Fig:TemperatureLine}.}
\label{Fig:table}
\end{figure}

The overall temperature-dependence of $\Delta G$ of a single Coulomb blockaded quantum dot is summarized in Fig.~\ref{Fig:table}.

\subsubsection{Correction to the conductance due to many Coulomb blockaded puddles\label{sub:QualitMany}}

In this section we show how  many quantum dots (or charge puddles) affect the resistivity of a long edge. It is assumed that the presence of many puddles does not lead to appreciable bulk conductivity at low temperatures. 
For doping-induced puddles this requires low enough dopant density; the crossover value  of the dopant density is determined by the properties of the heterostructure~\cite{gergel78} and is given in Eq.~(\ref{eq:DonorDensityCharacteristic}) in Subsection \ref{sub:Potential-fluctuations-in}.

The backscattering of edge electrons off a quantum dot is inelastic
and therefore incoherent. It means that the conductance correction
from several puddles is additive, and the self-averaging resistivity $\varrho$ of a long edge
 is obtained by independently summing over single-puddle contributions,  $\varrho=n_{p}\Lambda\langle \Delta G\rangle/G_{0}^{2}$, see the derivation of Eq.~(\ref{eq:resistivityNoECIntermediate}). Here $n_{p}$ is the number of puddles per unit area in the bulk, $\Lambda\sim v/E_{g}$ is the penetration depth of the electron wave function into the bulk (determined by bulk band gap and the Dirac velocity $v$ in HgTe), 
and $\langle \Delta G\rangle$ is the single-puddle conductance correction averaged over the puddle parameters and the number of particles in it. This last average can be done by keeping the dot levels fixed and averaging over the chemical potential, or, equivalently, the gate voltage. 

We consider the most interesting case of low temperatures, $T\ll\delta$, where bulk conductivity can be safely ignored.
In the presence of charging $E_{C}\gtrsim\delta$, the average of $\Delta G$ over number of particles is dominated by odd occupations (see Fig. \ref{FigKondoComparison}a).
For averaging we need to take into account properly the gate voltage dependence of the Kondo temperature $T_{K}$ and $\Delta G$.
 The valley is symmetric, so it suffices to consider the
gate voltage interval from the bottom of the valley to the peak separating dot charge states 1 and 2, $E_{C}>E_{+}>\Gamma$.
We also need to sum over different distances $d$ between the edge and the dot.
This sum can be converted into an integral over the edge-dot tunneling rate $\Gamma$, which decays exponentially with the distance $d$ between the dot and the edge, $\Gamma(d)\sim e^{-2d/\Lambda}$.   
 The upper limit for $\Gamma$ is the level spacing $\delta$, as follows from our assumption of narrow levels.
\footnote{The lower limit is non-zero for a finite width sample where $d<W$. Here we however assume a wide enough sample, so that the lower limit can be taken smaller than any other scale\label{fn:samplewidth}} 
As was already mentioned, the average $\Delta G$ is dominated by the odd valleys and more specifically by those gate voltages where $T>T_K(E_+)$ is valid. In this interval the logarithmic corrections to the bare value of $\rho J_{ij}$ are small and we can use the first term in Eq.~(\ref{eq:QualitDeltaGoddNonRGd}). The average conductance is then
\begin{equation}
\langle \Delta G\rangle \sim\int_{\Gamma}^{E_C} \frac{dE_+}{E_C} \int_{0}^{\delta}\frac{d\Gamma}{\Gamma}\frac{\Gamma^{2}}{g^{2}(\delta+E_{+})^2} \Theta[T-T_K(E_+)]\,.\label{eq:QualitatDeltaGgvAvLowT}
\end{equation}
Inserting $\rho J_0 = 2\Gamma/\pi E_+$ [see Eq.~(\ref{eq:QualitJ0})] in the Kondo temperature $T_K\sim \delta e^{-\frac{1}{|J_{0}\rho|}}$ allows us to write the step function $\Theta$ as a condition on level width: $\Gamma < \frac{\pi E_+}{2\ln(\delta/T)}$. 
If this condition is stricter than $\Gamma<\delta$, then it is the step function rather than upper limit $\delta$ that determines the domain of integration over $\Gamma$.  
We will confine our considerations to that limit, realized at low temperatures, $T\ll\delta e^{-\frac{\pi E_{C}}{2\delta}}$.
 We then get from Eq.~(\ref{eq:QualitatDeltaGgvAvLowT}) the average $\Delta G$ at low temperatures,
\begin{equation}
\left\langle \Delta G\right\rangle \sim\frac{1}{g^{2}}\frac{1}{\ln^{2}(\delta/T)}\,,\quad T\ll\delta e^{-\frac{\pi E_{C}}{2\delta}}\,.\label{eq:QualitGavLowT}
\end{equation}

Application of standard methods of disordered semiconductors theory~\cite{gergel78,shklovskii84} to a doped heterostructure allows one to conclude that typically $E_C\sim\delta$ and $g \sim 1$ at weak doping. This condition is sufficient to yield a substantial fraction of puddles carrying an odd charge.
 Equation~(\ref{eq:QualitGavLowT}) then yields a long-edge resistivity with a weak temperature dependence [Eq.~(\ref{eq:varrhoEcEqdelta})]
\begin{equation}
\varrho\sim\frac{n_{p}\Lambda}{G_{0}}\frac{1}{\ln^{2}(\delta/T)},\quad T\ll\delta \,.
\label{eq:QualitvarrhoEcEqdelta}
\end{equation}
This equation is one of the main results of this paper, and concludes the qualitative part. We would like to point out that the above result differs from that given in Ref. \onlinecite{vayrynen13} ($\varrho \propto T^3$), where the odd occupations were not accounted for. We now turn to detailed calculations, starting with the evaluation of $\Delta G$ in the weak-interaction limit (i.e., in the absence of Coulomb blockade). 

\section{Electron backscattering in the absence of charging effects\label{sec:BSinAbsenceOfEC}}


We consider a time-reversal (TR) symmetric Hamiltonian of a helical
edge coupled to a quantum dot. The coupling is assumed to be via a
point contact with tunneling matrix elements $t_{n}$; generalization
to a line contact is given in Appendix \ref{Adx:Tunneling-through-an}.
For a TR-symmetric point contact the matrix elements $t_{n}$ can
be chosen to be real with a rotation of the spin quantization axis of the
dot (Appendix \ref{Adx:Tunneling-through-an}). 
More specifically, the rotation aligns the dot spin quantization axis at the point contact to be parallel with that of the edge electrons at the Fermi level. 

The finite-size quantum
dot is assumed to be in the metallic regime (dimensionless conductance
$g\gg1$) with a set of discrete random energy levels  whose distribution
is described by random matrix theory (RMT) (with mean level-spacing
$\delta$). The free part of the Hamiltonian then reads

\begin{align}
H_{0}= & -iu\sum_{\gamma}\gamma\int dx\psi_{\gamma}^{\dagger}(x)\partial_{x}\psi_{\gamma}(x)+\sum_{n,\gamma}\varepsilon_{n}c_{n\gamma}^{\dagger}c_{n\gamma}\label{eq:FreeHamiltonian}\\
+ & \sum_{n,\gamma}t_{n}c_{n\gamma}^{\dagger}\psi_{\gamma}(0)+\text{H.c.}\,,\nonumber 
\end{align}
where the point contact is at $x=0$ on the edge, and $\gamma=\pm1\equiv L,R$ labels the Kramers pairs in the
dot and the edge (left- and right-movers). Note that the free Hamiltonian
is diagonal in this label. The discrete levels $\varepsilon_{n}$
of the quantum dot are measured from the Fermi energy. The tunneling
between the edge and the dot results in a finite lifetime to the eigenstates of the dot. 
We denote the decay rate, or level width, of level $n$ by $\Gamma_{n}$.
Interactions in the dot are described by the Hamiltonian
\begin{equation}
\hat{U}=\frac{1}{2}\sum_{n_{i}}\sum_{\gamma_{i}}U_{n_{1}\gamma_{1}n_{2}\gamma_{2};n_{3}\gamma_{3}n_{4}\gamma_{4}}c_{n_{1}\gamma_{1}}^{\dagger}c_{n_{2}\gamma_{2}}^{\dagger}c_{n_{4}\gamma_{4}}c_{n_{3}\gamma_{3}}\,.\label{eq:IntradotInteraction}
\end{equation}
By TR symmetry, the interaction matrix element satisfies  $U_{n_{1}\gamma_{1}n_{2}\gamma_{2};n_{3}\gamma_{3}n_{4}\gamma_{4}} = \gamma_{1}\gamma_{2}\gamma_{3}\gamma_{4} U^{*}_{n_{1}\overline{\gamma_{1}}n_{2}\overline{\gamma_{2}};n_{3}\overline{\gamma_{3}}n_{4}\overline{\gamma_{4}}}$, denoting $\overline{\gamma}=-\gamma$ for Kramers indices. 
Note that the TR symmetry does not prevent $U$ from having matrix elements that do not conserve the total spin of the two electrons. 
The typical value of the squared matrix element is given in Appendix \ref{sec:Averaging-over-disorder}. 
In this section we consider weak charging effects, \textit{i.e.}, the limit where all non-zero matrix elements of interaction are small.
The perturbative treatment of the entire interaction Eq.~(\ref{eq:IntradotInteraction})
is possible when the matrix elements are small compared to $\Gamma_{n}$.


In absence of $\hat{U}$, we define the free retarded dot-dot Green function as
\begin{equation}
G_{nm}^{R}(t-t')=-i\theta(t-t')\langle\{c_{n\gamma}(t),c_{m\gamma}^{\dagger}(t')\}\rangle\,.
\end{equation}
It satisfies the Dyson equation
\begin{equation}
G_{nm}^{R}(\omega)=G_{nm}^{(0)R}(\omega)+\sum_{n'n''}G_{nn'}^{(0)R}(\omega)\Sigma_{n'n''}^{R}(\omega)G_{n''m}^{R}(\omega)\,,
\end{equation}
where $G_{nm}^{(0)R}(\omega)=\delta_{nm}(\omega-\varepsilon_{n}+i0)^{-1}$
is the dot Green function in absence of tunneling, and $\Sigma_{nm}^{R}(\omega)=\sum_{k}t_{n}t_{m}(\omega-E_{\gamma, k}+i0)^{-1}$
is the self-energy. 
Here $E_{\gamma, k}= \gamma u k $ is the edge state dispersion relation. From TR symmetry, $E_{\gamma, k}=E_{-\gamma, -k}$, it follows that the self-energy is independent of $\gamma$.
 The imaginary part of the self-energy broadens the dot levels,
\begin{equation}
\Gamma_{nm}\equiv\sqrt{\Gamma_{n}\Gamma_{m}}=-2\text{Im}\,\Sigma_{nm}^{R}(\omega)=2\pi \rho\, t_{n}t_{m}\,,
\end{equation}
where $\rho = \sum_k \delta(E-E_k)$ is the single particle density of states, which is assumed to be energy-independent. (By TR-symmetry, the density of states is the same for both Kramers pairs, and we can leave out the index $\gamma$.)
As a matrix in the levels space, the solution to Dyson equation is then
\begin{equation}
\mathbf{G}^{R}(\omega)=(\omega\boldsymbol{1}-\boldsymbol{\varepsilon}+\frac{i}{2}\boldsymbol{\Gamma})^{-1}\,.\label{eq:exactGreenFn}
\end{equation}
where $(\boldsymbol{\varepsilon})_{nm}=\varepsilon_{n}\delta_{nm}$.

We are interested in how the coupling of the helical edge to the quantum
dot (and interaction within) affects the conductance of the edge.
To this end, we need to consider scattering processes $|E_{1}\gamma_{1},E_{2}\gamma_{2}\rangle\rightarrow|E_{3}\gamma_{3},E_{4}\gamma_{4}\rangle$
between exact left- and right-propagating eigenstates of the single-particle Hamiltonian.
In the Born approximation the amplitude for these processes is $\langle E_{3}\gamma_{3},E_{4}\gamma_{4}|\hat{U}|E_{1}\gamma_{1},E_{2}\gamma_{2}\rangle$.
The corresponding scattering cross section can be written as a sum
over dot-eigenstates, 
\begin{align}
 & S_{\gamma_{1}\gamma_{2};\gamma_{3}\gamma_{4}}(E_{1},E_{2};E_{3},E_{4})=\frac{2}{\pi^{3}}\sum_{m_{i},n_{i}}\left[\prod_{i=1}^{4}\text{Im}G_{n_{i}m_{i}}^{R}(E_{i})\right]\nonumber \\
 & \qquad\times U_{m_{1}\gamma_{1}m_{2}\gamma_{2};m_{3}\gamma_{3}m_{4}\gamma_{4}}^{*}U_{n_{1}\gamma_{1}n_{2}\gamma_{2};n_{3}\gamma_{3}n_{4}\gamma_{4}}\,.\label{eq:S}
\end{align}

The correction $\Delta G$ to the edge conductance can be expressed in
terms of the cross section. Inelastic backscattering due to $\hat{U}$
reduces the steady-state current $I=I_{0}-\Delta I$ from its ideal
value $I_{0}=G_{0}V$ by $\Delta I = \Delta G V$,  
\begin{align}
\Delta I & =-e\sum_{\gamma_{i}}\Delta N_{\gamma_{1}\gamma_{2};\gamma_{3}\gamma_{4}}\int dE_{1}dE_{2}dE_{3}dE_{4}\nonumber \\
 & \times S_{\gamma_{1}\gamma_{2};\gamma_{3}\gamma_{4}}(E_{1},E_{2};E_{3},E_{4})\delta(E_{1}+E_{2}-E_{3}-E_{4})\nonumber \\
 & \times\big[f_{\gamma_{1}}(E_{1})f_{\gamma_{2}}(E_{2})(1-f_{\gamma_{3}}(E_{3}))(1-f_{\gamma_{4}}(E_{4}))\nonumber \\
 & -f_{\gamma_{3}}(E_{3})f_{\gamma_{4}}(E_{4})(1-f_{\gamma_{1}}(E_{1}))(1-f_{\gamma_{2}}(E_{2}))\big]\,.\label{eq:CurrentGeneral}
\end{align}
Here $V$ is the source-drain voltage, $\Delta N_{\gamma_{1}\gamma_{2};\gamma_{3}\gamma_{4}}=(\gamma_{3}+\gamma_{4}-\gamma_{1}-\gamma_{2})/2$
counts the net number of backscattered particles, $f_{\gamma_{i}}(E)=1/[e^{(E+\gamma_{i}eV/2)/T}\!+1]$
is the Fermi function shifted by $\pm eV/2$. A two-electron process
allows backscattering of one or two electrons. We will denote these
contributions as $\Delta G_{1,2}$: 
\begin{equation}
\Delta G=\Delta G_{1}+\Delta G_{2}\,.\label{eq:DeltaGEqualsDeltaG1plusDeltaG2}
\end{equation}

When the tunneling is weak and individual levels are well-defined,
$\Gamma_{nm}\ll|\varepsilon_{n}-\varepsilon_{n\pm1}|$ , the dot-dot
Green function can be calculated approximately by expanding Eq.~(\ref{eq:exactGreenFn}).
The leading order approximation in $\Gamma/\delta$ for diagonal and
off-diagonal parts of $\mathbf{G}^{R}$ is 
\begin{align}
G_{nn}^{R}(\omega)= & \frac{1}{\omega-\varepsilon_{n}+\frac{i}{2}\Gamma_{n}}\,,\label{eq:GreenFnNarrowDiag}\\
G_{nm}^{R}(\omega)= & -\frac{i}{2}\frac{\Gamma_{nm}}{(\omega-\varepsilon_{n}+\frac{i}{2}\Gamma_{n})(\omega-\varepsilon_{m}+\frac{i}{2}\Gamma_{m})},\, n\neq m\,.\label{eq:GreenFnNarrowOffDiag}
\end{align}
These equations are valid for all real frequencies $\omega$.

\subsection{The correction to the conductance at low temperature,  $T\ll\delta$ \label{sub:LowTwithoutEC}}

Let us first consider low temperatures, $T\ll\delta$. In linear response,
the energies of the external states in Eq.~(\ref{eq:S}) are restricted
by the Fermi functions and energy conservation to be within $T$ of
the Fermi energy. Therefore the backscattering current $\Delta I$
depends strongly on the position of the Fermi level which is controlled
by an external gate in experiments. This leads to peaks and valleys
in the conductance $\Delta G$ as a function of Fermi level position.
Peaks correspond to a level $n$ close to the Fermi level and their widths are determined by the temperature or the level widths, $|\varepsilon_{n}|\sim\max(T,\,\Gamma_{n})$. 
Here the level energy is measured from the Fermi energy. Note that
since charging effects are neglected in this section, the ground state
of the quantum dot is non-degenerate and always has zero total spin.

Consider first the valley conductance correction $\Delta G^{\text{valley}}$. In this case the energies of
external states in Eq.~(\ref{eq:S}) are far from the resonances.
Using the Green functions of Eqs.~(\ref{eq:GreenFnNarrowDiag}) and
(\ref{eq:GreenFnNarrowOffDiag}) the cross section for a process $\gamma\gamma\rightarrow\gamma_{3}\gamma_{4}$
is to leading order 

\begin{widetext}

\begin{equation}
\begin{aligned}S_{\gamma\gamma;\gamma_{3}\gamma_{4}}(E_{1},E_{2},E_{3},E_{4})= & \frac{1}{8\pi^{3}}\sum_{m_{i},n_{i}}\text{Im}G_{n_{3}m_{3}}^{R}(E_{3})\text{Im}G_{n_{4}m_{4}}^{R}(E_{4})(E_{1}-E_{2})^{2}\Gamma_{n_{1}m_{1}}\Gamma_{n_{2}m_{2}}\frac{(\varepsilon_{m_{1}}-\varepsilon_{m_{2}})(\varepsilon_{n_{1}}-\varepsilon_{n_{2}})}{\varepsilon_{m_{1}}^{2}\varepsilon_{m_{2}}^{2}\varepsilon_{n_{1}}^{2}\varepsilon_{n_{2}}^{2}}\\
\times & U_{m_{1}\gamma m_{2}\gamma;m_{3}\gamma_{3}m_{4}\gamma_{4}}^{*}U_{n_{1}\gamma n_{2}\gamma;n_{3}\gamma_{3}n_{4}\gamma_{4}}\,.
\end{aligned}
\end{equation}
The factor $(E_{1}-E_{2})^{2}$ results from antisymmetrizing over the
indices $(m_{1},m_{2})$, $(n_{1},n_{2})$ in the Green functions,
by using the fermionic property of the interaction matrix elements.
For backscattering of one particle ($\gamma_{3}\neq\gamma_{4}$ above),
which can be done in four ways, 
 one gets a conductance correction 
\begin{equation}
\begin{aligned}\Delta G_{1}^{\text{valley}}/G_{0}= & \frac{8\pi}{15}T^{4}\left|\sum_{n_{i}}U_{n_{1}\gamma n_{2}\gamma;n_{3}\gamma n_{4}\overline{\gamma}}\frac{\sqrt{\Gamma_{n_{1}}\Gamma_{n_{2}}\Gamma_{n_{3}}\Gamma_{n_{4}}}(\varepsilon_{n_{1}}-\varepsilon_{n_{2}})}{\varepsilon_{n_{1}}^{2}\varepsilon_{n_{2}}^{2}\varepsilon_{n_{3}}\varepsilon_{n_{4}}}\right|^{2}\,.\end{aligned}
\label{eq:G1-particlevalleyNocharging}
\end{equation}
For backscattering of two particles ($\gamma_{3}=\gamma_{4}=\overline{\gamma}$
above) we can also antisymmetrize over the indices $(m_{3},m_{4})$, $(n_{3},n_{4})$
resulting in an extra factor $(E_{3}-E_{4})^{2}$ giving 2 extra powers
of temperature in the conductance. 
We get
\begin{equation}
\begin{aligned}\Delta G_{2}^{\text{valley}}/G_{0}= & \frac{2\pi^{3}}{35}T^{6}\left|\sum_{n_{i}}U_{n_{1}\gamma n_{2}\gamma;n_{3}\overline{\gamma}n_{4}\overline{\gamma}}\sqrt{\Gamma_{n_{1}}\Gamma_{n_{2}}\Gamma_{n_{3}}\Gamma_{n_{4}}}\frac{(\varepsilon_{n_{1}}-\varepsilon_{n_{2}})(\varepsilon_{n_{3}}-\varepsilon_{n_{4}})}{\varepsilon_{n_{1}}^{2}\varepsilon_{n_{2}}^{2}\varepsilon_{n_{3}}^{2}\varepsilon_{n_{4}}^{2}}\right|^{2}\,.\end{aligned}
\label{eq:G2-particlevalleyNocharging}
\end{equation}
\end{widetext}The two-particle backscattering Eq.~(\ref{eq:G2-particlevalleyNocharging})
is thus suppressed by a factor $\sim T^{2}/\delta^{2}$ compared to single-particle contribution, Eq.~(\ref{eq:G1-particlevalleyNocharging}), at low temperatures.

Consider now the peak conductance correction $\Delta G^{\text{peak}}$. Let one of the levels, $\varepsilon_{1}$,
be near the Fermi energy but far from other levels, $|\varepsilon_{1}|\ll\delta$.
The cross section for $\gamma\gamma\rightarrow\gamma_{3}\gamma_{4}$
is dominated by processes that involve the level $\varepsilon_{1}$,\begin{widetext}

\begin{equation}
\begin{aligned}S_{\gamma\gamma;\gamma_{3}\gamma_{4}}(E_{1},E_{2},E_{3},E_{4})= & \frac{\pi}{2\pi^{4}}\sum_{m_{i},n_{i}}\text{Im}G_{n_{3}m_{3}}^{R}(E_{3})\text{Im}G_{n_{4}m_{4}}^{R}(E_{4})\frac{\Gamma_{n_{2}m_{2}}\Gamma_{1}}{\varepsilon_{n_{2}}\varepsilon_{m_{2}}}\frac{(E_{1}-E_{2})^{2}}{((E_{1}-\varepsilon_{1})^{2}+\frac{1}{4}\Gamma_{1}^{2})((E_{2}-\varepsilon_{1})^{2}+\frac{1}{4}\Gamma_{1}^{2})}\\
\times & U_{1\gamma m_{2}\gamma;m_{3}\gamma_{3}m_{4}\gamma_{4}}^{*}U_{1\gamma n_{2}\gamma;n_{3}\gamma_{3}n_{4}\gamma_{4}}\,.
\end{aligned}
\end{equation}
For one-particle backscattering we can choose $n_{3}=n_{4}=m_{3}=m_{4}=1$
 above to take full advantage of the resonant level. For two-particle backscattering
we can at most set $n_{3}=m_{3}=1$, leading to $\Delta G_{2}\ll\Delta G_{1}$
in Eq.~(\ref{eq:DeltaGEqualsDeltaG1plusDeltaG2}), as we will see below.

In the limit of relatively high temperature, $T\gg\Gamma$, the resonance condition allows us to replace $E_{i}\rightarrow\varepsilon_{1}$
(for all $i=1,\dots,4$ in the case of one and $i=1,\dots,3$ in the case
of two backscattered particles) in the Fermi functions of Eq.~(\ref{eq:CurrentGeneral}).
The Lorentzians are then easily integrated-over and $\Delta G$ at the peak ($|\varepsilon_{1}|\ll\delta$) is
\begin{equation}
\Delta G_{1}^{\text{peak}}/G_{0}=\frac{1}{2T}\frac{1}{\cosh^{4}(\varepsilon_{1}/2T)}\left|\sum_{n\neq1}\frac{\sqrt{\Gamma_{n}}}{\varepsilon_{n}}U_{1\gamma n\gamma;1\gamma1\overline{\gamma}}\right|^{2}\,,\quad T\gg\Gamma\,,\label{eq:DeltaG1PeakHighTNoEC}
\end{equation}
for one-particle backscattering, and 
\begin{equation}
\begin{aligned}\Delta G_{2}^{\text{peak}}/G_{0}= & \frac{1}{\pi}\frac{1}{\cosh^{2}(\varepsilon_{1}/2T)}\left|\sum_{n,m\neq1}\frac{\Gamma_{nm}}{\varepsilon_{n}\varepsilon_{m}}U_{1\gamma n\gamma;1\overline{\gamma}m\overline{\gamma}}\right|^{2}\end{aligned}
\,,\quad T\gg\Gamma\,,
\end{equation}
for two-particle backscattering.

In the opposite limit, $T\ll\Gamma$, the denominators of the cross
section depend weakly on the energies $E_{1,\dots,4}$. 
We find for backscattering of one particle,
\begin{equation}
\Delta G_{1}^{\text{peak}}/G_{0}=\frac{8\pi}{15}T^{4}\frac{\Gamma_{1}^{3}}{(\varepsilon_{1}^{2}+\frac{1}{4}\Gamma_{1}^{2})^{4}}\left|\sum_{n\neq1}\frac{\sqrt{\Gamma_{n}}}{\varepsilon_{n}}U_{1\gamma n\gamma;1\gamma1\overline{\gamma}}\right|^{2}\,,\quad T\ll\Gamma\,,
\label{eq:DeltaG1PeakLowTNoEC}
\end{equation}
 and for backscattering of two particles, 
\begin{equation}
\begin{aligned}\Delta G_{2}^{\text{peak}}/G_{0}= & \frac{32\pi^{3}}{35}T^{6}\frac{\Gamma_{1}^{2}}{(\varepsilon_{1}^{2}+\frac{1}{4}\Gamma_{1}^{2})^{4}}\left|\sum_{n,m\neq1}\frac{\Gamma_{nm}}{\varepsilon_{n}\varepsilon_{m}}U_{1\gamma n\gamma;1\overline{\gamma}m\overline{\gamma}}\right|^{2}\end{aligned}
\,,\quad T\ll\Gamma\,.
\end{equation}
\end{widetext} 
Comparing the one- and two-particle contributions in Eq.~(\ref{eq:DeltaGEqualsDeltaG1plusDeltaG2}), we see that $\Delta G_{1}^{\text{peak}}/\Delta G_{2}^{\text{peak}}\propto T\min(\Gamma,T)/\delta^{2}$.
At low temperatures the effects of two-particle backscattering are
negligible compared to those coming from one-particle backscattering.

The above results are valid for a single quantum dot and the conductance
correction is dominated by the one-particle backscattering processes. Averaging
over energy levels (see Appendix~\ref{sec:Averaging-over-disorder}), and replacing level widths  by their typical values,
we obtain from Eqs.~(\ref{eq:DeltaG1PeakHighTNoEC}) and (\ref{eq:DeltaG1PeakLowTNoEC}) the interpolation
\begin{align}
\frac{\Delta G^{\text{peak}}}{G_{0}} & \sim\frac{r_{1}}{2}\frac{1}{g^{2}}\frac{1}{\cosh^{4}(\varepsilon_{1}/2T)}\frac{\Gamma}{T}\Theta(T-\Gamma)\nonumber \\
 & +\frac{27}{16}r_{1}\frac{1}{g^{2}}\frac{T^{4}\Gamma^{4}}{(\varepsilon_{1}^{2}+\frac{1}{4}\Gamma^{2})^{4}}\Theta(\Gamma-T)\,.\label{eq:TypicalDeltaG1PeakLowTNoEC}
 \end{align}
 Similarly from Eq.~(\ref{eq:G1-particlevalleyNocharging}) we find
 \begin{equation}
\frac{\Delta G^{\text{valley}}}{G_{0}}\sim  \frac{27}{16}r_{2}\frac{1}{g^{2}}\left(\frac{\Gamma T}{\delta^{2}}\right)^{4}\,,\label{eq:TypicalDeltaG1ValleyLowTNoEC}
\end{equation}
where $\delta$ and $\Gamma$ are, respectively, the mean level spacing
and level width. 
We used $\langle U^{2}\rangle \sim \delta^2 / g^2$ for the average
interaction matrix elements, assuming screened Coulomb interaction and strong spin-orbit interaction,~\cite{Note8} see Appendix~\ref{sub:AverageU}. 
Here $g=E_T/\delta$ is the dimensionless conductance of the dot, $E_T$ being its Thouless energy.
The factors $r_{1}$ and $r_{2}$ are coming from level statistics, see
Appendix~\ref{sec:AveragesOverLevels}. The average of the above
results, Eqs.~(\ref{eq:TypicalDeltaG1PeakLowTNoEC}) and (\ref{eq:TypicalDeltaG1ValleyLowTNoEC}), over the Fermi level position, $\Delta G^{\text{av}}$,
is dominated by the peaks, and we get 
\begin{equation}
\frac{\Delta G^{\text{av}}}{G_{0}}\sim\frac{3}{2}r_{1}\frac{1}{g^{2}}\frac{\Gamma}{\delta}\Theta(T-\Gamma)+\frac{27}{2}r_{1}\frac{1}{g^{2}}\frac{T^{4}}{\Gamma^{3}\delta}\Theta(\Gamma-T)\,.\label{eq:DeltaGavLowTNoEC}
\end{equation}

\subsection{Crossover to higher temperatures, $T\gg\delta$\label{sub:HighTempWithoutEC}}

In the above we looked at low temperatures, $T\ll\delta$. Next, we
will study higher temperatures, where direct tunneling gives the dominant
contribution to backscattering, and peaks and valleys seen in the
previous subsection are washed out. We will see below that the crossover happens at $T\sim\delta/\ln(\delta/\Gamma)$. This is somewhat
similar to conventional quantum dot transport where the low temperature
transport is dictated by virtual elastic and inelastic processes (so-called
cotunneling), while thermally activated direct tunneling takes over
at higher temperatures.\cite{glazman05} In the case of a helical
edge coupled to a quantum dot the picture remains qualitatively the
same, except that elastic cotunneling is absent (see previous subsection).
We will now investigate what is the conductance correction $\Delta G$
due to direct tunneling.

Direct tunneling amounts to using only diagonal Green functions, Eq.~(\ref{eq:GreenFnNarrowDiag}) in the backscattering cross section,
Eq.~(\ref{eq:S}). One gets a correction to the linear conductance
\begin{widetext}
\begin{align}
 & \frac{\Delta G}{G_{0}}=\frac{1}{T}\sum_{n_{i}}\!\frac{\sum_{i=1\cdots4}\Gamma_{n_{i}}}{(\varepsilon_{n_{1}}+\varepsilon_{n_{2}}-\varepsilon_{n_{3}}-\varepsilon_{n_{4}})^{2}+\frac{1}{4}(\sum_{i=1\cdots4}\Gamma_{n_{i}})^{2}}\nonumber \\
 & \times\left\{ \left|U_{n_{1}Ln_{2}Ln_{3}Ln_{4}R}\right|^{2}\right.\left((2-f_{1}-f_{2}-f_{3}+f_{4})f_{1}f_{2}(1-f_{3})(1-f_{4})+(f_{1}+f_{2}+f_{3}-f_{4})f_{3}f_{4}(1-f_{1})(1-f_{2})\right)\nonumber \\
 & \qquad\left.+2\left|U_{n_{1}Ln_{2}Ln_{3}Rn_{4}R}\right|^{2}(2-f_{1}-f_{2}+f_{3}+f_{4})f_{1}f_{2}(1-f_{3})(1-f_{4})\vphantom{\left|U_{n_{1}Ln_{2}Rn_{3}Rn_{4}R}\right|^{2}}\right\} \,,\label{eq:DeltaGDirectTunneling}
\end{align}
\end{widetext}where $f_{i}\equiv f_{\gamma_{i}}(\varepsilon_{i})|_{V=0}$.
There are two distinct types of contributions to $\Delta G$. The
first one involves transitions within a pair of levels, $\{n_{3},n_{4}\}=\{n_{1},n_{2}\}$, 
and takes the full advantage of the resonant tunneling processes.
The other one involves more levels, $\{n_{3},n_{4}\}\neq\{n_{1},n_{2}\}$, 
and gains importance as the temperature increases due to the broadening
of the available phase space. At temperatures much lower than the average level
spacing, $T\ll\delta$, it is enough to consider only the two levels
nearest to Fermi energy. Then Eq.~(\ref{eq:DeltaGDirectTunneling})
reduces to 
\begin{align}
 & \frac{\Delta G}{G_{0}}=\frac{1/2T}{\cosh^{2}(\varepsilon_{1}/2T)\cosh^{2}(\varepsilon_{2}/2T)}\frac{1}{\Gamma_{1}+\Gamma_{2}}\nonumber \\
 & \times\left(\left|U_{1L2L1L2R}\right|^{2}+\left|U_{1L2L1R2L}\right|^{2}+2\left|U_{1L2L1R2R}\right|^{2}\right)\,.\label{eq:two-level}
\end{align}
Comparison with Eq.~(\ref{eq:DeltaG1PeakLowTNoEC}) shows that the
crossover from cotunneling to direct-tunneling occurs at $T\sim\delta/\ln(\delta/\Gamma)$.

At higher temperatures, $T\gg\delta$ (but still 
$T\ll E_{T}$ so that RMT-description is valid), many levels contribute in Eq.~(\ref{eq:DeltaGDirectTunneling}).
Using average values for dot parameters and screened Coulomb interaction, Eq.~(\ref{eq:UaverageCoulomb}), we find 
\begin{align}
\frac{\Delta G}{G_{0}}= & \frac{3c_{\gamma}}{2\pi^{2}}\frac{T}{g^{2}\Gamma}+4\pi c_{\gamma}\frac{T^{2}}{g^{2}\delta^{2}}\,.\label{eq:DeltaGavHighTScreenedCoulomb}
\end{align}
The first term in Eq.~(\ref{eq:DeltaGavHighTScreenedCoulomb}) comes from processes involving only a pair of levels in Eq.~(\ref{eq:DeltaGDirectTunneling}), while the second term involves four levels. 
It is instructive to interpret the first
term: To estimate
the backscattering current $\Delta I$, we note that the levels participating
in the backscattering processes are located within an energy strip
$\sim T$ around the Fermi level. The number of levels admitting,
say, two right-movers is $\sim(T/\delta)^{2}$. The rate of scattering
for the two-level process in the dot is $w\sim(\delta/g)^{2}\nu$, with
$\nu\sim1/\Gamma$ being the density of states for a tunnel-broadened
level. Finally, the imbalance $\sim eV/T$ between the numbers of
right- and left-movers is determined by the applied bias voltage $V$.
Collecting all of the above factors, we find 
\begin{equation}
\Delta I\sim\frac{eV}{T}\frac{T^{2}}{\delta^{2}}\frac{(\delta/g)^{2}}{\Gamma}\,,
\end{equation}
yielding, up to a numerical constant the first (two-level) term of
Eq.~(\ref{eq:DeltaGavHighTScreenedCoulomb}). The multi-level backscattering
(the last term) starts to dominate over the 2-level backscattering
at temperatures $T\sim\delta^{2}/\Gamma$. Its form is closely related
to the electron relaxation rate in the dot,\cite{sivan94} ${\tau_{e-e}^{-1}(T)\sim T^{2}/ g^{2} \delta}$,
as we discuss in Subsection \ref{sub:Kinetic-equation}.

Equations (\ref{eq:TypicalDeltaG1PeakLowTNoEC}), (\ref{eq:DeltaGavLowTNoEC}),
(\ref{eq:two-level}), (\ref{eq:DeltaGavHighTScreenedCoulomb}) contain seemingly
divergent contributions in the limit $\Gamma\rightarrow0$. These
terms describe resonant tunneling from the helical edge into the dot,
and the limiting factor to backscattering is the intra-dot interaction
$U$: Electrons tunnel frequently into the quantum dot but they also
leave fast and only those few who stayed long enough get backscattered
by $U$. In the limit $\Gamma\rightarrow0$ tunneling becomes weaker
and all tunneled electrons have time to scatter multiple times in
the dot. In this limit perturbation theory in $U$ breaks down
(as indicated by the divergences). Indeed, the above results were
obtained in the Born approximation, valid when the backscattering
matrix elements (those corresponding to processes with $\Delta N_{\gamma_{1}\gamma_{2};\gamma_{3}\gamma_{4}}\ne0$)
are small with respect to the level widths $\Gamma_{n}$ and form
the bottleneck for backscattering. In the opposite limit, $\Gamma\ll\sqrt{\langle U^{2}\rangle}$,
the bottleneck shifts to the tunneling in and out of the dot, and
the backscattering rate saturates at a value independent of $\langle U^{2}\rangle$.
The full crossover behavior is complicated in general; for a toy model
with only the two-electron backscattering matrix element present,
Eq.~(\ref{eq:two-level}) is generalized by the replacement (see Appendix
\ref{Adx:AppendixToymodel}): 
\begin{equation}
\left|U_{1L2L1R2R}\right|^{2}\to\frac{\frac{1}{4}(\Gamma_{1}+\Gamma_{2})^{2}\left|U_{1L2L1R2R}\right|^{2}}{\frac{1}{4}(\Gamma_{1}+\Gamma_{2})^{2}+\left|U_{1L2L1R2R}\right|^{2}}.\label{eq:ToymodelReplacement}
\end{equation}
Similarly, the validity of the first term in Eq.~(\ref{eq:DeltaGavHighTScreenedCoulomb})
requires $\Gamma\gg\sqrt{\langle U^{2}\rangle}$; in the opposite
limit of smaller $\Gamma$ this term would be replaced by a term $\sim T\Gamma/\delta^{2}$.

The Born approximation in the interaction may fail even if $\Gamma\gg\sqrt{\langle U^{2}\rangle}$
as one raises the temperature. At high temperatures the levels broaden
due to the large phase space available for electron-electron scattering.
For screened Coulomb interaction, the level broadening due to the
electron-electron scattering, $\tau_{e-e}^{-1}(E)\sim E^{2}/g^{2}\delta$,
exceeds the many-particle level spacing, $\sim\delta^{3}/T^{2}$, at
$T\gg g^{1/2}\delta$. In this ``high-temperature'' regime one may
replace the dot spectrum by a continuum.\cite{altshuler97} Such a 
replacement allows us to develop a kinetic equation approach and evaluate $\Delta G$ in the regimes beyond the one described by Eq.~(\ref{eq:DeltaGavHighTScreenedCoulomb}).

\subsection{The kinetic equation\label{sub:Kinetic-equation}}

In the kinetic equation approach it is useful to think of $\Delta G$ as the conductance of a quantum dot tunnel-coupled to two (left and right) fictitious spin-polarized leads, see Fig. \ref{Fig:3resistors}.
The leads model the helical edge, if we assume that
the left and right leads are reservoirs of respectively $L$- and
$R$-particles. The Kramers labels $R,\, L$ are 
conserved in tunneling into and out of the dot.
 For a non-zero steady-state current through the dot to exist,
there needs to be inelastic intra-dot spin-relaxation which converts
$R$- and $L$-particles into each other. 

Let $p_{n\gamma}$ denote the distribution function inside the dot
($\gamma$ labels the degenerate Kramers pairs, $n$ is the single-particle
level) and $f_{L,R}$ for the left and right leads. We
refer to the index $\gamma$ as just ``spin'' of the state. We write
$p_{n\gamma}$ as a sum of the equilibrium part $p_{n}^{(0)}$ (the Fermi distribution) and a small
deviation, $p_{n\gamma}=p_{n}^{(0)}+\delta p_{n\gamma}$. Without interaction in the
dot we have the linear rate equation 
\begin{equation}
\frac{dp_{n\gamma}}{dt}=\Gamma_{n\gamma}\left(f_{\gamma}(\varepsilon_{n})-p_{n\gamma}\right),\quad\gamma=L,R
\end{equation}
where $\Gamma_{n\gamma}=2\pi|t_{n\gamma}|^{2}\rho$, with $\rho=\sum_{k}\delta(E_{k}-\varepsilon)$
the edge density of states, which is assumed to be independent of
energy. In this case in steady state it is easy to see that the current
into the dot vanishes. To facilitate  comparison with conventional
quantum dot tunneling results, we will keep the index $\gamma$ in
the tunneling rates,\textit{ i.e.}, left and right leads have in general different
tunneling rates. In the physical system these rates are equal as dictated
by TR symmetry (see Appendix \ref{Adx:Tunneling-through-an}). 

\begin{widetext}The dot relaxation can be treated by the Fermi golden
rule. 
The inelastic contribution to the rate equation is 
\begin{align}
\left(\frac{d}{dt}p_{n\gamma}\right)_{e-e}= & 2\pi\sum_{n_{i},\gamma_{i}}|U_{n_{3}\gamma_{3}n_{4}\gamma_{4};n\gamma n_{2}\gamma_{2}}|^{2}\left[(1-p_{n\gamma})(1-p_{n_{2}\gamma_{2}})p_{n_{3}\gamma_{3}}p_{n_{4}\gamma_{4}}-p_{n\gamma}p_{n_{2}\gamma_{2}}(1-p_{n_{3}\gamma_{3}})(1-p_{n_{4}\gamma_{4}})\right]\nonumber \\
\times & \delta(\varepsilon_{n}+\varepsilon_{n_{2}}-\varepsilon_{n_{3}}-\varepsilon_{n_{4}}) \,;\label{eq:CollisionTermDiscrete}
\end{align}
\end{widetext}it vanishes in equilibrium $p_{n\gamma}\rightarrow p_{n}^{(0)}$.
The above equation becomes meaningful when averaged over realizations
of disorder, i.e., dot levels.\cite{sivan94} We define the average distribution function as $p_{\gamma}(E)=\langle\sum_{n}\delta(E-\varepsilon_{n})p_{n\gamma}\rangle/\nu_{0}$
where $\nu_{0}$ is the average density of levels 
 and $\langle\dots\rangle$ denotes
averaging over disorder. For convenience, we will also denote the
equilibrium distribution $p_{n}^{(0)}$ by $f$, where $f$ is
a Fermi function. Assuming small deviation from equilibrium distribution
at small bias, we can linearize the collision term Eq.~(\ref{eq:CollisionTermDiscrete})
to get

\begin{align}
 & \left(\frac{d}{dt}\delta p_{\gamma}(E_{1})\right)_{e-e}=-\tau_{e-e}^{-1}(E_{1})\delta p_{\gamma}(E_{1})\nonumber \\
 & +\sum_{\delta}\int dE_{2}K_{\gamma\delta}(E_{1},E_{2})\delta p_{\delta}(E_{2})\,.\label{eq:CollisionTermAveraged}
\end{align}
In the above equation we have split the averages
of products in Eq.~(\ref{eq:CollisionTermDiscrete}) into products
of averages, which is justified,\cite{altshuler97} if the level
broadening $\tau_{e-e}^{-1}(E)\sim E^{2}/(g^{2}\delta)$ exceeds the
many-particle level spacing, $\sim\delta^{3}/E^{2}$. This condition
leads to the constraint $T\gg g^{1/2}\delta$. Furthermore, we ignored
terms with $\{n,n_{2}\}=\{n_{3},n_{4}\}$ in Eq.~(\ref{eq:CollisionTermDiscrete}).
These formally diverging terms are regularized by the level broadening.
The broadening is provided by tunneling (at rate $\sim\Gamma$) and
by scattering to other levels in the dot. The scattering processes
with $\{n,n_{2}\}=\{n_{3},n_{4}\}$ are responsible for the first
term in Eq.~(\ref{eq:DeltaGavHighTScreenedCoulomb}). Ignoring them
limits the applicability of the kinetic equation (derived below) to
temperatures $T\gtrsim\delta^{2}/\Gamma$, see the discussion following
Eq.~(\ref{eq:DeltaGavHighTScreenedCoulomb}).

The two terms on the right-hand side of Eq.~(\ref{eq:CollisionTermAveraged})
are ``out'' and ``in'' contributions. In the ``out'' part, $\tau_{e-e}^{-1}(E_{1})$
gives the inverse lifetime of the state with energy $E_{1}$; it is
independent of spin because of TR symmetry. For the ``in'' term TR symmetry dictates that  $K_{\gamma\delta}=K_{\overline{\gamma}\overline{\delta}}$,
so that this kernel consists of only two independent elements, $K_{\gamma\delta}=K_{+}\delta_{\gamma\delta}+K_{-}(1-\delta_{\gamma\delta})$,
which characterize forward- and backscattering respectively. The explicit
expressions for $\tau_{e-e}^{-1}$ and $K_{\gamma\delta}$ obtained
from Eq.~(\ref{eq:CollisionTermDiscrete}) are (here $f$ is the equilibrium
Fermi function of the dot and energies are measured from the Fermi
level):\begin{widetext} 
\begin{equation}
\tau_{e-e}^{-1}(E_{1})=\pi\int dE_{2}dE_{3}dE_{4}\left\{ [1-f(E_{2})]f(E_{3})f(E_{4})+f(E_{2})[1-f(E_{3})][1-f(E_{4})]\right\} \sum_{\gamma_{i}}W_{\gamma_{1}\gamma_{2};\gamma_{3}\gamma_{4}}(E_{1},E_{2};E_{3},E_{4})\,,
\end{equation}
and
\begin{align}
K_{\gamma\delta}(E_{1},E_{2})= & 4\pi\int dE_{3}dE_{4}\left\{ [1-f(E_{1})][1-f(E_{3})]f(E_{4})+f(E_{1})f(E_{3})[1-f(E_{4})]\right\} \sum_{\alpha\beta}W_{\gamma\alpha;\delta\beta}(E_{1},E_{3};E_{2},E_{4})\nonumber \\
- & 2\pi\int dE_{3}dE_{4}\left\{ [1-f(E_{1})]f(E_{3})f(E_{4})+f(E_{1})[1-f(E_{3})][1-f(E_{4})]\right\} \sum_{\alpha\beta}W_{\gamma\delta;\alpha\beta}(E_{1},E_{2};E_{3},E_{4})\,.
\end{align}
Here
\begin{equation}
W_{\gamma_{1}\gamma_{2};\gamma_{3}\gamma_{4}}(E_{1},E_{2};E_{3},E_{4})=\langle\nu_{0}^{-1}\sum_{n_{i}}|U_{n_{3}\gamma_{3}n_{4}\gamma_{4};n_{1}\gamma_{1}n_{2}\gamma_{2}}|^{2}\prod_{i=1}^{4}\delta(E_{i}-\varepsilon_{n_{i}})\rangle\delta(E_{1}+E_{2}-E_{4}-E_{3})
\end{equation}
has the symmetries 
\begin{equation}
W_{\gamma_{1}\gamma_{2};\gamma_{3}\gamma_{4}}(E_{1},E_{2};E_{3},E_{4})=W_{\gamma_{2}\gamma_{1};\gamma_{3}\gamma_{4}}(E_{2},E_{1};E_{3},E_{4})=W_{\gamma_{1}\gamma_{2};\gamma_{4}\gamma_{3}}(E_{1},E_{2};E_{4},E_{3})=W_{\gamma_{3}\gamma_{4};\gamma_{1}\gamma_{2}}(E_{3},E_{4};E_{1},E_{2})\,,
\end{equation}
\end{widetext} and TR symmetry $W_{\gamma_{1}\gamma_{2};\gamma_{3}\gamma_{4}}=W_{\overline{\gamma_{1}}\overline{\gamma_{2}};\overline{\gamma_{3}}\overline{\gamma_{4}}}$.

For the special case of screened Coulomb interaction [Eq.~(\ref{eq:UaverageCoulomb})] we have 
\begin{equation}
W_{\gamma_{1}\gamma_{2};\gamma_{3}\gamma_{4}}(E_{1},E_{2};E_{3},E_{4})=\frac{c_{\gamma}}{8\pi^{2}}\frac{\delta(E_{1}+E_{2}-E_{4}-E_{3})}{g^{2}\delta}\,,
\end{equation}
and the scattering rate becomes \cite{sivan94} 
\begin{equation}
\tau_{e-e}^{-1}(E)=\frac{c_{\gamma}}{\pi g^{2}\delta}(\pi^{2}T^{2}+E^{2})\,.\label{eq:ScatteringrateScrndClmb}
\end{equation}

The number of particles and total energy are conserved, as is manifested in the following relations between the kernel $K_{\pm}$ and lifetime
$\tau_{e-e}$, 
\begin{align}
\tau_{e-e}^{-1}(E_{2})= & \int dE_{1}[K_{+}(E_{1},E_{2})+K_{-}(E_{1},E_{2})]\label{eq:NumberConservation}\,,\\
\tau_{e-e}^{-1}(E_{2})E_{2}= & \int dE_{1}E_{1}[K_{+}(E_{1},E_{2})+K_{-}(E_{1},E_{2})]\,.\label{eq:EnergyConservation}
\end{align}
In addition,
\begin{equation}
\tau_{\pm}^{-1}(E_{1})f'(E_{1})=\int dE_{2}K_{\pm}(E_{1},E_{2})f'(E_{2})\,,\label{eq:3rdIdentity}
\end{equation}
where $\tau_{\pm}^{-1}(E_{2})=\int dE_{1}K_{\pm}(E_{1},E_{2})$. Hence the collision term, Eq.~(\ref{eq:CollisionTermAveraged}), vanishes in equilibrium, as it should. Note
that $\tau_{e-e}^{-1}=\tau_{+}^{-1}+\tau_{-}^{-1}$, and $\tau_{-}^{-1}$
($\tau_{+}^{-1}$) is the rate of backscattering (forwardscattering).

Including the collision term we can write the full disorder-averaged
rate equation 
\begin{flalign}
\frac{dp_{\gamma}(E)}{dt}= & \Gamma_{\gamma}(E)[f_{\gamma}(E)-p_{\gamma}(E)]+\left(\frac{d}{dt}p_{\gamma}(E)\right)_{e-e}\,,\label{eq:FullAveragedRateEq}
\end{flalign}
where $\Gamma_{\gamma}=\langle\Gamma_{n\gamma}\rangle$ is generally
energy-dependent, and the last term is the intra-dot relaxation,
\begin{align}
\left(\frac{d}{dt}p_{\gamma}(E)\right)_{e-e} &=  \sum_{\gamma_{2}}\int dE_{2}\left[K_{\gamma\gamma_{2}}(E,E_{2})\delta p_{\gamma_{2}}(E_{2})\right. \label{eq:CollisionTermRateEq} \\
 & \left.-K_{\gamma_{2}\gamma}(E_{2},E)\delta p_{\gamma}(E)\right]\,;\quad p_{\gamma}=p^{(0)}+\delta p_{\gamma}\,. \nonumber
\end{align}
 In the steady state we have $dp_{\gamma}/dt=0$ in Eq.~(\ref{eq:FullAveragedRateEq}), hence 
\begin{align}
 & \Gamma_{\gamma}(E)[f_{\gamma}(E)-p_{0}(E)] =\Gamma_{\gamma}(E)\delta p_{\gamma}(E) \label{eq:SteadyStateEqForpGamma}\\
 & -\sum_{\gamma_{2}}\int dE_{2}\left[K_{\gamma\gamma_{2}}(E,E_{2})\delta p_{\gamma_{2}}(E_{2})-K_{\gamma_{2}\gamma}(E_{2},E)\delta p_{\gamma}(E)\right]\,.\nonumber
\end{align}

The current into the dot is
\begin{equation}
I=I_{L}=e\nu_{0}\Gamma_{L}\int dE[f_{L}(E)-p_{L}(E)]\,.
\label{eq:ILeftKinEqNoEC}
\end{equation}
It follows from particle number conservation, Eq.~(\ref{eq:NumberConservation}),
that $I_{L}=-I_{R}$. It is useful to write the current in the symmetric
way 
\begin{align}
 & I=\frac{1}{2}(I_{L}-I_{R})\\
 & =\!\frac{e\nu_{0}}{2}\! \int \!\! dE\left\{ \Gamma_{L}[f_{L}(E)-p_{L}(E)]-\Gamma_{R}[f_{R}(E)-p_{R}(E)]\right\} .\nonumber 
\end{align}
Inserting the steady state equations (\ref{eq:SteadyStateEqForpGamma}) in the current, we get
\begin{equation}
I=e\nu_{0}\int dE_{2}\tau_{-}^{-1}(E_{2})[\delta p_{L}(E_{2})-\delta p_{R}(E_{2})] \,.\label{eq:CurrentInKineticEq}
\end{equation}
We see that the current is determined by the rate of backscattering
$\tau_{-}^{-1}$ and the deviation of the dot distribution function from that in equilibrium. Next, we will solve the steady-state
equation (\ref{eq:SteadyStateEqForpGamma}) for the difference $\delta p_{L}-\delta p_{R}$ in the case of equal tunneling rates, $\Gamma_{L}=\Gamma_{R}$, and strong spin-orbit scattering, $\tau_{so}\delta\ll1$, corresponding
to symplectic RMT ensemble assumed in the previous subsection.

With strong spin-orbit coupling the kernels for backward and forward
scattering are equal, $K_{+}=K_{-}$. As the tunneling rates are identical
for the two spin species, subtracting Eq.~(\ref{eq:SteadyStateEqForpGamma})
with $\gamma=R$ from that with $\gamma=L$ leads to 
\begin{flalign}
\delta p_{L}(E)-\delta p_{R}(E)= & \frac{\Gamma}{\Gamma+\tau_{e-e}^{-1}(E)}[f_{L}(E)-f_{R}(E)]\,.
\end{flalign}
 The conductance is 
\begin{equation}
\Delta G=\frac{e^{2}}{8T}\nu_{0}\int dE\frac{\tau_{e-e}^{-1}(E)\Gamma}{\Gamma+\tau_{e-e}^{-1}(E)}\frac{1}{\cosh^2 \frac{E}{2T}},\label{eq:GkinEqSmalltausoEqualtunnelingrates}
\end{equation}
where 
 $\nu_{0}=\delta^{-1}$. The
main contribution to the integral comes from $|E|\lesssim T$. For
screened Coulomb interaction, Eq.~(\ref{eq:ScatteringrateScrndClmb}),
the conductance is
\begin{equation}
\Delta G=\frac{e^{2}}{2}\frac{\Gamma}{\delta}Y(\frac{T^{2}}{g^{2}\Gamma\delta}),\label{eq:GkinEqSmalltausoEqualtunnelingratesScrndCmb}
\end{equation}
where $Y$ is a dimensionless integral, 
\[
Y(y)=\frac{1}{4}\int dx\frac{\frac{c_{\gamma}}{\pi}y(\pi^{2}+x^{2})}{1+\frac{c_{\gamma}}{\pi}y(\pi^{2}+x^{2})}\frac{1}{\cosh^2 \frac{x}{2}}\,.
\]
The low- and high-temperature limits of Eq.~(\ref{eq:GkinEqSmalltausoEqualtunnelingratesScrndCmb}) are
\begin{equation}
\Delta G/G_{0}=\begin{cases}
2\pi c_{\gamma}T^{2}/3g^{2}\delta^{2}\,, & T^{2}/g^{2}\delta\ll \Gamma \,,\\
\Gamma/2\delta\,, & T^{2}/g^{2}\delta\gg \Gamma\,.
\end{cases}\label{eq:KineticConductanceFinal}
\end{equation}

\begin{figure}[h]
\includegraphics[width=0.9\columnwidth]{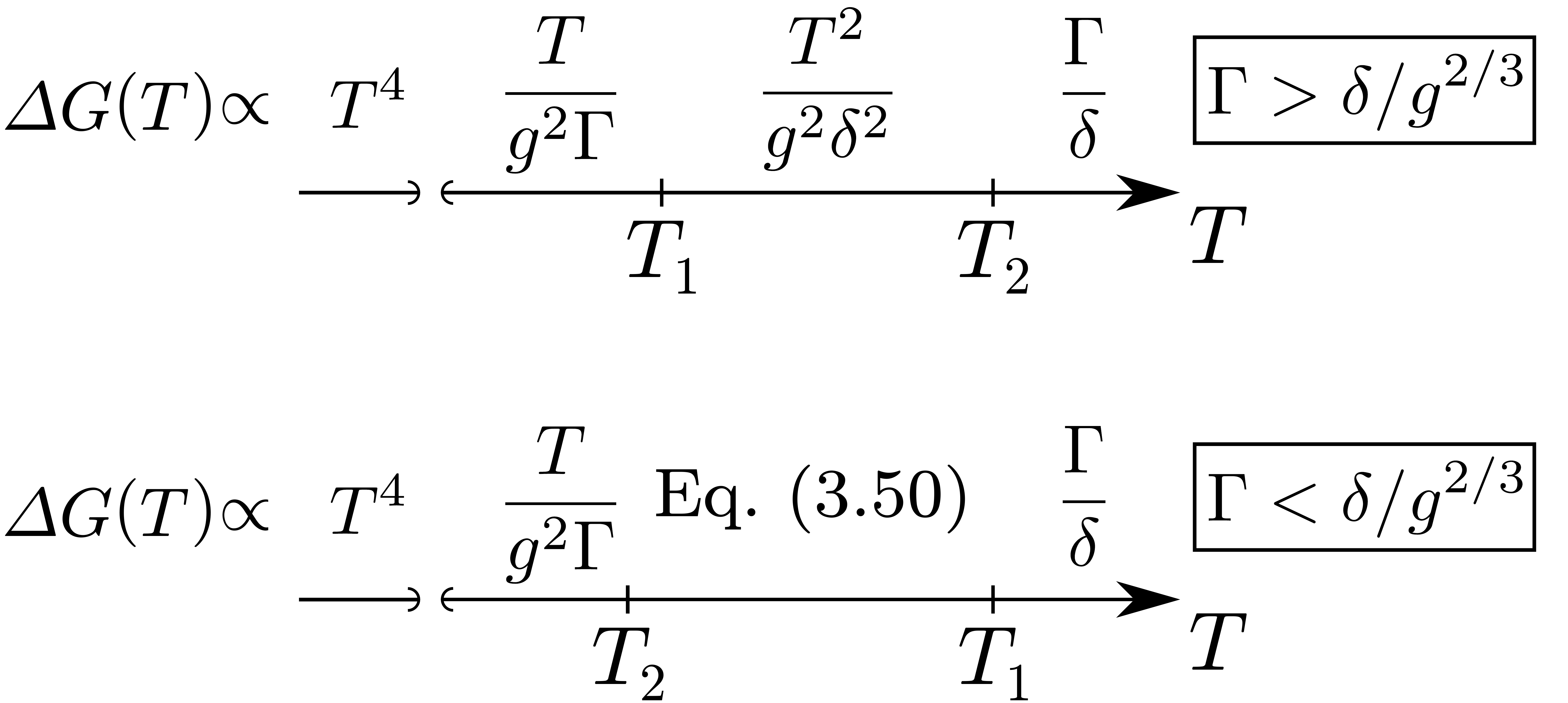}\caption{
The crossover from two-level dominant contribution in Eq.~(\ref{eq:DeltaGavHighTScreenedCoulomb}) to the high-temperature asymptote of Eq.~(\ref{eq:KineticConductanceFinal}). The order of the characteristic temperatures $T_1\sim \delta^2/\Gamma$ and $T_2 \sim g \sqrt{\Gamma \delta}$ depends on the parameter $g^{2/3}\Gamma/\delta$.} 
\label{Fig:TemperatureLine}
\end{figure}

Now we are ready to discuss the temperature
dependence of $\Delta G$ in a broader interval, covered by Eqs.~(\ref{eq:DeltaGavHighTScreenedCoulomb})
and (\ref{eq:KineticConductanceFinal}). As we already mentioned after
Eq.~(\ref{eq:DeltaGavHighTScreenedCoulomb}), the crossover from the
two-level contribution to the four-level dominated one in backscattering occurs at $T_{1}\sim\delta^{2}/\Gamma$.
 This latter contribution crosses over from
$\propto T^{2}$ to $T$-independent value at $T_{2}\sim g\sqrt{\Gamma\delta}$,
see Eq.~(\ref{eq:KineticConductanceFinal}) and Fig. \ref{Fig:TemperatureLine}. At $\Gamma\gtrsim\delta/g^{2/3}$
these two crossovers follow one after another, as $T_{1}<T_{2}$.
At $\Gamma\lesssim\delta/g^{2/3}$, the two characteristic temperatures
change their order, $T_{2}<T_{1}$, see Fig. \ref{Fig:TemperatureLine}. Under this condition, the broadening
of levels induced by interaction may affect the two-level contributions
described by the first term of Eq.~(\ref{eq:DeltaGavHighTScreenedCoulomb}).
The behavior of $\Delta G(T)$ can be assessed by the proper modification
of that term's interpretation given after Eq.~(\ref{eq:DeltaGavHighTScreenedCoulomb}).
Level broadening diminishes the density of states $\nu$ entering
in the transition rate $w$ from $\nu\sim1/\Gamma$ to $\nu\sim1/[\Gamma+\tau_{e-e}^{-1}(T)]$.
As the result, we find an interpolation 
\begin{equation}
\frac{\Delta G}{G_{0}}\sim\frac{T}{g^{2}[\Gamma+\tau_{e-e}^{-1}(T)]}\,,
\end{equation}
 which at $T\lesssim T_{2}$ matches, up to numerical factors, the first term in Eq.~(\ref{eq:DeltaGavHighTScreenedCoulomb}), 
and at $T\gtrsim T_{1}$ the high-temperature asymptote of Eq.~(\ref{eq:KineticConductanceFinal}).

In conclusion, above we investigated the temperature dependence of the conductance correction $\Delta G$ in the absence of charging effects. It has a universal asymptote, $\Delta G \propto T^4$, at low temperatures, and saturates to a constant value $\Delta G \sim \Gamma/\delta$ at fairly high temperatures, $T\gtrsim\delta$. We also saw that $\Delta G(T)$ varies substantially in between. There is no intermediate, parametrically large temperature interval where $\Delta G$ can be fairly approximated by a constant.
This is in contrast with a Coulomb blockaded quantum dot where, due to Kondo effect, there is a broad range of temperatures $T\lesssim\delta$ characterized by a weak $T$-dependence, as we will see in the next section.

\section{Coulomb blockade of the electron backscattering\label{sec:CoulombBlockade}}

In the previous section we considered the effect of weak interaction
in the quantum dot. It means that even the largest (and universal~\cite{aleiner02}) part of interaction (\ref{eq:IntradotInteraction}), \emph{i.e.},
charging energy, $E_{C}=\langle U_{n_{1}\gamma_{1}n_{2}\gamma_{2};n_{1}\gamma_{1}n_{2}\gamma_{2}}\rangle$, is small compared to level spacing $\delta$. Whether this is the case in a real-world experiment depends on the parameters in the measurement setup (see Subsection~\ref{sub:Potential-fluctuations-in} for more details). 
Generally, the capacitance of a quantum dot of linear size $w$ and distance $\ell_g$ from a gate can be approximated as $C\sim\kappa w \max(1,w/\ell_g)$, where $\kappa$ is the dielectric constant.
Similarly, the level spacing $\delta$ depends on the size of the dot, but also on the bulk band structure. We consider here a massive Dirac spectrum appropriate for HgTe near the band gap.~\cite{bernevig06} 
We denote the effective mass by $m^{*}=E_g /2v^2$, with $E_g$ being the band gap and $v$ the Dirac velocity.
We find that the average level spacing is $\delta\sim v/w^{2} \max(m^{*}v,k_F)$, interpolated here between the linear 
 ($k_F \gg m^{*}v$) and quadratic 
  ($k_F \ll m^{*}v$) parts of spectrum. (Here $k_F$ is the Fermi momentum.) 
Since $E_{C}=e^{2}/2C$, we get for the relative strength of charging energy,
\begin{equation}
E_{C}/\delta\sim \frac{\min(w,\ell_g)}{a_B} \max\left(1,\frac{k_F}{m^{*}v}\right)\,.
\label{eq:EcOverdelta}
\end{equation}
Here we introduced the effective Bohr radius $a_B = 2v /\alpha E_g$, where $\alpha = e^2/\kappa v$ is the effective fine-structure constant.
From Eq.~(\ref{eq:EcOverdelta}) we see that if the puddles are large and well separated from the gate electrode, $w,\ell_g > a_B$, the charging interaction becomes important.\footnote{
For example, for HgTe with $E_g \approx 20\, \text{meV}$, $v \approx {5.5\times 10^5\, \text{m/s}}$, and $\alpha \approx 0.3$ one finds~\cite{gusev13} $a_B \approx 120\, \text{nm}$.}
We therefore think that it is relevant to discuss the case where the condition $E_{C}\ll\delta$ of Section~\ref{sec:BSinAbsenceOfEC} is not met (see also Subsection~\ref{sub:Potential-fluctuations-in}).

In this section we turn to study strong charging effects, $E_{C}\gg\delta$,
corresponding to a large universal part of the interaction $U$ compared
to the level spacing.  Replacing these large diagonal matrix elements
by their random-matrix averages leads to the so-called universal Hamiltonian
of the isolated dot,\cite{aleiner02} 
\begin{equation}
H_{\text{dot}}=\sum_{n\gamma}\varepsilon_{n}c_{n\gamma}^{\dagger}c_{n\gamma}+E_{C}(\hat{N}-N_{g})^{2}\,,\label{eq:HdotWithEc}
\end{equation}
where $E_{C}=e^{2}/2C$ is the charging energy of the dot, $C$ being
its total capacitance. The tunable parameter $N_{g}=C_{g}V_{g}/e$
is controlled by the gate voltage $V_{g}$ applied between the metal gate
and the quantum well, where $C_{g}$ is the dot-gate capacitance. In
its ground state the dot is populated by the nearest-integer-to-$N_{g}$
number of electrons $N_{0}$. The charging energy required to add
an electron into the dot is $E_{+}=2E_{C}(N_{0}+\frac{1}{2}-N_{g})$, while the cost of removing one is $E_{-}=2E_{C}(N_{g}-N_{0}+\frac{1}{2})$.

Integer values of $N_{g}$ correspond to Coulomb blockade valley centers where the energy costs of adding or removing an electron are equal, $E_{\pm}=E_{C}$.
 In the even valleys ($N_{0}$ even) the correction to the helical edge conductance scales as $\Delta G\propto T^{4}$ at $T\lesssim \delta$.
 In the odd valleys ($N_{0}$ odd) $\Delta G$ exhibits a very different behavior because of a degenerate ground state due to the electron  spin.
  We will see that the degeneracy gives rise to an emergent scale $T_K$ (the Kondo temperature, $T_K\ll \delta$), and that the conductance correction $\Delta G$ becomes logarithmic in temperature at $T>T_{K}$.
At half-integer values of $N_{g}=N_0+1/2$, the dot states with $N_0$ and $N_0+1$ electrons are approximately degenerate.
This gives rise to a large conductance correction (a peak in $\Delta G$ vs. $N_g$ dependence), which has only a weak dependence on temperature at $T \gg \Gamma$.

Let us denote by $H_{0}$ the Hamiltonian of the decoupled helical edge and quantum dot, $H_{0}=H_{\text{edge}}+H_{\text{dot}}$.
We treat the non-universal part of the interaction (denoted by $U$ in this
section) and the tunneling between the edge and the dot as perturbations,

\begin{align}
 & H_{\text{tun}}+U\nonumber \\*
 & =\sum_{n,\gamma}t_{n}\left(c_{n\gamma}^{\dagger}\psi_{\gamma}(0)+\psi_{\gamma}^{\dagger}(0)c_{n\gamma}\right)\label{eq:PerturbationCoulombBlockade}\\
 & +\frac{1}{4}\sum_{n_{i},\gamma_{i}}U_{\{n_{i}\gamma_{i}\}}c_{n_{1}\gamma_{1}}^{\dagger}c_{n_{2}\gamma_{2}}^{\dagger}c_{n_{4}\gamma_{4}}c_{n_{3}\gamma_{3}}\,.\nonumber 
\end{align}
At low temperatures, $T\ll\delta$, and away from the peaks, the backscattering
current is dominated by high-order tunneling processes. 
 Generally the rate $r_{i\rightarrow f}$ of transition
from an initial state $|i\rangle$ to a final state $|f\rangle$ is
\begin{equation}
r_{i\rightarrow f}=2\pi\left|\langle f|\mathcal{T}(E_{i})|i\rangle\right|^{2}w_{i}\delta(E_{i}-E_{f})\,.\label{eq:RateTmatrixGeneral}
\end{equation}
Here $w_{i}$ is the thermal probability of the initial state $|i\rangle$.
It factorizes to parts corresponding to the dot, and the left and right Kramers states of the edge, ${w_{i}=w_{i}^{dot}w_{i}^{L}w_{i}^{R}}$.
The $T$-matrix $\mathcal{T}(E)$ is suited for high-order perturbation
theory, and satisfies 
\begin{equation}
\mathcal{T}(E)=(H_{\text{tun}}+U)+(H_{\text{tun}}+U)G_{0}(E)\mathcal{T}(E)\,,\label{eq:TmatrixDysonEq}
\end{equation}
where we denote $G_{0}(E)=(E-H_{0})^{-1}$. The solution can be written
as 
\begin{equation}
\mathcal{T}(E)=(H_{\text{tun}}+U)\sum_{n=0}^{\infty}[G_{0}(E)(H_{\text{tun}}+U)]^{n}\,.\label{eq:TmatrixUandHt}
\end{equation}

In the following sections we use this formalism to calculate the correction
to helical edge conductance, $\Delta G$, due to scattering off a
Coulomb-blockaded quantum dot. In Subsection \ref{sub:Odd-valleys-of}
we describe the conductance correction in the odd valley, and we show
that due to a degeneracy of the dot ground state, an effective exchange-type
interaction can be derived from the appropriate low-energy $T$-matrix.
The even valley of the Coulomb blockade is discussed in Subsection \ref{sub:Even-valley-conductance}.
There the ground state of the isolated dot has an even number of particles
and is thus unique. We use a low-energy $T$-matrix that accounts
for  the most important virtual processes. In Subsection \ref{sub:Peak-conductance.}
we find the conductance correction near the charge degeneracy point
(peak) between an even and an odd valley. The limit of high temperatures $T\gg\delta$ is discussed in Subsection \ref{sub:HighTCB}. In the same section we look at even higher temperatures, $T\gg E_{C}$, and make connection to Subsection \ref{sub:Kinetic-equation}.
 Finally, in Subsection \ref{sub:AverageCmbBckade} we calculate the average of the conductance correction over the gate voltage at low temperatures $T\ll \delta$.

\subsection{Odd Coulomb-blockade valleys \label{sub:Odd-valleys-of}}

In this section we discuss the low-temperature, $T\ll\delta$, backscattering
in the odd valleys where the gate voltage $V_{g}\propto N_{g}$ is
tuned so that the number of particles in the dot, $N_{0}$, is odd.
The ground state of the isolated dot is then doubly degenerate, corresponding
to the two Kramers states of the odd electron. These two states can
be viewed as those of a spin-1/2 particle, and a weak tunneling between
the edge and the dot results in an exchange coupling between the itinerant
electrons and this spin. The tensor of exchange coupling constants
is constructed in Subsection \ref{sub:Derivation-of-the}, by deriving
perturbatively the low-energy $T$-matrix. Because of  virtual processes
violating spin conservation in the dot, the exchange coupling tensor
generally cannot be diagonalized by a rotation of the dot effective
spin. We show in Subsection \ref{sub:The-correction-to} that this
\emph{anisotropic} exchange coupling results in backscattering of
electrons. To assess the correction to the conductance, we first use a
Bloch equation to calculate the steady-state expectation value of
the dot spin in Subsection \ref{sub:The-Bloch-equation}. In Subsection
\ref{sub:validityOfPerturbation} we show that our perturbative approach
is valid at temperatures much larger than an emergent temperature
scale, the Kondo temperature $T_{K}$. Finally, as already mentioned,
the correction to the conductance introduced by a spin-carrying dot is
evaluated in Subsection \ref{sub:The-correction-to}. In the same section
we also show how to extend our results to near and below the Kondo
temperature, $T\gtrsim T_{K}$ and $T\ll T_{K}$, respectively.

\subsubsection{Derivation of the effective Hamiltonian.\label{sub:Derivation-of-the}}

In this section we derive the low-energy Hamiltonian of the tunnel-coupled
interacting edge-dot system. We start with the full Hamiltonian $H=H_{0}+H_{\text{tun}}+U$,
where $H_{0}$ is the Hamiltonian of the decoupled edge and dot {[}charging
energy included, Eq.~(\ref{eq:HdotWithEc}){]}, $U$ is the non-universal
part of the intra-dot interaction, and $H_{\text{tun}}$ is the tunnel
coupling between the edge and the dot. The weak perturbations $H_{\text{tun}}$
and $U$ can create high-energy excitations above the ground state
of $H_{0}$. We perturbatively project out these high-energy states
(of energies $\gtrsim D$), leaving us with a new Hamiltonian valid
in a strip of energies $|E-E_{F}|\ll D$ around the edge Fermi level.
At low temperatures $T\ll D$ this new Hamiltonian accurately describes
the exchange interaction between the edge and the dot. To classify
these high-energy states, we start by discussing the ground state
and excitation spectrum of $H_{0}$.

By tuning the gate voltage so that $N_{g}$ is close to an odd integer
$N_{0}$, the ground state of the dot has a single electron on the
highest occupied level $\varepsilon_{1}$ (note that $\varepsilon_{1}<0$, 
as the energy is measured from the Fermi level). This ground state
is doubly degenerate because of the Kramers degeneracy of that level, and we denote the two ground states by $|\uparrow \rangle = |L \rangle$ and $|\downarrow \rangle = |R \rangle$. The perturbations $H_{\text{tun}}$ and $U$ create excitations from the ground state of $H_{0}$ describing
the decoupled edge-dot system. The respective excitations correspond to removal/addition of an electron from/into the dot, and to creation of an electron-hole pair in the dot.
The energies of these excitations are of order $E_{\pm}={2E_{C}|N_{g}-(N_{0}\pm\frac{1}{2})|}$ and $\delta$,
respectively, which prompts us to take $D\lesssim\min(E_{\pm},\delta)$.
Then at energies within the band $D$ such excitations only occur
in virtual processes, allowing a simplification of the $T$-matrix
(\ref{eq:TmatrixUandHt}).

The low-energy $T$-matrix can be derived perturbatively from the
Dyson equation (\ref{eq:TmatrixDysonEq}) by separating the high- and
low-energy states.~\cite{anderson70} For this we introduce projectors
$P_{\gamma}$, and $P_{0,\pm1}$. The first one, $P_{\gamma}$, projects the dot to its ground state with spin $\gamma$.
The three other projectors project to the high-energy subspace:
$P_{0}$ projects to states with at least one particle-hole excitation above the dot ground state, but no change in dot particle number.
Projectors $P_{\pm1}$ project to states where the dot has an excess (+1) or deficit (-1) of one particle (and possibly particle-hole excitations). 
To lowest order in tunneling we do not need to consider virtual dot states with more than one particle added or removed. 
Projecting Eq.~(\ref{eq:TmatrixDysonEq})
into the low-energy subspace and casting the resulting equation in the form
of Eq.~(\ref{eq:TmatrixDysonEq}) with a modified perturbation, $H_{\text{eff}}$,
and considering only the first order in $U$ and second order in tunneling
$H_{\text{tun}}$, we get \begin{widetext}
\begin{flalign}
(H_{\text{eff}})_{\gamma'\gamma}= & U_{\gamma'\gamma}+\sum_{p=\pm1}(H_{\text{tun}})_{\gamma'p}\frac{1}{E_{1}-(H_{\text{dot}})_{pp}}(H_{\text{tun}})_{p\gamma}\nonumber \\
+ & \sum_{p=\pm1}(H_{\text{tun}})_{\gamma'p}\frac{1}{E_{1}-(H_{\text{dot}})_{pp}}U_{pp}\frac{1}{E_{1}-(H_{\text{dot}})_{pp}}(H_{\text{tun}})_{p\gamma}\nonumber \\
+ & \sum_{p=\pm1}U_{\gamma'0}\frac{1}{E_{1}-(H_{\text{dot}})_{00}}(H_{\text{tun}})_{0p}\frac{1}{E_{1}-(H_{\text{dot}})_{pp}}(H_{\text{tun}})_{p\gamma}\nonumber \\
+ & \sum_{p=\pm1}(H_{\text{tun}})_{\gamma'p}\frac{1}{E_{1}-(H_{\text{dot}})_{pp}}(H_{\text{tun}})_{p0}\frac{1}{E_{1}-(H_{\text{dot}})_{00}}U_{0\gamma}\,.\label{eq:HeffWithProjectors}
\end{flalign}
\end{widetext}Here $E_{1}$ is the energy of the dot in its ground
state, and we denote $(H)_{ab}=P_{a}HP_{b}$ where $P_{a}$ is one
of the projectors $P_{\gamma}$, $P_{0}$, $P_{\pm1}$. We also used
the fact that at low temperatures $T\ll D$ we can neglect in the
denominators the energies $\sim T$ of the edge excitations in comparison
with those of the virtual states. From TR symmetry it follows
that the first term gives just a constant shift in energy, $U_{\gamma'\gamma}=U_{LL}\delta_{\gamma'\gamma}$,
and we will ignore it from now on. The second term is independent
of the interaction $U$, and gives the known isotropic anti-ferromagnetic
exchange interaction, see Eq.~(\ref{eq:J0Abbrev}).
The remaining terms in Eq.~(\ref{eq:HeffWithProjectors}) bring in the exchange anisotropy.

Considering only the exchange terms, we can write the effective Hamiltonian as 
\begin{equation}
H_{\text{eff}}=\sum_{i,j=x,y,z}S_{i}J_{ij}s_{j}\,,\label{eq:Heffective}
\end{equation}
where $\mathbf{S}$ is the spin-1/2 operator of the dot, and $\mathbf{s}$ is the edge spin density at the point contact, 
\begin{align}
\mathbf{s}= & \frac{1}{2}\sum_{\alpha,\beta}\psi_{\alpha}^{\dagger}(0)\boldsymbol{\sigma}_{\alpha\beta}\psi_{\beta}(0)\,.
\end{align}
We find that the exchange tensor obtained from Eq.~(\ref{eq:HeffWithProjectors})
can be written as $\mathbf{J}=\mathbf{R}(J_{0}+\delta\mathbf{J})$,
where $\mathbf{R}$ is a rotation matrix acting on the dot spin, and
$\delta\mathbf{J}$ is a lower triangular matrix (in the $x,\, y,\, z$
basis). In the basis of rotated dot spin, $\mathbf{S}\rightarrow\mathbf{R}\mathbf{S}$,
we get  
\begin{equation}
\mathbf{J}=J_0 \mathbf{1}+\left(\begin{array}{ccc}
\delta J_{xx} & 0 & 0\\
\delta J_{yx} & \delta J_{yy} & 0\\
\delta J_{zx} & \delta J_{zy} & \delta J_{zz}
\end{array}\right)\,.\label{eq:JtensorRotated}
\end{equation}
The full expressions for components $J_{0}$ and $\delta J_{ij}$
in terms of dot parameters are given in Table \ref{table-Jcomponents}
of Appendix \ref{sec:Components-of-the}.
 The leading contribution to $J_{0}$ is the familiar exchange coupling of the Anderson Hamiltonian,\cite{anderson66}
\begin{equation}
J_{0}=2t_{1}^{2}(\frac{1}{E_{-}-\varepsilon_{1}}+\frac{1}{E_{+}+\varepsilon_{1}})+\dots\,,
\label{eq:J0Abbrev}
\end{equation}
where $\dots$ denotes higher order corrections that are smaller than the main
term by a factor $U/\min(E_{\pm},\delta)$ {[}see Eq.~(\ref{eq:J0Full})
in Appendix \ref{sec:Components-of-the}{]}. As an example of the
anisotropic terms we give, 
\begin{flalign}
\delta J_{yx}= & -4\sum_{n,m<1}\frac{t_{m}t_{n}\,\text{Im}\,U_{nL1L;mR1R}}{(\varepsilon_{m}-E_{-})(\varepsilon_{n}-E_{-})}\nonumber \\
 & -4\sum_{n,m>1}\frac{t_{m}t_{n}\,\text{Im}\,U_{nL1L;mR1R}}{(\varepsilon_{m}+E_{+})(\varepsilon_{n}+E_{+})}\nonumber \\
 & +8\sum_{n<1}\sum_{m>1}\frac{t_{m}t_{n}\,\text{Im}\,U_{nL1L;mR1R}}{(\varepsilon_{n}-\varepsilon_{m})(\varepsilon_{n}-E_{-})}\nonumber \\
 & +8\sum_{n<1}\sum_{m>1}\frac{t_{m}t_{n}\,\text{Im}\,U_{nL1L;mR1R}}{(\varepsilon_{n}-\varepsilon_{m})(\varepsilon_{m}+E_{+})}\,.
 \label{eq:Jyx}
\end{flalign}
Generally the isotropy breaking terms $\delta J_{ij}$ are smaller
by a factor $U/\max(E_{\pm},\delta)$ compared to $J_{0}$.   We
will see below that the isotropic part of $\mathbf{J}$ will not contribute
to backscattering.

\subsubsection{The Bloch equation and its steady-state solutions.\label{sub:The-Bloch-equation}}

Passing a current along a helical edge leads to spin polarization
of a quantum dot coupled to the edge by the exchange interaction. The
spin polarization $\langle{\bf S\rangle}$, in turn, affects the backscattering.
In this section, we derive the Bloch equations~\cite{bloch46} for $\langle{\bf S\rangle}$.
We use the standard scheme for the derivation. First we use the Heisenberg
equation of motion to relate $d\langle{\bf S}\rangle/dt$ to the higher-order
correlators $\langle S_{i}s_{j}\rangle$ (here $s_{j}$ is the operator
of spin density of the helical edge at the point of contact). Second,
we express the correlators $\langle S_{i}s_{j}\rangle$ in terms $\langle S_{i}\rangle$
and $\langle s_{j}\rangle$ perturbatively, to the first order in
$J$. That procedure yields the spin relaxation rates to the
second (lowest non-vanishing) order of perturbation theory in $J$.
To illustrate, using Eqs.~(\ref{eq:Heffective}) and  (\ref{eq:JtensorRotated}) we get, for example,
\begin{align}
 & \frac{d}{dt}\left\langle S_{y}\right\rangle =(J_{0}+\delta J_{zz})\rho eV\left\langle S_{x}\right\rangle -(J_{0}+\delta J_{xx})\left\langle S_{z}:s_{x}:\right\rangle \nonumber \\
 & +\left\langle S_{x}(\delta J_{zx}:s_{x}:+\delta J_{zy}:s_{y}:+(J_{0}+\delta J_{zz}):s_{z}:)\right\rangle \,,\label{eq:dSy/dt}
\end{align}
where $:s_{i}:=s_{i}-\langle s_{i}\rangle$ has zero average, and
the first term in Eq.~(\ref{eq:dSy/dt}) comes from $\langle s_{i}\rangle=\delta_{iz}\rho eV$,
where $\rho$ is the edge density of states per spin, $\rho=\sum_{k}\delta(E_{k}-E)$.
Here we have chosen the spin quantization axis of the edge electrons at the Fermi level to be along $\mathbf{z}$. Expressions similar  to Eq.~(\ref{eq:dSy/dt}) hold for derivatives of $\langle S_{x}\rangle$ and $\langle S_{z}\rangle$. Next, we use lowest-order
perturbation theory in $J$ to evaluate the right-hand side in Eq.
(\ref{eq:dSy/dt}). For any operator ${\hat{O}_{kk'}=\sum_{\gamma\gamma'}\mathbf{O}_{kk'}^{\gamma\gamma'}\cdot\hat{\mathbf{S}}:\hat{d}_{k\gamma}^{\dagger}\hat{d}_{k'\gamma'}:}$
we have 
\begin{flalign}
\left\langle \hat{O}_{kk'}(t)\right\rangle = & i\int_{-\infty}^{t}dt'e^{-i(E_{k}-E_{k'}+i0)(t'-t)}\nonumber \\
\times & \left\langle [\hat{H}_{\text{eff}}(t'),\hat{O}_{kk'}(t')]\right\rangle \nonumber \\
\approx & -\frac{\left\langle [\hat{H}_{\text{eff}}(t),\hat{O}_{kk'}(t)]\right\rangle }{(E_{k}-E_{k'}+i0)}\,.\label{eq:QuarticAveraging}
\end{flalign}
 Using the above formula we get, for example, that
\begin{flalign}
\left\langle S_{x}:s_{y}:\right\rangle = & -\frac{1}{8}\pi(J_{0}+\delta J_{xx})\rho^{2}eV\nonumber \\
 & -\frac{1}{4}(2\pi \, \delta J_{zy}\rho^{2}T-\delta J_{zx}\mathcal{I})\left\langle S_{y}\right\rangle \nonumber \\
 & +\frac{1}{4}(2\pi(J_{0}+\delta J_{yy})\rho^{2}T-\delta J_{yx}\mathcal{I})\left\langle S_{z}\right\rangle \,,\label{eq:Sxsy}
\end{flalign}
where 
\begin{align}
\mathcal{I}= & \sum_{kk'}\frac{f_{k'L}(1-f_{kR})-f_{k'R}(1-f_{kL})}{E_{k}-E_{k'}}\nonumber \\
\approx & 2eV\rho^{2}\ln\frac{T}{D}\,.\label{eq:IntegralWithoutB}
\end{align}
The last equality is the expansion of $\mathcal{I}$ for small bias
$V$, and $D$ is the bandwidth. The logarithmic term here is a variety of a Kondo correction.~\cite{hewson97} We will ignore these corrections in this section. The justification for this is given in the next section. In Subsection \ref{sub:The-correction-to} we show how the Kondo effect is accounted for.

Combining Eqs.~(\ref{eq:dSy/dt}) and (\ref{eq:Sxsy}), and similar
equations for other components of $\langle{\bf S}\rangle$ and pair
averages, we find Bloch equation in the form
\begin{equation}
\frac{d}{dt}\left\langle \mathbf{S}\right\rangle =\mathbf{h}\times\left\langle \mathbf{S}\right\rangle -\boldsymbol{\gamma}\left\langle \mathbf{S}\right\rangle +\mathbf{c}\,,\label{eq:BlochEqWithoutB}
\end{equation}
with
\begin{equation}
\mathbf{h}=J_{0}\rho eV\mathbf{z} + \delta J_{zz}\rho eV\mathbf{z}\,,\label{eq:BlochEqNoBh}
\end{equation}
where $\mathbf{z}$ is the unit vector along the spin quantization
axis of the edge electrons at the Fermi level,
\begin{align}
\mathbf{c}&=\frac{\pi}{4}J_{0}\rho^{2}eV J_{0}\mathbf{z} \\
&+\frac{\pi}{4}J_{0}\rho^{2}eV[(\delta J_{xx}+\delta J_{yy})\mathbf{z}-\delta J_{zx}\mathbf{x}-\delta J_{zy}\mathbf{y}]\,, \nonumber
\label{eq:BlochEqNoBc}
\end{align}
and, \begin{widetext}
\begin{equation}
\boldsymbol{\gamma}=\pi J_{0}^2\rho^{2}T\mathbf{1} +\frac{\pi}{2}J_{0}\rho^{2}T\left(\begin{array}{ccc}
2(\delta J_{yy}+\delta J_{zz}) & -\delta J_{yx} & -\delta J_{zx}\\
-\delta J_{yx} & 2(\delta J_{xx}+\delta J_{zz}) & -\delta J_{zy}\\
-\delta J_{zx} & -\delta J_{zy} & 2(\delta J_{xx}+\delta J_{yy})
\end{array}\right)\,.\label{eq:BlochEqNoBGamma}
\end{equation}
\end{widetext} Here $\mathbf{h},\,\mathbf{c},$ and $\boldsymbol{\gamma}$
are given to linear order in both the bias voltage and the deviation $\delta J_{ij}$ from isotropic exchange.

The steady-state expectation value of $\mathbf{S}$ is obtained by
setting the left-hand-side of Eq.~(\ref{eq:BlochEqWithoutB}) to zero.
For small bias $|\mathbf{h}|,|\mathbf{c}|\propto V$, and thus $\mathbf{S} \propto V$ as well, hence to linear order in the bias voltage the cross-product term in Eq.~(\ref{eq:BlochEqWithoutB}) can be neglected and we can approximate the
steady-state value of the spin as $\left\langle \mathbf{S}\right\rangle =\boldsymbol{\gamma}^{-1}\mathbf{c}$.
Expanding to first order around the isotropic $\mathbf{J}$, we have
\begin{equation}
\left\langle \mathbf{S}\right\rangle =\frac{eV}{4T}\mathbf{z}-\frac{eV}{8TJ_{0}}(\delta J_{zx}\mathbf{x}+\delta J_{zy}\mathbf{y})\,.\label{eq:SpinSteadyState}
\end{equation}
The first term here is independent of $J_0$. This is a consequence of equilibrium being established between the dot and the edge with unequal populations of left and right movers at $\delta J\to 0$, see Eq.~(\ref{eq:Gibbs}) and the related discussion. The two other terms in Eq.~(\ref{eq:SpinSteadyState}) have kinetic origin associated with the 
anisotropic part of $J_{ij}$.

\subsubsection{Validity of perturbation theory, the Kondo temperature $T_{K}$\label{sub:validityOfPerturbation}}

The perturbatively evaluated ``bare'' exchange constant, Eq.~(\ref{eq:J0Abbrev}), is small, $J_0\rho\ll 1$. In the conventional magnetic impurity problem, the renormalized exchange constant increases with the energy (or temperature) being lowered below $D$, due to the Kondo effect.~\cite{hewson97} This is also the case here, as can be seen from the first two corrections in powers of $J_0$ to $\langle S_{z}\rangle$. This average can be evaluated by thermodynamic perturbation theory applied to the Gibbs distribution Eq.~(\ref{eq:Gibbs}). Similar to the conventional Kondo problem, the corrections to $\langle S_{z}\rangle$ in Eq.~(\ref{eq:SpinSteadyState}) are small when $J_0 \rho \ll 1$ and~\cite{hewson97}
\begin{equation}
J_{0}\gg J_{0}^{2}\rho\ln\frac{D}{T}\,.
\end{equation}
This equation defines the Kondo temperature
$T_{K}$, 
\begin{equation}
T\gg De^{-\frac{1}{|J_{0}\rho|}}\sim T_{K}\,,\label{eq:TbiggerthanTK}
\end{equation}
which depends on the gate voltage. To lowest order in the non-universal
part of intra-dot interaction we have [see Eq.~(\ref{eq:J0Abbrev})]
\begin{equation}
|J_{0}\rho|=\frac{1}{\pi}\Gamma_{1}(\frac{1}{E_{-}-\varepsilon_{1}}+\frac{1}{E_{+}+\varepsilon_{1}})\,,
\label{eq:rhoJ0}
\end{equation}
($\Gamma_{1}=2\pi t_{1}^{2}\rho$ is the width of the level $\varepsilon_{1}$.)
The Kondo temperature can be written as (we absorb $\pm\varepsilon_{1}$ into $E_{\pm}$)
\begin{equation}
T_{K}\sim\min(\delta,E_{\pm})\exp[-\frac{\pi E_{-}E_{+}}{2\Gamma_{1}E_{C}}]\,.\label{eq:KondoTempDef}
\end{equation}
We always have $E_{-}E_{+}\leq E_{C}^{2}$ which gives a lower bound,
\begin{equation}
T_{K}^{\text{min}}=\delta\exp(-\pi E_{C}/2\Gamma_{1})\,,\label{eq:TkMinDefinition}
\end{equation}
for the Kondo temperature, reached in the middle of the Coulomb blockade
valley. As one moves closer to a peak so that $E_{+}<\delta$, the
Kondo temperature becomes $T_{K}\sim E_{+}\exp(-\pi E_{+}/\Gamma_{1})$,
and has its maximum $T_{K}^{\text{max}}\sim\Gamma_{1}$ at $E_{+}=\Gamma_{1}/\pi$.
 
Our perturbation theory in $J_{ij}$ is thus valid when ${D\gg T\gg T_{K}}$, where both $T_{K}$ and $D=\min(E_{\pm},\delta)$ depend on gate voltage (through $E_{\pm}$). Near the peak $D\sim T_K \sim \Gamma$, and the assumption $J_0 \rho \ll 1$ fails, as can also be seen from Eq.~(\ref{eq:rhoJ0}). 
Likewise, the required condition on temperature cannot be satisfied at any gate voltage if $T<T_{K}^{\text{min}}$. At such low temperatures, $T\ll T_{K}$, the logarithmic corrections to $J_{0}$ become large. 
 In this strong coupling regime, one can write
a phenomenological Fermi liquid model as we demonstrate in the next subsection.\cite{nozieres74} 
The correction to the conductance in both the perturbative and non-perturbative regimes is discussed there.

\subsubsection{The correction to the  conductance\label{sub:The-correction-to}}

The coupling of the edge electrons to the spin of the dot $\mathbf{S}$
modifies the ideal conductance $G_{0}$ of the edge. The correction
$\Delta G$ to the conductance, is calculated, \emph{e.g.}, from the change
in the number of left moving electrons on the edge in the steady state.
Denoting $N_{\gamma}=\sum_{k}d_{k\gamma}^{\dagger}d_{k\gamma}$,
the backscattering current is $\Delta I=-\frac{1}{2}e\langle\frac{d}{dt}(N_{L}-N_{R})\rangle$.
It is convenient to add to the current a term $\langle\frac{d}{dt}S_{z}\rangle$
which has vanishing time-average since it is the time-derivative of
a bounded operator. Upon this addition, the current takes the form
\begin{equation}
\Delta I=-e\left\langle \frac{d}{dt}[\frac{1}{2}(N_{L}-N_{R})+S_{z}]\right\rangle \,.
\end{equation}
The advantage of the modification is in the fact that the operator
in the square brackets is an integral of motion if the exchange coupling
is isotropic; it's time variation is associated only with $\delta J\neq 0$:
\begin{align}
\Delta I= & e\left\langle S_{x}[(\delta J_{yy}-\delta J_{xx}):s_{y}:+\delta J_{yx}:s_{x}:]\right\rangle \nonumber \\
 & +e\left\langle S_{y}[(\delta J_{yy}-\delta J_{xx}):s_{x}:-\delta J_{yx}:s_{y}:]\right\rangle \nonumber \\
 & +e\left\langle S_{z}(\delta J_{zy}:s_{x}:-\delta J_{zx}:s_{y}:)\right\rangle \,. \label{eq:integralofmotion}
\end{align}

We can use Eq.~(\ref{eq:Sxsy}) and its companions to express the
averages in $\Delta I$ in terms of $\langle\mathbf{S}\rangle$. Inserting
the steady-state solution (\ref{eq:SpinSteadyState}) we get a correction
to the conductance 
\begin{align}
 & \frac{\Delta G}{G_{0}}=\frac{\pi\rho^{2}}{4}\left[(\delta J_{xx}-\delta J_{yy})^{2}+\delta J_{yx}^{2}+\frac{1}{4}(\delta J_{zx}^{2}+\delta J_{zy}^{2})\right]\,,\nonumber \\
 & \hphantom{=\frac{\pi\rho^{2}}{4}(\delta J_{xx}-\delta J_{yy})^{2}+\delta J_{yx}^{2}+(\delta J_{zx}^{2}+\delta J_{zy}^{2})}T\gg T_{K}\,.\label{eq:DeltaGOddValleyAboveTKNonRGd}
\end{align}
Here the components $\delta J_{ij}$ are given in Table \ref{table-Jcomponents}
of Appendix \ref{sec:Components-of-the}. They depend on the dot parameters
as well as on the gate voltage, and generally grow in magnitude towards the
peaks. We can extend the above formula to lower temperatures, $T\sim T_{K}$,
by using the standard renormalization group technique, see Appendix~\ref{sec:RG-for-anisotropic}.

As one moves towards low temperatures, the logarithmic corrections
to $\mathbf{J}$ become non-negligible, see Subsection \ref{sub:validityOfPerturbation}.
The anisotropic components of $\mathbf{J}$ (which enter $\Delta G$)
are irrelevant perturbations, as shown by the renormalization group
analysis carried out in Appendix \ref{sec:RG-for-anisotropic}. Accordingly
$\Delta G$ decreases with $T$:
\begin{align}
 & \Delta G/G_{0}=\frac{\pi\rho^{2}}{4}\left(\frac{\ln\frac{T}{T_{K}}}{\ln\frac{D}{T_{K}}}\right)^{2}\nonumber \\
 & \times\left[(\delta J_{xx}-\delta J_{yy})^{2}+\delta J_{yx}^{2}+\frac{1}{4}(\delta J_{zx}^{2}+\delta J_{zy}^{2})\right]\,,\label{eq:DeltaGOddValleyAboveTK}\\
 & \hphantom{\delta J_{xx}-\delta J_{yy})^{2}+\delta J_{yx}^{2}+(\delta J_{zx}^{2}+\delta J_{zy}^{2})}T>T_{K}\,.\nonumber 
\end{align}
Here $\delta J_{ij}$ is the ``bare'' exchange coupling and the
renormalization (``dressing'') is given by the logarithmic factor,
see Appendix \ref{sec:RG-for-anisotropic}. This factor is of
order unity at high temperatures, $T\gg T_{K}$,  but becomes small as $T$ approaches  $T_{K}$.

Below the Kondo temperature, $T\ll T_{K}$, the impurity spin $\mathbf{S}$
is strongly coupled to the itinerant electrons. To asses the correction
to the conductance at such temperatures, we can write a phenomenological
Fermi liquid Hamiltonian.\cite{nozieres74}
Assuming TR symmetry and neglecting Luttinger liquid effects,
there is no relevant perturbation that can cause backscattering in
this model. The least irrelevant term of that type is the one-particle
backscattering \cite{schmidt12,lezmy12} 
\begin{equation}
H_{\lambda}=\lambda[\psi_{L}^{\dagger}\partial_{x}\psi_{L}^{\dagger}\psi_{L}\psi_{R}
-\psi_{R}^{\dagger}\psi_{L}^{\dagger}\psi_{R}\partial_{x}\psi_{R}]
+\text{h.c.}\,,
\end{equation}
evaluated at the tunneling contact $x=0$. 
The correction to the conductance due to $H_{\lambda}$ is (see Section \ref{sec:Qualitative}) 
\begin{equation}
\Delta G/G_{0}\sim \lambda^{2} \rho^{4} T^{4},\quad T\ll T_{K}\,.\label{eq:TypicalDeltaGOddValleyBelowTk}
\end{equation}
By matching Eqs.~(\ref{eq:TypicalDeltaGOddValleyBelowTk}) and  (\ref{eq:DeltaGOddValleyAboveTK})
at $T\sim T_{K}$, we get an estimate for the phenomenological
parameter $\lambda$. Combining the two limits leads to the following
interpolation, 
\begin{align}
 & \frac{\Delta G}{G_{0}}=\frac{\pi\rho^{2}}{4}\left[(\delta J_{xx}-\delta J_{yy})^{2}+\delta J_{yx}^{2}+\frac{1}{4}(\delta J_{zx}^{2}+\delta J_{zy}^{2})\right]\nonumber \\
 & \times\frac{1}{\ln^{2}\frac{D}{T_{K}}}\left[\Theta(T_{K}-T)c\frac{T^{4}}{T_{K}^{4}}+\Theta(T-T_{K})\ln^{2}\frac{T}{T_{K}}\right]\,,\label{eq:TypicalDeltaGOddValleyCombined}
\end{align}
where the numerical coefficient $c\sim 1$. This equation is
valid at low temperatures, $T\ll D=\min(E_{\pm},\delta)$, where our
low-energy effective theory, Eq.~(\ref{eq:Heffective}), is justifiable.

Averaging over the dot parameters is detailed in Appendix \ref{sec:Averaging-over-disorder}.
Here we give the results for screened Coulomb interaction, ${\left\langle U^2\right\rangle \sim\delta^2/g^2}$.
Near either of the peaks, ${\min(E_{+},E_{-})\ll\delta}$, we arrive
at \begin{subequations}
\begin{align}
 & \langle\Delta G\rangle/G_{0}\sim\frac{1}{\ln^{2}\frac{D}{T_{K}}}\frac{\Gamma^{2}}{g^{2}\delta^{2}}\left[\Theta(T_{K}-T)c\frac{T^{4}}{T_{K}^{4}} \right.\nonumber \\
 & \left.+\Theta(T-T_{K})\ln^{2}\frac{T}{T_{K}}\right]\,,\quad T\ll D,\,\,D=\min(E_+,E_-)\,.\label{eq:TypicalDeltaGOddValley0}
\end{align}
Farther from the peaks ($E_{\pm}\gg\delta$) we get 
\begin{align}
 & \langle\Delta G\rangle/G_{0}\sim\frac{1}{\ln^{2}\frac{\delta}{T_{K}}}\frac{\Gamma^{2}}{g^{2}}\left[\frac{\ln\frac{E_{+}}{\delta}}{E_{+}^{2}}-\frac{\ln\frac{E_{+}E_{-}}{\delta^{2}}}{E_{+}E_{-}}+\frac{\ln\frac{E_{-}}{\delta}}{E_{-}^{2}}\right]\nonumber \\
 & \times\left[\Theta(T_{K}-T)c\frac{T^{4}}{T_{K}^{4}}+\Theta(T-T_{K})\ln^{2}\frac{T}{T_{K}}\right]\,,\quad T\ll \delta\,.\label{eq:TypicalDeltaGOddValley1}
\end{align}
The corrections to Eq.~(\ref{eq:TypicalDeltaGOddValley1}) are of order $\Gamma^{2}/g^{2}E_{\pm}^{2}$,
so that close to the middle of the valley ($|E_{+}-E_{-}|\ll E_{C}$,
$E_{\pm}\approx E_{C}$)
\begin{align}
 & \langle\Delta G\rangle/G_{0}\sim\frac{1}{\ln^{2}\frac{\delta}{T_{K}}}\frac{\Gamma^{2}}{g^{2}E_{C}^{2}}\nonumber \\
 & \times\left[\Theta(T_{K}-T)c\frac{T^{4}}{T_{K}^{4}}+\Theta(T-T_{K})\ln^{2}\frac{T}{T_{K}}\right]\,,\quad T\ll D\,.\label{eq:TypicalDeltaGOddValley2}
\end{align}
\end{subequations}Combining the above three formulas, the conductance
correction in the entire range between one of the peaks and the middle of the odd valley (gate voltages
$N_{0}\leq N_{g}\leq N_{0}+1/2$) can be accurately described by 
\begin{flalign}
 & \langle\Delta G\rangle/G_{0}\sim\frac{1}{\ln^{2}\frac{D}{T_{K}}}\frac{\Gamma^{2}}{g^{2}\max(E_{+}^{2},\delta^{2})}\label{eq:TypicalDeltaGOddValleyApprx}\\
 & \times\left[\Theta(T_{K}-T)c\frac{T^{4}}{T_{K}^{4}}+\Theta(T-T_{K})\ln^{2}\frac{T}{T_{K}}\right]\,,\quad T\ll D\,.\nonumber 
\end{flalign}
The derivation of Eqs.~(\ref{eq:TypicalDeltaGOddValley0})--(\ref{eq:TypicalDeltaGOddValleyApprx})
assumes that the dot is in the Kondo rather than the mixed valence regime
($T_{K}\ll\Gamma$). Since $T_K$ is a function of the gate voltage [see Eq.~(\ref{eq:KondoTempDef})], the condition on $E_{\pm}$ is $\min(E_+,E_-) > \Gamma$. The discussion of the opposite limit, $T_{K}\sim\Gamma$,
is deferred to Subsection (\ref{sub:Peak-conductance.}), where we study
the conductance correction at the peak.

\subsection{Even valley conductance\label{sub:Even-valley-conductance}}

Let us now discuss the conductance correction in the even valley,
where, unlike in the previous section, the dot ground state is unique.
Here the isolated quantum dot has an even number of electrons, $N_{0}$
for definiteness, in its ground state. At low temperatures $T\ll\delta$
we can neglect thermal excitations of the dot. When $U$, the non-universal part of the interaction, is weak, the elementary processes leading to backscattering
are the same as in Subsection \ref{sub:LowTwithoutEC} -- two electrons
scatter inelastically off the dot and in the process at least one
of them flips its spin. To lowest order in tunneling and in the interaction
$U$, these processes have amplitudes that are fourth order in tunneling
and first order in $U$ [c.f. Eqs.~(\ref{eq:G1-particlevalleyNocharging}),
(\ref{eq:G2-particlevalleyNocharging}) 
for the $E_C =0$ counterpart]. The corresponding amplitude, obtained from Eq.~(\ref{eq:TmatrixUandHt}), is of 5th order in the perturbations and
contains four energy denominators which are combinations of $\delta$,
$E_{\pm}$, and $E_{C}$, according to the excitation energy of the
corresponding virtual state (respectively, they are: creation of an
electron-hole pair, removal/addition of an electron, removal/addition
of two or more electrons). See also Table \ref{table-Tmatrix} in Appendix \ref{sec:Detailed-derivation-of}.

We will first consider gate voltages deep in the valley where $E_{\pm}\gg\delta$
(see next subsection). Closer to the peak, $\min(E_{+},E_{-})\sim\delta$,
backscattering is dominated by a different virtual process. In Subsection
\ref{sub:Cond-near-the-Pk-Even} we estimate the conductance correction
at the peak when approached from the even valley side.

\subsubsection{Conductance deep in the even valley.}

Away from the Coulomb blockade peaks, $E_{\pm}\gg\delta,\, T$, the
main contribution to the scattering amplitude comes from terms in Eq.~(\ref{eq:TmatrixUandHt}) with two denominators
of order $\delta$ so that in two of the virtual states the dot has
$N_{0}$ particles. The tunnelings then appear in ``in-out''
or ``out-in'' combinations, where ``in-out'' stands for insertion of one 
electron into the dot, followed by a removal of another (or vice versa for ``out-in''). Such tunneling terms
can be conveniently written as blocks $\tilde{H}=PH_{\text{tun}}G_{0}(E)H_{\text{tun}}P$,
where $P$ is a projector onto the low-energy subspace, that is, the
subspace of states with $N_{0}$ particles in the dot. There are two of these
blocks in the fifth-order contribution to $T$-matrix, and they can
appear in three combinations relative to the interaction $U$. As
an example, one of the three terms is, \begin{widetext}
\begin{flalign}
 & \langle f|\tilde{H}G_{0}(E_{i})\tilde{H}G_{0}(E_{i})U|i\rangle\nonumber \\
= & \left[\frac{1}{E_{-}}+\frac{1}{E_{+}}\right]^{2}\langle i|n_{k_{1}\gamma_{1}}n_{k_{2}\gamma_{2}}(1-n_{k_{3}\gamma_{3}})(1-n_{k_{4}\gamma_{4}})|i\rangle\sum_{n_{1}n_{2}n_{3}n_{4}}U_{\{n_{i}\gamma_{i}\}}t_{n_{1}}t_{n_{2}}t_{n_{3}}t_{n_{4}}\nonumber \\
 & \times\left[\frac{\Theta_{-n_{1}}\Theta_{-n_{2}}\Theta_{n_{3}}\Theta_{n_{4}}}{\Delta E_{1}+\Delta E_{2}-\Delta E_{3}-\Delta E_{4}}\frac{1}{\Delta E_{3}-\Delta E_{1}}+(1\leftrightarrow2)+(3\leftrightarrow4)\right]\,,\label{eq:ExampleTmatrixTerm}
\end{flalign}
\end{widetext}where $\Delta E_{i}=\varepsilon_{n_{i}}-E_{k_{i}}$,
$n_{k\gamma}=d_{k\gamma}^{\dagger}d_{k\gamma}$, and we considered
an arbitrary initial state $|i\rangle$ and a final state ${\langle f|=\langle i|d_{k_{4}\gamma_{4}}d_{k_{3}\gamma_{3}}d_{k_{1}\gamma_{1}}^{\dagger}d_{k_{2}\gamma_{2}}^{\dagger}}$
with no excitation left in the dot. We also abbreviated the step function,
$\Theta_{\pm n}=\Theta(\pm\varepsilon_{n}\mp\varepsilon_{1})$, where level 1 is the lowest unoccupied level, $\varepsilon_{1} + E_+ >0$.

The above equation is valid when $E_{\pm}\gg\delta$ as discussed
in the previous paragraph. When this condition is not satisfied and
$\delta\sim E_{\pm}$, the projector $P$ can no longer be used
to separate the high- and low-energy virtual states, and more terms
need to be taken into account in the expansion of the $T$-matrix. Thus,
when averaging (\ref{eq:ExampleTmatrixTerm}) over disorder we need
to insert a cut-off $\text{min}(E_{+},E_{-})$ on the level energies
$\varepsilon_{n}$, see Eq.~(\ref{eq:TypicalDeltaG1ValleyLowTWithECCmb})
and Appendix \ref{sec:Averaging-over-disorder}. 

The total backscattering current $\Delta I$ is obtained by summing
the rates (\ref{eq:RateTmatrixGeneral}) that cause backscattering.
The factors ${\langle i|n_{k_{i}\gamma_{i}}|i\rangle}$ (being either
zero or one) summed over initial states with weight $w_{i}$ yield
Fermi functions ${f_{k\gamma}=1/[e^{(E_{k}+\gamma eV/2)/T}+1]}$,
and we are led to an equation for $\Delta I$ similar to Eq.~(\ref{eq:CurrentGeneral})
(see Appendix \ref{sec:Detailed-derivation-of} for a detailed derivation).
Likewise, we get $\Delta G_{1}\propto T^{4}$ and $\Delta G_{2}\propto T^{6}$
as in Eqs.~(\ref{eq:G1-particlevalleyNocharging}) and (\ref{eq:G2-particlevalleyNocharging}),
so that $\Delta G_{2}\ll\Delta G_{1}$ at low temperatures. Neglecting
the two-particle contribution  we get \begin{widetext}
\begin{align}
\Delta G/G_{0}= & \frac{8\pi}{15}T^{4}\left[\frac{1}{E_{-}}+\frac{1}{E_{+}}\right]^{4}\left|\sum_{n_{1}\neq n_{2}}\sum_{n_{3}n_{4}}(U_{n_{1}Ln_{2}L;n_{3}Ln_{4}R}-U_{n_{1}Ln_{2}L;n_{3}Rn_{4}L})\sqrt{\Gamma_{n_{1}}\Gamma_{n_{3}}\Gamma_{n_{2}}\Gamma_{n_{4}}}\right.\nonumber \\
\times & \left\{ \frac{\Theta_{n_{1}}\Theta_{n_{2}}\Theta_{-n_{3}}\Theta_{-n_{4}}+\Theta_{-n_{1}}\Theta_{-n_{2}}\Theta_{n_{3}}\Theta_{n_{4}}}{(\varepsilon_{n_{1}}+\varepsilon_{n_{2}}-\varepsilon_{n_{3}}-\varepsilon_{n_{4}})}\frac{2\varepsilon_{n_{3}}-(\varepsilon_{n_{1}}+\varepsilon_{n_{2}})}{(\varepsilon_{n_{3}}-\varepsilon_{n_{1}})^{2}(\varepsilon_{n_{3}}-\varepsilon_{n_{2}})^{2}}(\varepsilon_{n_{1}}-\varepsilon_{n_{2}})\right.\nonumber \\
 & \left.\left.+\Theta_{-n_{1}}\Theta(\varepsilon_{n_{2}})\Theta_{n_{3}}\Theta_{-n_{4}}\frac{(\varepsilon_{n_{1}}-\varepsilon_{n_{2}}-\varepsilon_{n_{3}}+\varepsilon_{n_{4}})}{(\varepsilon_{n_{4}}-\varepsilon_{n_{2}})^{2}(\varepsilon_{n_{3}}-\varepsilon_{n_{1}})^{2}}\right\} \right|^{2}\,.\label{eq:DeltaG1EvenValley}
\end{align}
\end{widetext} 

When averaging $\Delta G$ over disorder, the sums over dot energy
levels diverge at the upper limit. This is a result of approximating
some of the energy denominators by $E_{\pm}$ in the $T$-matrix,
see the paragraph below Eq.~(\ref{eq:ExampleTmatrixTerm}). The divergence
is logarithmic and is cut-off by $\text{min}(E_{+},E_{-})$, see Appendix
\ref{sec:Averaging-over-disorder}. For screened Coulomb interaction
we have, 
\begin{align}
 & \langle\Delta G\rangle/G_{0}\nonumber \\
 & \sim\frac{\Gamma^{4}}{g^{2}\delta^{4}}T^{4}\left[\frac{1}{E_{-}}+\frac{1}{E_{+}}\right]^{4}\ln\frac{\text{min}(E_{+},E_{-})}{\delta}\,.\label{eq:TypicalDeltaG1ValleyLowTWithECCmb}
\end{align}

The above equation was derived for the two-level dot in Section \ref{sec:Qualitative}, Eq.~(\ref{eq:QualitDeltaGevenLowT}).
It is also the generalization of the earlier Eq.~(\ref{eq:TypicalDeltaG1ValleyLowTNoEC})
to a dot with charging energy $E_{C}\gg\delta$. In the Coulomb blockade
valley, changing the occupation of the dot requires a large energy
$E_{\pm}\sim E_{C}$. Consequently, two of the virtual states in a
lowest-order backscattering process have large denominators $\propto E_{C}$
instead of $\propto\delta$ as in the non-interacting case. Therefore
Eq.~(\ref{eq:TypicalDeltaG1ValleyLowTWithECCmb}), up to prefactors
(the logarithm is of order unity for $E_{C}\gtrsim\delta$), amounts
to replacing the amplitude $\propto1/\delta^{4}$ in Eq.~(\ref{eq:TypicalDeltaG1ValleyLowTNoEC})
by $\propto1/E_{C}^{2}\delta^{2}$. Our result (\ref{eq:TypicalDeltaG1ValleyLowTWithECCmb})
diverges at $N_{g}=N_{0}\pm\frac{1}{2}$ corresponding to the peaks
adjacent to the valley with $N_0$ electrons. As one moves close to the
peaks, eventually $\min(E_{+},E_{-})\sim\delta$ and our perturbation
theory breaks down {[}see the paragraph below Eq.~(\ref{eq:ExampleTmatrixTerm}){]}.
In this limit one needs to take into account additional contributions
to the $T$-matrix, which we deal with in the next subsection.

\subsubsection{Conductance near the peak\label{sub:Cond-near-the-Pk-Even}}

In the previous subsection we assumed that gate voltage is tuned
away from the charge degeneracy points of the quantum dot (peaks).

Let us now consider backscattering close to the peak separating the dot states of $N_{0}$ and $N_{0}+1$ electrons, with $N_{0}$ even (the peak separating $N_{0}$ and $N_{0}-1$ states is treated similarly). 
We will assume that we are on the even valley side of
the peak so that the energy required for adding (rather than removing) an electron is small, $0<{E_{+}+\varepsilon_{1}}\ll\delta$. Here $\varepsilon_{1}$
is the  lowest unoccupied level in the dot with $N_{0}$ electrons.
The correction to the conductance is calculated in the $T$-matrix formalism,
as was done in the previous subsection. We saw in Subsection \ref{sub:Odd-valleys-of}
that the correction to the conductance $\Delta G$ is largest in the odd valley
(rather than in the even). Hence, close to the peak on the even
side, most backscattering is caused by virtual tunneling to the odd-number
state.  For this reason, the backscattering close to the peak is
similar to that in the odd valley, described in Subsection \ref{sub:Odd-valleys-of}.
Close to the peak, the lowest energy excitation above the dot ground
state is to add an electron to the level $\varepsilon_{1}$. This excitation has energy $E_{+}+\varepsilon_{1}$ (disregarding the energy deficit of the hole created in the helical edge). 
 As a result, the main contribution to the $T$-matrix comes from virtual processes 
 in which an edge electron is first scattered to level $\varepsilon_1$, thus increasing the dot population to the odd value $N_0+1$, followed by scattering of another edge electron off the dot using the interaction $H_{\text{eff}}$ [Eq.~(\ref{eq:HeffWithProjectors})], and finalized by a tunneling of an electron from the dot back to the  edge.
  For initial and final states $|i\rangle$ and $|f\rangle$ the amplitude of such a process,
${\langle f|H_{\text{tun}}G_{0}H_{\text{eff}}G_{0}H_{\text{tun}}|i\rangle}$,
contains two large factors $G_{0}\sim1/\max(E_{+}+\varepsilon_{1},T)$.

For a generic final state $\langle f|={\langle i|d_{k_{4}\gamma_{4}}d_{k_{3}\gamma_{3}}d_{k_{1}\gamma_{1}}^{\dagger}d_{k_{2}\gamma_{2}}^{\dagger}}$,
the $T$-matrix $H_{\text{tun}}G_{0}H_{\text{eff}}G_{0}H_{\text{tun}}$
yields an amplitude
\begin{flalign}
 & \langle f|\mathcal{T}(E_{i})|i\rangle\nonumber \\
 & =\langle f|d_{k_{3}\gamma_{3}}^{\dagger}d_{k_{4}\gamma_{4}}^{\dagger}d_{k_{2}\gamma_{2}}d_{k_{1}\gamma_{1}}|i\rangle\Theta(E_{+}+\varepsilon_{1})t_{1}^{2}\nonumber \\
 & \sum_{n,m}\frac{t_{n}t_{m}}{(\varepsilon_{n}-\varepsilon_{m})}\left[\frac{\Theta_{-m}}{(\varepsilon_{1}-\varepsilon_{m})}-\frac{\Theta_{-n}}{(\varepsilon_{1}-\varepsilon_{n})}\right]\nonumber \\
 & \left\{ \frac{U_{1\gamma_{4}n\gamma_{3};1\gamma_{1}m\gamma_{2}}}{(\varepsilon_{1}+E_{+}-E_{k_{4}})(\varepsilon_{1}+E_{+}-E_{k_{1}})}\right.\nonumber \\
 & -\frac{U_{1\gamma_{4}n\gamma_{3};1\gamma_{2}m\gamma_{1}}}{(\varepsilon_{1}+E_{+}-E_{k_{4}})(\varepsilon_{1}+E_{+}-E_{k_{2}})}\nonumber \\
 & -\frac{U_{1\gamma_{3}n\gamma_{4};1\gamma_{1}m\gamma_{2}}}{(\varepsilon_{1}+E_{+}-E_{k_{3}})(\varepsilon_{1}+E_{+}-E_{k_{1}})}\nonumber \\
 & \left.+\frac{U_{1\gamma_{3}n\gamma_{4};1\gamma_{2}m\gamma_{1}}}{(\varepsilon_{1}+E_{+}-E_{k_{3}})(\varepsilon_{1}+E_{+}-E_{k_{2}})}\right\} \,.\label{eq:TmatrixElementEvenValley}
\end{flalign}
Here we replaced $E_{k_{i}},\, E_{+}+\varepsilon_{1}\rightarrow0$
in the denominators that are large $\sim\delta$, which is justified
at low temperatures and close to the peak, $T,\, E_{+}+\varepsilon_{1}\ll\delta$.
For one-particle backscattering, we get a conductance correction 
\begin{flalign}
 & \Delta G_{1}/G_{0}\nonumber \\
 & =\pi\frac{\Theta(E_{+}+\varepsilon_{1})}{64\pi^{4}}\frac{\Gamma_{1}^{2}}{T}\int dE_{1}dE_{2}dE_{3}dE_{4}\prod_{i=1}^{4}\frac{1}{\cosh\frac{E_{i}}{2T}} \nonumber \\
 & \times\frac{(E_{1}-E_{2})^{2}\delta(E_{1}+E_{2}-E_{3}-E_{4})}{(\varepsilon_{1}+E_{+}-E_{1})^{2}(\varepsilon_{1}+E_{+}-E_{2})^{2}}\nonumber \\
 & \times\left|\sum_{n,m}\frac{\sqrt{\Gamma_{n}\Gamma_{m}}}{\varepsilon_{n}-\varepsilon_{m}}\left( \frac{\Theta_{-m}}{\varepsilon_{1}-\varepsilon_{m}}-\frac{\Theta_{-n}}{\varepsilon_{1}-\varepsilon_{n}} \right) \right.\nonumber \\
 & \times\left.\left(\frac{U_{1LnL;1RmL}}{\varepsilon_{1}+E_{+}-E_{4}}-\frac{U_{1LnL;1LmR}}{\varepsilon_{1}+E_{+}-E_{3}}\right)\right|^{2}\,.\label{eq:DeltaG1pkWithEc}
\end{flalign}
The corresponding 2-particle contribution is given in Appendix \ref{sub:2-particle-near-Pk},
Eq.~(\ref{eq:DeltaG2pkWithEc}). The above integrals in $\Delta G_{1}$
are well-defined only at low temperatures and far enough from the peak,
$\varepsilon_{1}+E_{+}\gg T$, where the activation factors limit
the integration, and it is justified to replace $E_{i}\rightarrow0$
in the denominators. We then get 
\begin{flalign}
 & \Delta G_{1}/G_{0}=\frac{4\pi}{15}\frac{\Theta(E_{+}+\varepsilon_{1})T^{4}\Gamma_{1}^{2}}{(\varepsilon_{1}+E_{+})^{6}}\nonumber \\
 & \times\left|\sum_{n,m}\frac{\sqrt{\Gamma_{n}\Gamma_{m}}}{\varepsilon_{n}-\varepsilon_{m}}\left(\frac{\Theta_{-m}}{\varepsilon_{1}-\varepsilon_{m}}-\frac{\Theta_{-n}}{\varepsilon_{1}-\varepsilon_{n}}\right)\right.\nonumber \\
 & \times\left.\left(U_{1LnL;1RmL}-U_{1LnL;1LmR}\right)\vphantom{\left|U_{n_{1}L}\right|^{2}}\right|^{2}\,,\quad\varepsilon_{1}+E_{+}\gg T\,.\label{eq:DeltaG1NearPkLowT}
\end{flalign}
The contribution from 2-particle backscattering is subleading by a
factor $T^{2}/(\varepsilon_{1}+E_{+})^{2}$, see Eq.~(\ref{eq:DeltaG2NearPkLowT}).
Averaging over disorder with screened Coulomb interaction, we obtain
the leading contribution 
\begin{flalign}
 & \Delta G/G_{0}\sim\frac{\Gamma^{4}}{g^{2}(\varepsilon_{1}+E_{+})^{6}\delta^{2}}T^{4}\,,\quad\varepsilon_{1}+E_{+}\gg\Gamma, T\,.\label{eq:TypicalDeltaGEvenPkLowTWithEC}
\end{flalign}
The limitations on $\varepsilon_{1}+E_{+}$ come from the conditions of applicability of perturbation theory in tunneling.
Far from the peak, $\varepsilon_{1}+E_{+}\sim\delta$, Eq.~(\ref{eq:DeltaG1NearPkLowT}) matches our previous result, Eq.~(\ref{eq:TypicalDeltaG1ValleyLowTWithECCmb}), valid deeper in the even valley.

At higher temperatures, $T\gg\varepsilon_{1}+E_{+}$, the integration domain in Eq.~(\ref{eq:DeltaG1pkWithEc}) is no longer limited by temperature.
To regularize the integrals, the width of the resonant level $\varepsilon_{1}$ needs
to be taken into account. Near the peak, tunneling
between the helical edge and level $\varepsilon_{1}$ gives rise to
 a non-zero imaginary part $i\Gamma_{1}$ in the energy denominators originating from $G_{0}(E)$ in Eq.~(\ref{eq:TmatrixElementEvenValley}).
 Now at $T\gg\Gamma_{1}$ in $\Delta G_{1}$ two of the integrals
over energies are restricted by $\Gamma_{1}$ rather than $T$. The
integration is conveniently done by using the Fourier transform of
the energy conserving $\delta$-function. We get  
\begin{flalign}
 & \Delta G_{1}/G_{0}=\frac{1}{4\pi}\Theta(E_{+}+\varepsilon_{1})\nonumber \\
 & \times\sum_{\gamma}\left|\sum_{n,m}\frac{\sqrt{\Gamma_{n}\Gamma_{m}}}{\varepsilon_{n}-\varepsilon_{m}}(\frac{\Theta_{-m}}{\varepsilon_{1}-\varepsilon_{m}}-\frac{\Theta_{-n}}{\varepsilon_{1}-\varepsilon_{n}})U_{1LnL;1\overline{\gamma}m\gamma}\right|^{2}\,,\label{eq:DeltaG1NearPkHighT}\\
 & \hphantom{\left|\times\frac{\sqrt{\Gamma_{n}\Gamma_{m}}}{\varepsilon_{n}-\varepsilon_{m}}[\frac{\Theta(-\varepsilon_{m})}{\varepsilon_{1}-\varepsilon_{m}}-\frac{\Theta(-\varepsilon_{n})}{\varepsilon_{1}-\varepsilon_{n}}]\right|^{2}}T\gg\Gamma_{1},\,\varepsilon_{1}+E_{+}\,,\nonumber 
\end{flalign}
and $\Delta G_{2}$ is of the same order, see Eq.~(\ref{eq:DeltaG2NearPkHighT}).
Averaging over disorder, one has
\begin{flalign}
 & \Delta G/G_{0}\sim\frac{\Gamma^{2}}{g^{2}\delta^{2}}\,,\quad T\gg\Gamma,\,\varepsilon_{1}+E_{+}\,.\label{eq:TypicalDeltaGEvenPkHighTWithEC}
\end{flalign}
Since this result is independent of the ratio $\Gamma/(\varepsilon_{1}+E_{+})$,
we anticipate that it accurately describes $\Delta G$ 
at the peak at relatively high temperatures (but still $T\ll \delta$). The above equation coincides with
Eq.~(\ref{eq:TypicalDeltaGEvenPkLowTWithEC}) when $T\sim\varepsilon_{1}+E_{+}\sim\Gamma$.

\subsection{Peak conductance \label{sub:Peak-conductance.}}

To estimate the conductance correction at the charge degeneracy point,
we can extrapolate the results of Subsections \ref{sub:Odd-valleys-of}
and \ref{sub:Even-valley-conductance} to gate voltages near the peak.
Denoting by $\Gamma_{1}$ the width of the resonant level at the peak,
at high temperatures $T\gg\Gamma_{1}$ we can approach the peak from
the even valley side. This was discussed at the end of the previous
section where we got  
\begin{equation}
\Delta G^{\text{peak}}/G_{0}\sim\frac{\Gamma^{2}}{g^{2}\delta^{2}}\,,\quad\Gamma\ll T\ll\delta\,.\label{eq:TypicalDeltaGPeakHighTWithEC}
\end{equation}

At low temperatures $T\ll\Gamma_{1}$ we can approach the peak from
the even valley side by replacing $\varepsilon_{1}+E_{+}\rightarrow\Gamma$
in Eq.~(\ref{eq:TypicalDeltaGEvenPkLowTWithEC}) of previous section.
This yields 
\begin{equation}
\Delta G^{\text{peak}}/G_{0}\sim\frac{T^{4}}{g^{2}\delta^{2}\Gamma^{2}}\,,\quad T\ll\Gamma\,.\label{eq:TypicalDeltaGPeakLowTWithEC}
\end{equation}
We get the same result by extrapolating the odd valley result, Eq.
(\ref{eq:TypicalDeltaGOddValleyApprx}), to the mixed valence regime
$T_{K}\sim\Gamma$. Near the peak replacing $T_{K},\, E_{+}\to\Gamma$
in (\ref{eq:TypicalDeltaGOddValleyApprx}) we obtain Eq.~(\ref{eq:TypicalDeltaGPeakLowTWithEC}).

\subsection{Correction to the conductance at high temperatures $T\gg\delta$\label{sub:HighTCB}}

As we saw in Subsections \ref{sub:Odd-valleys-of} and \ref{sub:Even-valley-conductance},
at $T\ll\delta$ there is a drastic difference in temperature dependence
of $\Delta G$ between the even and odd valleys. In even ones $\Delta G\propto T^{4}$,
see Eq.~(\ref{eq:TypicalDeltaG1ValleyLowTWithECCmb}). In odd valleys
the dependence is weak, $\Delta G\propto\ln^{2}(T/T_{K})$ as long
as $T$ exceeds the exponentially-small $T_{K}$, see Eqs.~(\ref{eq:TypicalDeltaGOddValleyCombined})
and (\ref{eq:KondoTempDef}). The distinction between the valleys
gradually disappears at $T\gtrsim\delta$, once spin degrees of freedom
of the dot become thermally excited, irrespective the parity of electron
number. At the same time, thermal fluctuations of the electron number
remain suppressed as long as $T\ll E_{C}$. Then $\Delta G$ is dominated
by a virtual process (inelastic co-tunneling) which we describe at
length in Appendix \ref{sec:Cotunneling}. The main result there is
that $\Delta G$ crosses over from the $T$-independent
odd valley result to a quadratic dependence on temperature, see Eq.~(\ref{eq:KineticConductanceFinalWithCot}). As temperature rises towards $E_{C}$,
direct tunneling becomes important, first as an activated
contribution to $\Delta G$. At even higher temperatures, $T\gtrsim E_{C}$,
charging effects become irrelevant, and we may use the results of
Subsection~\ref{sub:Kinetic-equation}. Therefore, at $T\gtrsim\max(E_{C},T_{2})$
the temperature dependence of $\Delta G$ saturates at ${\Delta G\sim\Gamma/\delta}$, see Eq.~(\ref{eq:KineticConductanceFinal}) and Fig. \ref{Fig:TemperatureLine}.
The precise crossover between the co-tunneling-dominated and direct
tunneling -dominated regimes at $\delta\ll T\ll E_{C}$ depends
on specific dot parameters; we will leave aside the discussion of the detailed
behavior of $\Delta G$ in this interval.

\subsection{Average of $\Delta G$ over the dot chemical potential at low temperatures $T\ll \delta$\label{sub:AverageCmbBckade}}

The results derived in the previous subsections, \ref{sub:Odd-valleys-of}--\ref{sub:Peak-conductance.},
fully describe the low-temperature conductance correction due to a
Coulomb blockaded quantum dot for any gate voltage. 
Having in mind the resistivity of a long edge (see Section~\ref{sec:LongEdge} below), we will, in this section, average $\Delta G$ over the gate voltage, across multiple peaks and valleys. 
We will focus on low temperatures, $T\ll\delta$,
where $\Delta G$ is strongly gate voltage dependent. It is useful
to do the averaging in the following way: First one averages over
all the valleys keeping fixed the position of gate voltage relative to closest
peaks. For example, in the odd valley this amounts to averaging over
dot levels in Eq.~(\ref{eq:TypicalDeltaGOddValleyCombined}) for given
$E_{\pm}$. This average is given by  Eq.~(\ref{eq:TypicalDeltaGOddValleyApprx}).
After this first step of averaging, one   averages the resulting
conductance over different values of $E_{+}$ in the valley. 

\begin{figure}
\includegraphics[width=0.9\columnwidth]{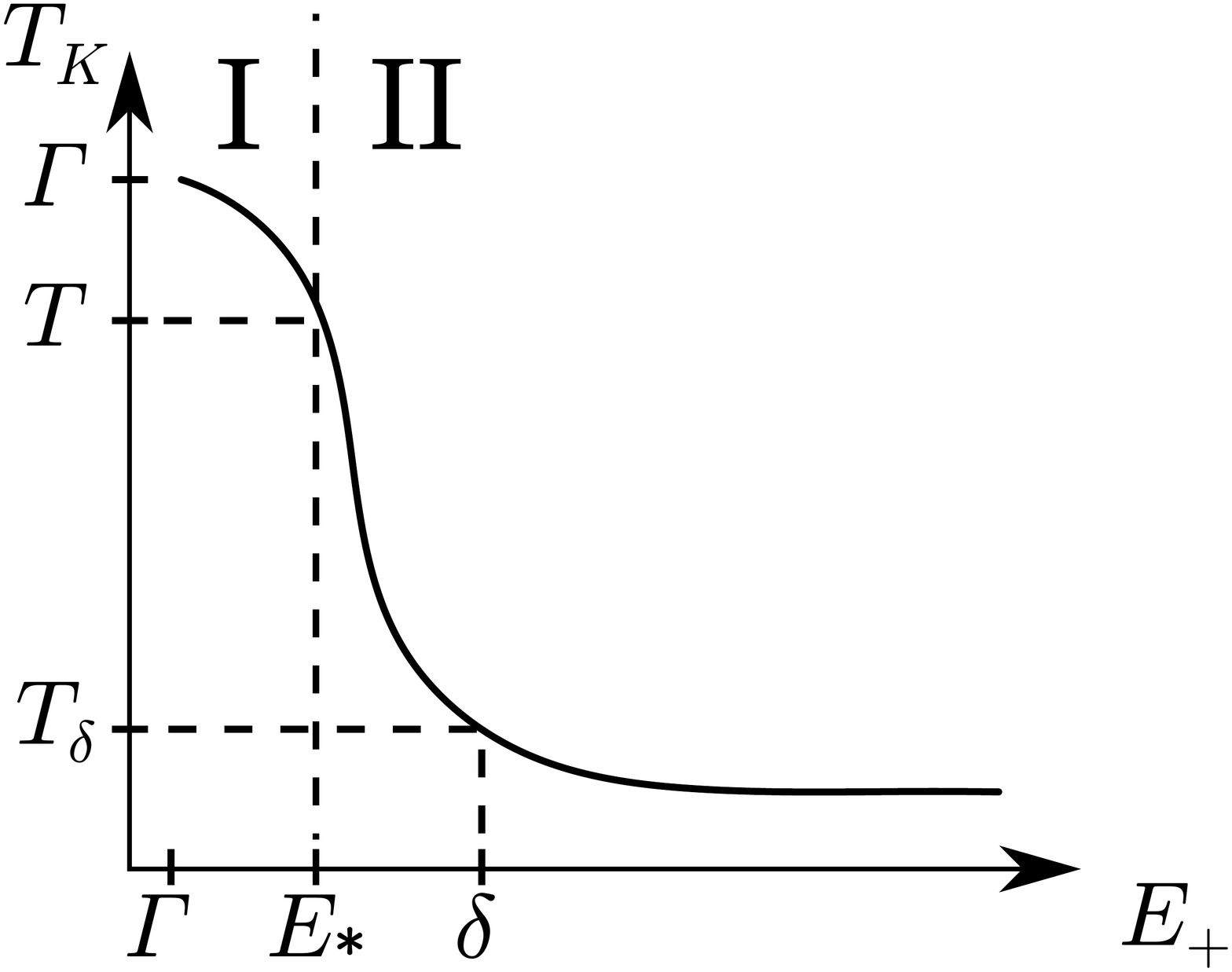}\caption{Graph of the Kondo temperature as a function of gate voltage, $T_{K}(E_{+})$.
Here $\Gamma> T > T_{\delta}$. At these temperatures the main contribution to the gate-voltage-averaged $\Delta G$ comes from the regime around $E_+ \approx \delta$. In region I the dot spin is strongly-screened since $T<T_K(E_+)$. There $\Delta G \propto T^4$ from the first term in Eq.~(\ref{eq:TypicalDeltaGOddValleyApprx}). In region II we have $T>T_K(E_+)$, and the second term of Eq.~(\ref{eq:TypicalDeltaGOddValleyApprx}) is applicable. }
\label{FigTK}
\end{figure}

At any temperature $T\ll \delta$, the largest contribution to the average over gate voltage comes from the odd valley, Eq.~(\ref{eq:TypicalDeltaGOddValleyApprx}).
There are three characteristic temperature intervals, where the functional dependence of $\Delta G$ on $E_+$ is different. 
In the lowest interval, where temperature is lower than $T_{K}^{\text{min}}$ (Kondo temperature in the bottom of the valley, $E_+ = E_C$),  $\Delta G$ is given by the first term in Eq.~(\ref{eq:TypicalDeltaGOddValleyApprx}), and $\Delta G \propto T^4$ in the entire valley. 
As temperature rises above $T_{K}^{\text{min}}$, the domain of $\Delta G \propto T^4$ shrinks towards the charge degeneracy points. 
Then the second term, $\Delta G \propto \ln^2(T/T_K)$, in Eq.~(\ref{eq:TypicalDeltaGOddValleyApprx}) becomes applicable in the bottom of the valley and gives the main contribution to average $\Delta G$.
The next characteristic temperature is the Kondo temperature at $E_+ = \delta$. We denote this temperature by  $T_{\delta}$, and it is given by
\begin{equation}
T_{\delta}=\delta e^{-\pi\delta/\Gamma}\,.\label{eq:TdeltaDefinition}
\end{equation}
See also Fig.~\ref{FigTK}. 
Above this temperature the second term in Eq.~(\ref{eq:TypicalDeltaGOddValleyApprx}) is applicable even for $E_+ \lesssim \delta$, and the average $\Delta G$ mainly comes from gate voltages $E_+ \approx \delta$.
We will next start our quantitative description from this last temperature interval.

In this high-temperature interval, $\delta> T > T_{\delta}$, the functional dependence of $\Delta G$ on $E_+$ changes~\footnote{This is the gate voltage at which the lowest energy excitation of the dot changes between one with charge and one with spin. The nature of these excitations defines the form of denominators in the perturbative result for $\delta J_{ij}$, see Eq.~(\ref{eq:Jyx}) and Table~\ref{table-Jcomponents} of Appendix \ref{sec:Components-of-the}.} at $E_+\approx \delta$, as can be seen from the overall factor in Eq.~(\ref{eq:TypicalDeltaGOddValleyApprx}). 
Integration over $E_+$ around $E_+\approx \delta$ gives the main contribution to $\Delta G^{\text{av}}$,
\begin{equation}
\frac{\Delta G^{\text{av}}}{G_{0}}\sim\frac{\Gamma^{2}}{g^{2}E_{C}\delta}\,,\quad T_{\delta}\ll T\ll\delta\,.\label{eq:DeltaGavWithECTggGamma}
\end{equation}

When temperature decreases below the value $T_{\delta}$, the dot spin becomes strongly screened (and $\Delta G \propto T^4$) even for gate-voltages $E_+ \approx \delta$. The main contribution to the average $\Delta G$ comes from the gate voltage interval ${\delta\ll E_{+}<E_{C}}$ where the second term of Eq.~(\ref{eq:TypicalDeltaGOddValleyApprx}) is valid. 
 We find
\begin{equation}
\frac{\Delta G^{\text{av}}}{G_{0}}\sim\, \frac{1}{E_{C}}\int_{E_{*}(T)}^{E_{C}}dE_{+}\frac{\ln^{2}\frac{T}{T_{K}(E_{+})}}{\ln^{2}\frac{\delta}{T_{K}(E_{+})}}\frac{\Gamma^{2}}{g^{2}E_{+}^{2}} \,,
\label{eq:DeltaGavIntermediate1}
\end{equation}
where $E_{*}(T)$ is the solution of $T_{K}(E_{*})=T$ (see also Fig. \ref{FigTK}). 
At $T<T_{\delta}$, it can be written in terms of the parameter $\eta(T)=\ln\frac{\delta}{T}/\ln\frac{T}{T_{K}^{\text{min}}}$ as
\begin{equation}
E_{*}={E_{C}[1-(1+\eta)^{-1/2}]} \,.
\end{equation}
Changing the integration variable in Eq.~(\ref{eq:DeltaGavIntermediate1}) to  $\epsilon={(E_{C}-E_{+})/E_{C}}$, we get
\begin{equation}
\frac{\Delta G^{\text{av}}}{G_{0}}\sim \frac{\Gamma^{2}}{g^{2}E_{C}^{2}}\int_{0}^{(1+\eta)^{-1/2}}\frac{d\epsilon}{(1-\epsilon)^{2}}\left[\frac{(1+\eta)^{-1}-\epsilon^{2}}{1-\epsilon^{2}}\right]^{2}\,.
\label{eq:DeltaGavIntermediate2}
\end{equation}
The temperature dependence is now in the parameter $\eta(T)$. We evaluate the integral in Eq.~(\ref{eq:DeltaGavIntermediate2}) asymptotically around $\eta=\delta/E_{C}$ (at $T=T_{\delta}$) and $\eta\to \infty$ (at $T\to T_{K}^{\text{min}}$). The crossover between the two limits happens at a temperature $T_{1/2}=\sqrt{\delta T_{K}^{\text{min}}}$,
at which $\eta=1$. In the first limit, $T_{1/2}\ll T\ll T_{\delta}$, we find 
\begin{flalign}
\frac{\Delta G^{\text{av}}}{G_{0}}\sim\, & \frac{\Gamma^{2}}{g^{2}E_{C}\delta}\frac{\ln\frac{\delta}{T_{\delta}}}{\ln\frac{\delta}{T}}\,,\label{eq:DeltaGavWithECTggThalf}
\end{flalign}
which matches with Eq.~(\ref{eq:DeltaGavWithECTggGamma}) at $T=T_{\delta}$. In the opposite limit $T_{K}^{\text{min}}\ll T\ll T_{1/2}$ we get
\begin{align}
\frac{\Delta G^{\text{av}}}{G_{0}}\sim\, & \frac{\Gamma^{2}}{g^{2}E_{C}\delta}\frac{\ln\frac{\delta}{T_{\delta}}}{\ln\frac{\delta}{T_{1/2}}}\left(\frac{\ln\frac{T}{T_{K}^{\text{min}}}}{\ln\frac{\delta}{T_{K}^{\text{min}}}}\right)^{5/2}\,.\label{eq:DeltaGavWithECTggTKmin}
\end{align}

Finally, as $T$ becomes less than $T_{K}^{\text{min}}$ the dot spin
is always strongly screened and the first term
in Eq.~(\ref{eq:TypicalDeltaGOddValleyApprx}) needs to be used for $\Delta G$. The
main contribution to average conductance comes from the vicinity of the bottom of the valley where $T_{K}$ is the smallest, 
\begin{align}
\frac{\Delta G^{\text{av}}}{G_{0}}\sim\, & \int_{\delta}^{E_{C}}\frac{dE_{+}}{E_{C}}\frac{1}{\ln^{2}\frac{\delta}{T_{K}(E_{+})}}\frac{\Gamma^{2}}{g^{2}E_{+}^{2}}\frac{T^{4}}{T_{K}(E_{+})^{4}}\nonumber \\
\sim\, & \frac{\Gamma^{4}}{g^{2}E_{C}^{4}}\frac{T^{4}}{(T_{K}^{\text{min}})^{4}}\int_{0}^{E_{C}}\frac{dE_{+}}{E_{C}}\exp[-4\frac{\pi(E_{C}-E_{+})^{2}}{2\Gamma E_{C}}]\nonumber \\
\sim\, & \frac{\Gamma^{9/2}}{g^{2}E_{C}^{9/2}}\frac{T^{4}}{(T_{K}^{\text{min}})^{4}}\,,\quad T\ll T_{K}^{\text{min}}\,.\label{eq:DeltaGavWithECTllTKmin}
\end{align}
[When going to the last line in Eq.~(\ref{eq:DeltaGavWithECTllTKmin}), we extended the integration to $(-\infty,E_C)$, as $E_C / \Gamma \gg 1$.]
Since $\frac{\Gamma}{E_{C}}\sim1/\ln\frac{\delta}{T_{K}^{\text{min}}}$,
the above equation agrees with Eq.~(\ref{eq:DeltaGavWithECTggTKmin})
at $T\sim T_{K}^{\text{min}}$.

\section{Long edge\label{sec:LongEdge}}

In Sections \ref{sec:BSinAbsenceOfEC} and \ref{sec:CoulombBlockade}
we showed how a single quantum dot (with small or large charging energy)
modifies the conductance of a helical edge. In section \ref{sub:Potential-fluctuations-in}
below we show how doping may naturally create quantum dots, or \emph{puddles},
in heterostructures such as HgTe quantum wells. Our results from the
previous two sections describe the resistance of a short helical edge where only few puddles reside near the edge.

In this section we consider a long edge of length $L$ much larger
than the mean distance between puddles, $L\gg n_{p}^{-1/2}$, with
$n_{p}$ being the puddle density. Near a long edge there can be multiple
puddles contributing to the helical edge resistance. The contributions
add up incoherently since scattering off puddles is inelastic. Therefore, the resistance $1/G$ of a long edge scales linearly with the length $L$, and we can characterize the edge by its resistivity $\varrho=1/GL$.
We calculate $\varrho$ caused by puddles of small or large charging energy at low temperatures ($T\ll\delta$) in Subsections \ref{sub:LongEdgeNoEC} and \ref{sub:LongEdgeWithEC}.

We find below that the charging energy is moderate, $E_C \sim \delta$, for doping-induced puddles. Even in this limit, puddles with an odd occupation may be present. This leads to resistivity $\varrho(T)\propto1/\ln^2(\delta/T)$ dominated by puddles with an odd number of electrons.
 This weak temperature dependence of $\varrho(T)$ may shed light on recent experiments in HgTe quantum wells.

\subsection{Electrostatic potential fluctuations in a heterostructure\label{sub:Potential-fluctuations-in}}

In this section we discuss the origin of charge puddles in a doped HgTe quantum
well. Our goal is to estimate the parameters of the puddles (such
as their density and the average level spacing in a puddle) in terms of known parameters
(density of a dopant, band structure properties of the semiconductor).
The random positions of donors (or acceptors) create fluctuations
in the electrostatic potential at the quantum well. This results in
the formation of electron (or hole) puddles in regions of severe fluctuations. The fluctuations need to be sufficiently strong, so that the conduction (valence) band dips below (above) the Fermi level residing in the semiconductor band gap. The Fermi level position is tuned with an external gate electrode. 
We consider here thin enough HgTe quantum wells that are topological band insulators in the clean limit. Effects of charge disorder in the semimetallic regime (thick wells) have been recently studied in Ref.~\onlinecite{knap14}.

\begin{figure}
\includegraphics[width=0.9\columnwidth]{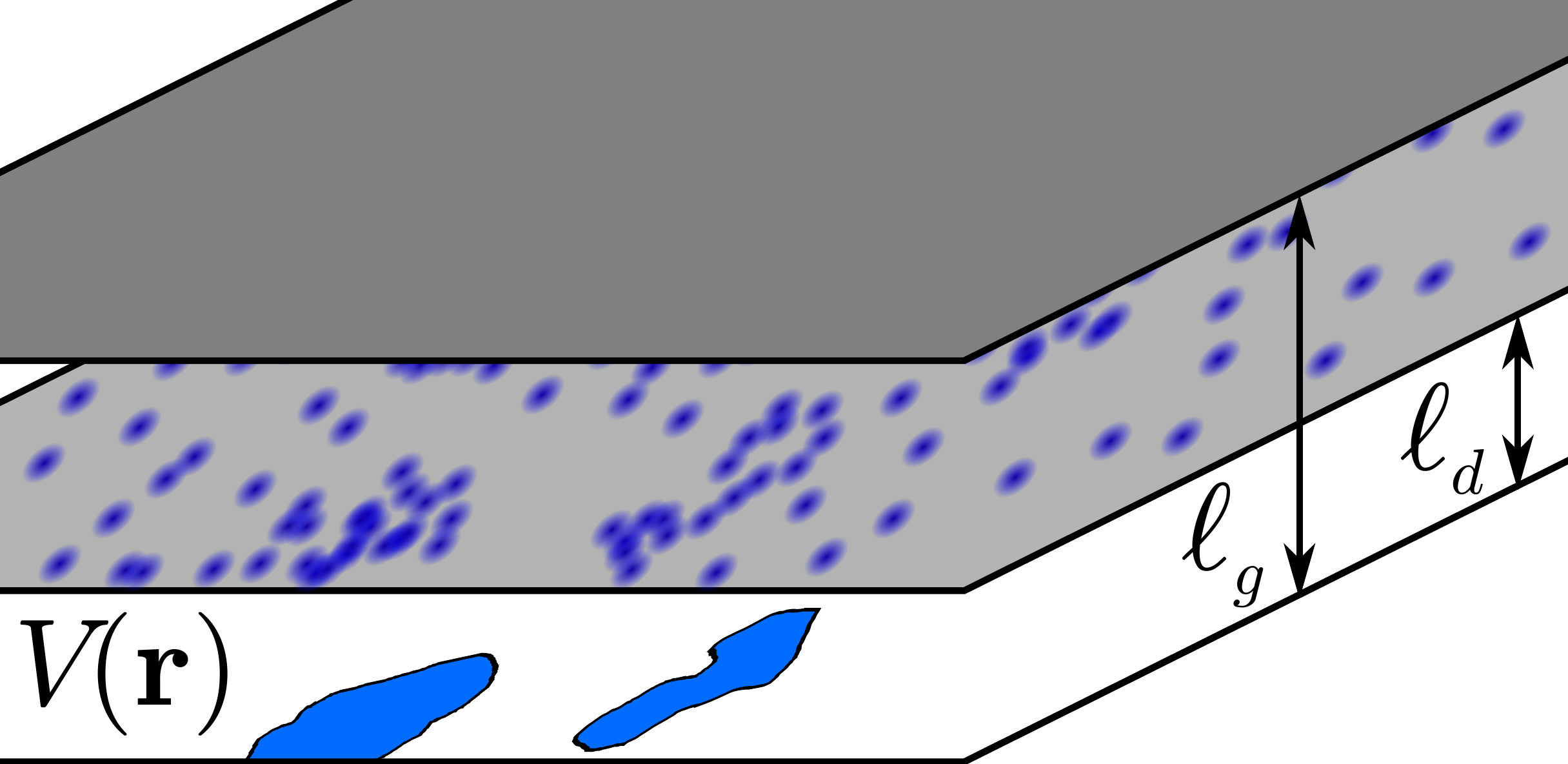}\caption{
(Color online) 
Schematic drawing of the heterostructure. The dark grey top layer
depicts the gate electrode, below which is the doping layer (light grey)
with a random distribution of dopants (blue dots) with density $n_{d}$.
The dopants induce a random potential $V(\mathbf{r})$ on the quantum
well (white) and charge puddles are formed (blue).}
\label{fig:QW}
\end{figure}
For definiteness we consider $n$-type doping from here on. In the experimental setup donors are located in a layer separated from the quantum well by a spacer of width $\ell_d$ (see Fig.~\ref{fig:QW}). 
Donors of average density $n_{d}$ induce a random potential $V(\mathbf{r})$ in the quantum well. It creates puddles by attracting carriers to those regions where the conduction band lies below the Fermi level. 
Ignoring the screening of $V(\mathbf{r})$ by these localized carriers, we have 
\begin{flalign}
 & V(\mathbf{r})=e\int d^{2}\mathbf{r}'\frac{\sigma(\mathbf{r}')}{\kappa}\nonumber \\
 & \times\left[\frac{1}{\sqrt{\ell_{d}^{2}+(\mathbf{r}'-\mathbf{r})^{2}}}-\frac{1}{\sqrt{(2\ell_{g}-\ell_{d})^{2}+(\mathbf{r}'-\mathbf{r})^{2}}}\right]\,,
 \label{eq:PotInQW}
\end{flalign}
where $\mathbf{r}$ lies in the plane of the heterostructure and the
last term comes from the induced charge on the gate electrode, located a
distance $\ell_{g}$ from the quantum well; $\ell_d \leq \ell_g$, see Fig.~\ref{fig:QW}. 
 We denoted the dielectric constant by $\kappa$ and the donor charge density by $\sigma$.
 Taking the latter to be $\delta$-correlated with variance $e^{2}n_{d}$, we get  
\begin{equation}
\langle\Delta V^{2}\rangle=V_{0}^{2}\ln\frac{\ell_{g}^{2}}{\ell_{d}(2\ell_{g}-\ell_{d})}\,,\label{eq:PotFluctuation}
\end{equation}
where $V_{0}=\sqrt{2\pi n_{d}}e^{2}/\kappa$ and $\Delta V=V-\langle V\rangle$.

We will now estimate the puddle parameters. At full depletion the
density of puddles depends on the ratio of $\sqrt{\langle\Delta V^{2}\rangle}$
and $E_{g}/2$. We define the characteristic density of donors, 
\begin{equation}
n_{0}=\frac{E_{g}^{2}\kappa^{2}}{8\pi e^{4}\ln[\ell_{g}^{2}/\ell_{d}(2\ell_{g}-\ell_{d})]}\,,
\label{eq:DonorDensityCharacteristic}
\end{equation}
at which $\sqrt{\langle\Delta V^{2}\rangle}=E_{g}/2$.

At high donor densities $n_{d}\gg n_{0}$ the band edge fluctuates
heavily and dips frequently below the Fermi level. This limit is
beyond the scope of this work, since inter-puddle tunneling can lead
to $\Gamma\gtrsim\delta$ and bulk conductivity $\sigma_{\text{bulk}}\gtrsim e^{2}/h$,
corresponding to a symplectic metallic phase rather than to a topological
insulator.\cite{fu12} We leave out an estimate of $\sigma_{\text{bulk}}$
since it depends strongly on the experimental parameters.

Let us next consider low donor density, $n_{d}\ll n_{0}$, so that
puddles can be made rare and small by adjusting the Fermi level 
a distance $\gtrsim V_{0}$ below the conduction band edge. 
Bulk conductivity is then activated in nature.
If temperature is low enough, conduction can become dominated by the edges.
Our next goal is to estimate the puddle density $n_{p}$ and the mean level spacing $\delta$. In our analysis we closely follow
Ref. \onlinecite{gergel78}.

The gate electrode is separated from the quantum well by a distance $\ell_g$. 
 Because of screening of the donors by the gate, the potential $V(r)$ is correlated in the quantum well on the same length scale $\ell_{g}$.
Puddles occur in rare regions of size $\sim\ell_{g}$, where the potential fluctuation exceeds the value $E_{g}/2$. 
The typical value by which
a rare fluctuation exceeds $E_{g}/2$ is obtained to leading
order in $V_{0}/E_{g}$ from the distribution $P(V)\propto\exp(-V^{2}/2V_{0}^{2})$
{[}we ignore the logarithm in Eq.~(\ref{eq:PotFluctuation}) considering
$\ell_{d}$ and $\ell_{g}$ of the same order of magnitude{]}.
We get 
\begin{equation}
\langle\Delta V-E_{g}/2\rangle_{\Delta V\geq E_{g}/2}=V_{0}^{2}/2E_{g}\,.\label{eq:PotFluctuationAbvEg}
\end{equation}
The carriers in the quantum well form a puddle of area $w^{2}$ to
smoothen out the dip $\sim V_{0}^{2}/2E_{g}$ caused by the donor density
fluctuation. Since $V_0^2 / E_g \ll E_g$, we approximate the Dirac spectrum by quadratic dispersion with mass $m^{*}=E_{g}/2v^{2}$ (see also the beginning of Section \ref{sec:CoulombBlockade}).
 Assuming weak interaction, $\alpha=e^{2}/\kappa v\ll1$,
we may use the Thomas-Fermi approximation for the description of the
puddle, $k_{F}^{2}/m^{*}\sim V_{0}^{2}/E_{g}$. Therefore the electron density in the puddle is $n_{e}\sim k_{F}^{2}\sim m^{*}V_{0}^{2}/E_{g}$.
Denoting by $N\sim n_{e}w^{2}$ the number of electrons in the puddle, their total electrostatic potential is $e^{2}N/\kappa w$, assuming $w\lesssim \ell_g$. This potential should compensate for the potential fluctuation, $e^{2}n_{e}w/\kappa\sim V_{0}^{2}/2E_{g}$. 
Using, in addition, the above-mentioned Thomas-Fermi estimate of $n_{e}$, we find the puddle linear size $w\sim a_{B}$, where $a_{B}=2v/\alpha E_{g}$ is the effective Bohr radius. Finally, since $\delta \sim 1/m^{*}w^2 \sim V_0^2/E_g N$, we get 
\begin{equation}
\delta\sim\alpha^{2}E_{g},\quad N\sim n_{d}/\alpha^{2}n_{0},\quad E_{C}/\delta\sim1\,.\label{eq:PuddleParams}
\end{equation}
The last relation is obtained from Eq.~(\ref{eq:EcOverdelta}) using our assumption of small puddles, $w \lesssim \ell_{g}$.
The puddle has a large number of electrons if $\alpha\ll\sqrt{n_{d}/n_{0}}\lesssim1$ (which we have assumed throughout this work, except for the qualitative discussion in Section \ref{sec:Qualitative}). 
Taking a nearly ballistic dot, where the Thouless energy 
 $E_{T}\lesssim v/w$, we can finally
estimate $g\lesssim v/w\delta\sim\alpha^{-1}$.

These are properties of a single puddle. To assess the puddle density
$n_{p}$, we first note that the distance to the gate $\ell_{g}$
is an effective screening radius, and we can divide the quantum well
into regions of area $\ell_{g}^{2}$ between which the potential $V(\mathbf{r})$ is uncorrelated. Puddles occur in
those rare places where the local potential fluctuation exceeds the
value $E_{g}/2$. The probability for this to happen is exponentially
small, $p \sim \sqrt{n_{d}/n_{0}}\exp(-n_{0}/2n_{d})$, as follows from the distribution $P(V)$. The total carrier number $N_{p}N$ of  $N_p$ puddles in the square $\ell_{g}\times\ell_{g}$ compensates for the corresponding fluctuation in the donor number, $N_{p}N\sim\sqrt{n_{0}\ell_{g}^{2}}$. Since only a fraction $p$ of the uncorrelated regions contains puddles, we obtain the puddle density 
\begin{equation}
n_{p}=\frac{p N_p}{\ell_g^2}\sim\frac{1}{\ell_{g}a_{B}}\sqrt{\frac{n_{0}}{n_{d}}}e^{-n_{0}/2 n_{d}}\,.\label{eq:DensityOfPuddles}
\end{equation}
In Eq.~(\ref{eq:PotInQW}) above we ignored the screening of potential fluctuations by the electron density $N n_{p}$ due to puddles. This density is to be compared with the typical fluctuation of the donor density on lateral scale $\sim \ell_{g}$ (the charge induced on the gate electrode smears out fluctuations on scales larger than $\ell_{g}$). Therefore the condition to ignore screening by carriers is $N n_{p} \ll\sqrt{n_{d}}/\ell_{g}$. Using the relation $\sqrt{n_0} \sim 1/\alpha^2 a_{B}$ [see Eq.~(\ref{eq:DonorDensityCharacteristic})] with Eqs.~(\ref{eq:PuddleParams}) and (\ref{eq:DensityOfPuddles}) we see that the condition is $p\ll \sqrt{n_{d}/n_{0}}$, which is consistent with our assumption of weak doping, $n_{d}\ll n_{0}$.

Finally, the bulk conductivity caused by rare puddles is negligible. The puddle-to-puddle
hopping conductivity is proportional to the tunneling probability $\exp[-(\Lambda^{2}n_{p})^{-1/2}]$
times the thermal activation factor $\exp(-\delta/T)$ (we used $E_{C}\sim\delta$ and denoted $\Lambda$ the penetration depth of the electron wavefunction into the bulk).
Both of these factors are small at temperatures $T\ll\delta$, making
it very difficult for edge electrons to find a percolation path through
the bulk. These reasons make the low-temperature bulk conductance
negligible compared with the conductance of the edges, even if the
latter one is reduced by inelastic backscattering. Next, we estimate
the resistivity of the edges induced by the backscattering of edge electrons by nearby puddles.

\subsection{Resistivity of a long edge\label{sub:LongEdge}}

We are now in a position to assess the applicability of the single-dot
theory developed in Sections \ref{sec:BSinAbsenceOfEC} and \ref{sec:CoulombBlockade}
to charge puddles created by random fluctuations of the donor density.
First of all, we used RMT for dots with $N\gg1$ electrons. The Thomas-Fermi
approximation utilized in derivation Eq.~(\ref{eq:PuddleParams})
also assumes $N\gg1$. This assumption is met in the rare-puddles
limit ($n_{d}\ll n_{0}$) only if $\alpha^{2}\ll1$ while $n_{d}/n_{0}\gg\alpha^{2}$.
Second, Sections \ref{sec:BSinAbsenceOfEC} and \ref{sec:CoulombBlockade}
consider the limits of $U\ll\delta$ and $E_{C}\gg\delta$, respectively,
while Eq.~(\ref{eq:PuddleParams}) predicts $E_{C}\sim\delta$ for
random puddles of charge. Therefore, the random-puddles case is on
the border-line of the descriptions developed in the aforementioned sections. In the following, we perform edge
resistivity estimate in two limits conforming with the assumptions
of Sections \ref{sec:BSinAbsenceOfEC} and \ref{sec:CoulombBlockade},
respectively. We will then identify, which of the model-based conclusions
hold under the realistic conditions described by Eq.~(\ref{eq:PuddleParams}).

\subsubsection{Resistivity of a long edge in the absence of charging effects\label{sub:LongEdgeNoEC}}

The single-dot backscattering in the absence of charging effects, $E_C \ll\delta$,
was discussed in Section \ref{sec:BSinAbsenceOfEC}. There, in deriving
Eqs.~(\ref{eq:TypicalDeltaG1PeakLowTNoEC}) and (\ref{eq:TypicalDeltaG1ValleyLowTNoEC}),
we replaced the dot level widths by their average values. 
This is what the conductance
correction from a typical puddle would be. To assess the conductance
of a long edge with many puddles, we need to average over the distribution
of level widths.  In the case of a long edge, there can be rare
instances of exceptional puddles which produce most of the backscattering.
This can be seen from Eq.~(\ref{eq:DeltaG1PeakLowTNoEC}) at low temperatures, $T\ll\Gamma$, when level 1 is close to Fermi energy, $|\varepsilon_{1}|\ll \delta$. (Here $\varepsilon_1$ is measured from the Fermi energy.) Averaging Eq.~(\ref{eq:DeltaG1PeakLowTNoEC}) over dot parameters, at fixed peak position $\varepsilon_1$ and width $\Gamma_1$, leads to~\footnote{The second term in Eq.~(\ref{eq:TypicalDeltaG1PeakLowTNoEC}) was obtained from Eq.~(\ref{eq:DeltaGAvPkRarePuddle1}) by replacing $\Gamma_1 \to \Gamma$, which, however, we will not do here} 
\begin{equation}
\frac{\Delta G^{\text{peak}}}{G_{0}} \sim \frac{1}{g^{2}}\frac{T^{4}\Gamma \Gamma_1^{3}}{(\varepsilon_{1}^{2}+\frac{1}{4}\Gamma_{1}^{2})^{4}}\Theta(\Gamma_1-T)\,.
\label{eq:DeltaGAvPkRarePuddle1}
\end{equation}

In absence of charging effects the gate-voltage averaged conductance correction, $\Delta G^{\text{av}}$, is dominated by the peaks, as indicated by strong $\varepsilon_1$-dependence in Eq.~(\ref{eq:DeltaGAvPkRarePuddle1}). 
Therefore at low temperatures $T\ll \Gamma$, we obtain $\Delta G^{\text{av}}$ by averaging Eq.~(\ref{eq:DeltaGAvPkRarePuddle1}) over the peak position $\varepsilon_{1}$, and the width $\Gamma_1$ of the resonant level. For the latter, we use the symplectic Porter-Thomas distribution\cite{haake01} $P(\Gamma_{1})\propto\Gamma_{1}e^{-2\Gamma_{1}/\Gamma}$. We then find from Eq.~(\ref{eq:DeltaGAvPkRarePuddle1}),
\begin{equation}
\frac{\Delta G^{\text{av}}}{G_{0}} \sim \int_{\Gamma_1 \geq T} d\Gamma_1 P(\Gamma_{1}) \int \frac{d \varepsilon_1}{\delta} \frac{1}{g^{2}}\frac{T^{4}\Gamma \Gamma_1^{3}}{(\varepsilon_{1}^{2}+\frac{1}{4}\Gamma_{1}^{2})^{4}}\,.
 \label{eq:DeltaGAvPkRarePuddle2}
\end{equation}
The integral over $\varepsilon_{1}$ in Eq.~(\ref{eq:DeltaGAvPkRarePuddle2}) can be done first and is dominated by $|\varepsilon_1| \lesssim \Gamma_1$. The remaining integral over $\Gamma_1$ is dominated by $\Gamma_1 \sim T$, since the integrand becomes large for small level widths. 
We see that most of the backscattering at low temperatures $T\ll \Gamma$ is caused by rare quantum dots with $\Gamma_{1}\sim T$ rather than typical ones with $\Gamma_{1}\sim\Gamma$. This was not captured in Eq.~(\ref{eq:TypicalDeltaG1PeakLowTNoEC}), which describes typical dots.

 The corresponding high-temperature limit is given in the first term in Eq.~(\ref{eq:TypicalDeltaG1PeakLowTNoEC}), and there the average over level width is trivial. Combining the limits of low and high temperatures, we get from the first term in Eq.~(\ref{eq:TypicalDeltaG1PeakLowTNoEC}), and Eq.~(\ref{eq:DeltaGAvPkRarePuddle2})
\begin{equation}
\frac{\Delta G^{\text{av}} }{G_{0}}\sim\frac{1}{g^{2}}\frac{\Gamma}{\delta}\Theta(T-\Gamma)+\frac{1}{g^{2}}\frac{T^{2}}{\Gamma\delta}\Theta(\Gamma-T)\,.
\label{eq:DeltaGAvCombined}
\end{equation}
Since the average level width $\Gamma$ depends on the distance between the dot and the edge, the above equation gives the conductance correction $\Delta G$ due to a puddle at a fixed distance from the edge.

\begin{figure}
\includegraphics[width=0.9\columnwidth]{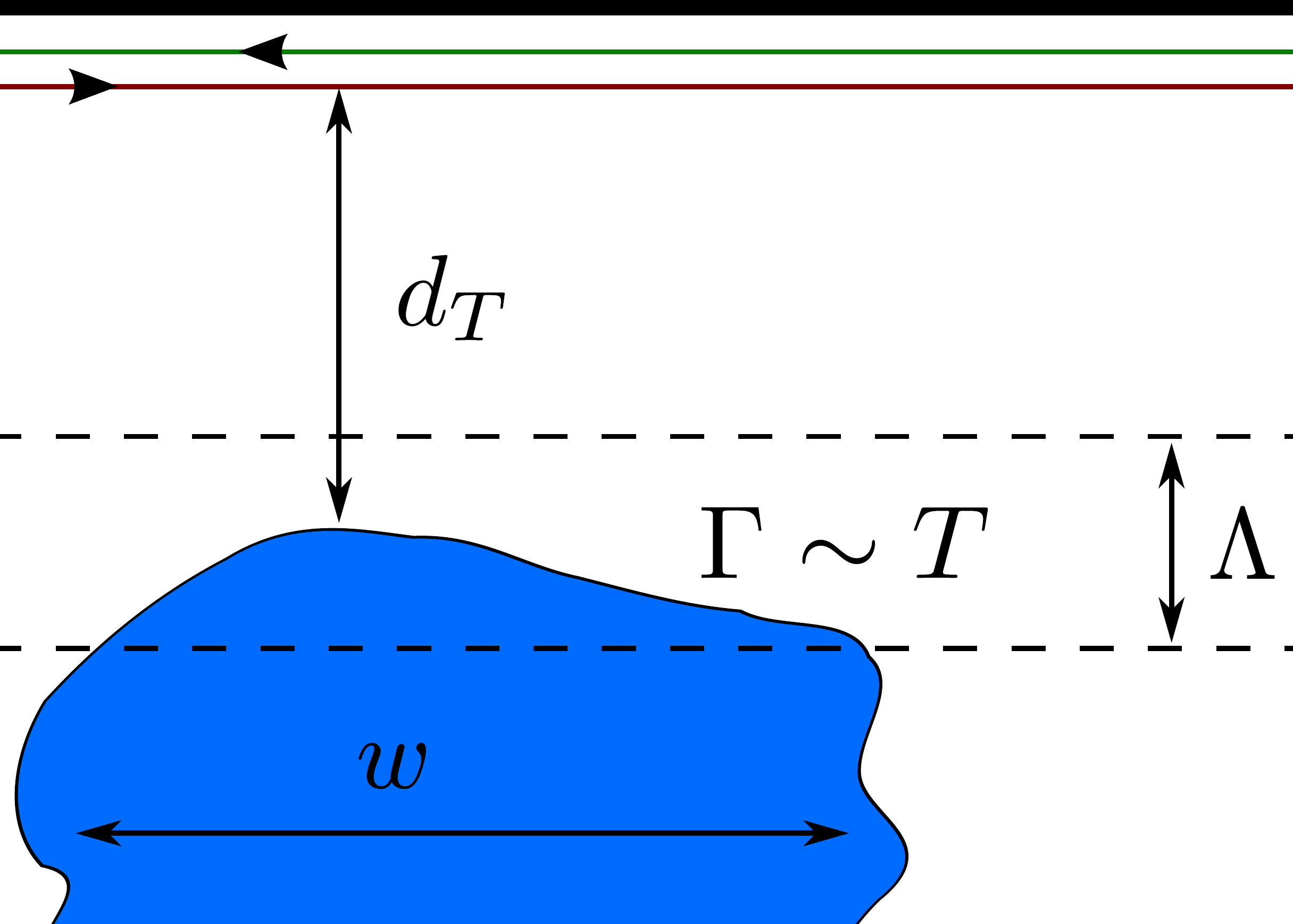}\caption{
(Color online) 
A helical edge with left- and right-propagating modes, (top) and a charge
puddle (bottom). 
In the absence of Coulomb blockade, puddles with an optimal level width $\Gamma \sim T$ yield the largest correction to the conductance. The level width is optimal in puddles at a distance $d_T$ from the edge. 
Those puddles whose distance from the edge is within $\Lambda$ of $d_T$ give the main contribution to resistivity $\varrho$ in  Eq.~(\ref{eq:resistivityNoECIntermediate})
 }
\label{fig:puddles}
\end{figure}

To estimate the long-edge resistivity $\varrho$, we will sum over the single-puddle corrections $\Delta G$ . Let us first consider an edge with just one puddle contributing to the edge resistance $R$. Then $R=(G_0 - \Delta G)^{-1}\approx R_0+\Delta R$ where $R_0 =G_0^{-1}= h/e^2$, and $\Delta R = \Delta G/G_0^2$ is a small resistance introduced by the puddle. 
On a long edge multiple puddles contribute to the edge resistance, yielding $\Delta R = \sum_i \Delta G_i/G_0^2$, where the sum is over all puddles, labeled by the index $i$.
Since scattering off puddles is inelastic and incoherent, the long edge is characterized by a self-averaging resistivity. 
Denoting the edge length by $L$, we then have 
\begin{equation}
\varrho= \frac{1}{L}\left\langle \sum_i \frac{\Delta G_i}{G_0^2} \right\rangle = \frac{n_p \Lambda}{G_0} \int_0^{\delta} \frac{d\Gamma}{\Gamma} \frac{\Delta G^{\text{av}} }{G_{0}}\,.
\label{eq:resistivityNoECIntermediate}
\end{equation}
We introduced the density of puddles $n_{p}$ [see Eq.~\ref{eq:DensityOfPuddles}] in converting the sum over the puddle positions into a two-dimensional integral. Since $\Delta G^{\text{av}}$ is independent of the position \emph{along} the edge [see Eq.~(\ref{eq:DeltaGAvCombined})], the integral over the length canceled against the pre-factor $L^{-1}$. 
The remaining integral over edge-dot distances, across the width of the sample, was converted into an integral over the average level width $\Gamma$. The level width decays exponentially in the distance $d$ between the edge and the dot, $\Gamma(d)\sim\exp(-2d/\Lambda)$, where $\Lambda\sim v/E_{g}$ is the penetration depth of the electron wave function into the bulk.

Inserting Eq.~(\ref{eq:DeltaGAvCombined}) into Eq.~(\ref{eq:resistivityNoECIntermediate}) yields
\begin{equation}
\varrho\sim\frac{n_{p}\Lambda}{G_{0}}\frac{T}{g^{2}\delta}\,,\quad T\ll\delta\,.
\label{eq:resistivityNoECFinal}
\end{equation}
The integral over $\Gamma$ in Eq.~(\ref{eq:resistivityNoECIntermediate}) was dominated by $\Gamma \sim T$. Thus the long edge resistivity in Eq.~(\ref{eq:resistivityNoECFinal}) mostly originates from those puddles that have $\Gamma \sim T$, see Fig.~\ref{fig:puddles}.

\subsubsection{Resistivity of a long edge in the Coulomb blockade regime \label{sub:LongEdgeWithEC}}

\begin{figure}
\includegraphics[width=1\columnwidth]{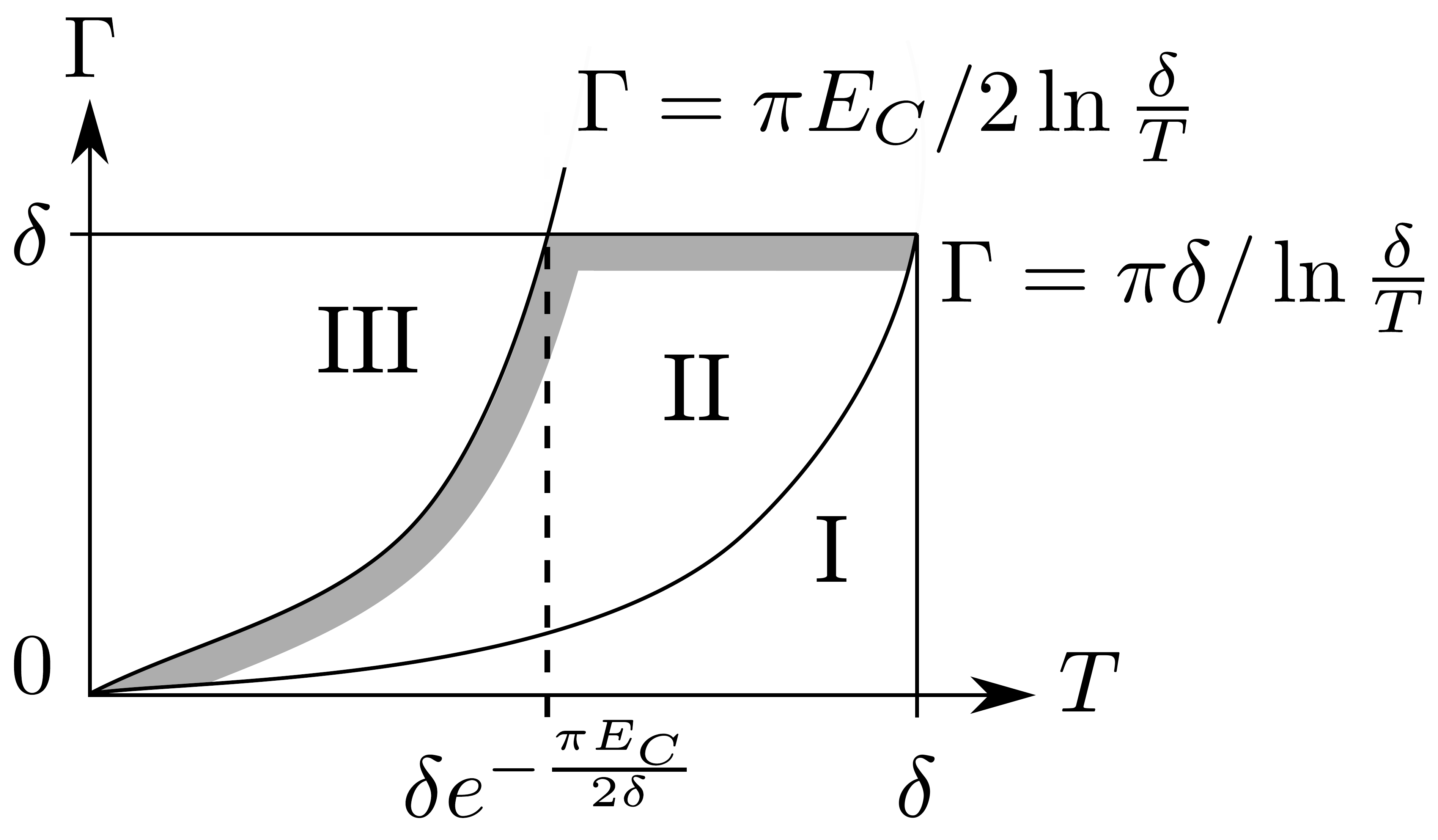}\caption{
The integration domain in Eq.~(\ref{eq:resistivityNoECIntermediate}) for Coulomb blockaded puddles.
Equation (\ref{eq:resistivityNoECIntermediate}) is valid also in the presence of large charging energy $E_C \gg \delta$. The domain of integration over $\Gamma$ in Eq.~(\ref{eq:resistivityNoECIntermediate}) can be divided into different regions determined by the temperature.
The integrand, $\propto \Delta G^{\text{av}}$, is given by Eqs.~(\ref{eq:DeltaGavWithECTggGamma}), 
(\ref{eq:DeltaGavIntermediate2}), and 
(\ref{eq:DeltaGavWithECTllTKmin}) in the regions I, II, and III, respectively. 
The dominant contribution to Eq.~(\ref{eq:resistivityNoECIntermediate}) at any temperature $T\ll \delta$ comes from the shaded part in region II, where $\Delta G^{\text{av}}$ is given by  Eq.~(\ref{eq:DeltaGavWithECTggThalf}).
Contribution from region III is never dominant in Eq.~(\ref{eq:resistivityNoECIntermediate}).}
\label{fig:integration}
\end{figure}

We will next calculate the long-edge resistivity $\varrho$ in the limit of large charging energy, $E_{C}\gg\delta$. 
The main contribution to the resistivity comes from puddles with an odd number of electrons. In those puddles the average of $\Delta G$ over level widths $\Gamma_n$ is dominated by their typical values, $\Gamma_n \sim \Gamma$ (unlike in the previous subsection).
Therefore, Eqs.~(\ref{eq:DeltaGavWithECTggGamma}), and
(\ref{eq:DeltaGavWithECTggThalf})--(\ref{eq:DeltaGavWithECTllTKmin}) 
give the conductance correction averaged over dot parameters and gate voltage, for a fixed edge-dot distance $d$. 
To find the long-edge resistivity, we need then to sum over possible distances. As explained in the end of the previous subsection, this sum can be converted into an integral over the average level width $\Gamma$.
This integral is given by Eq.~(\ref{eq:resistivityNoECIntermediate}) which is applicable also here, in the presence of large charging energy.
However, the integration over $\Gamma$ is more involved than in the previous subsection. We detail the integration in Fig.~\ref{fig:integration}. 

At temperatures $T \gg \delta e^{-\pi E_{C}/2\delta}$, the integration domain can be divided into two regions (I and II, see Fig.~\ref{fig:integration}) where $\Delta G^{\text{av}}$ is given by Eqs.~(\ref{eq:DeltaGavWithECTggGamma}) and (\ref{eq:DeltaGavIntermediate2}), respectively. The shaded part in region II, where $\Gamma \sim \delta$, gives the main contribution to the integral. 
Using Eq.~(\ref{eq:DeltaGavWithECTggThalf}) in Eq.~(\ref{eq:resistivityNoECIntermediate}) we find then \begin{subequations}
\begin{flalign}
\varrho \sim \frac{n_p \Lambda}{G_0} \frac{\delta}{g^{2}E_{C}}\frac{1}{\ln\frac{\delta}{T}}\,,\quad\delta e^{-\pi E_{C}/2\delta}\ll T\ll\delta\,.
\label{eq:GavLongEdgeWithEcHighT}
\end{flalign}

At lower temperatures  ${T\ll\delta e^{-\pi E_{C}/2\delta}}$, Eq.~(\ref{eq:DeltaGavIntermediate2}) is no longer applicable up to $\Gamma \sim \delta$, but only up to the smaller value $\Gamma \sim E_{C}/\ln\frac{\delta}{T}$, and region II shrinks accordingly, see Fig.~\ref{fig:integration}. Still, the main contribution to Eq.~(\ref{eq:resistivityNoECIntermediate}) comes from the upper end of this region, described by Eq.~(\ref{eq:DeltaGavWithECTggThalf}). We find
\begin{flalign}
\varrho \sim  \frac{n_p \Lambda}{G_0} \frac{1}{g^{2}}\frac{1}{\ln^{2}\frac{\delta}{T}}\,,\quad T\ll\delta e^{-\pi E_{C}/2\delta}\,,\label{eq:GavLongEdgeWithEc}
\end{flalign}
\end{subequations}
which matches with Eq.~(\ref{eq:GavLongEdgeWithEcHighT}) at $T\sim\delta e^{-\pi E_{C}/2\delta}$.

We see that in the Coulomb blockade case at any $T\ll \delta$, the resistivity of a long edge is weakly dependent on temperature, see Eqs.~(\ref{eq:GavLongEdgeWithEcHighT})-(\ref{eq:GavLongEdgeWithEc}). This is because most of the resistance originates from puddles with an odd number of electrons. In those puddles the ground state degeneracy induces a logarithmic-in-$T$ helical edge resistivity for a wide range of temperatures. 

The presence of odd-electron-number puddles is probable even if the
strong inequality between the charging energy and level spacing is
broken and even if the number of electrons in a puddle is not large,
i.e., if $E_{C}\sim\delta$ and $N\sim1$. In this case, we expect
the anisotropy of the ``bare'' exchange integral to become $\sim1$.
Taking into account that $g$ is also never a large parameter, we
arrive at the following estimate of edge resistivity 
\begin{equation}
\varrho\sim\frac{n_{p}\Lambda}{G_{0}}\frac{1}{\ln^{2}(\delta/T)},\quad T\ll\delta \,,
\label{eq:varrhoEcEqdelta}
\end{equation}
at low doping, $n_{d}\ll n_{0}$, see Eq.~(\ref{eq:DonorDensityCharacteristic}). This result was included in Section \ref{sec:Qualitative}, see Eq.~(\ref{eq:QualitvarrhoEcEqdelta}).
Equations~(\ref{eq:GavLongEdgeWithEcHighT}), (\ref{eq:GavLongEdgeWithEc}), and (\ref{eq:varrhoEcEqdelta}) above differ from that given in Ref. \onlinecite{vayrynen13}, where $\varrho \propto T^3$, since there the odd-occupied puddles were not accounted for.

\section{Analysis of existing experimental data\label{sec:Analysis}}

In this section we  connect our theory with existing experiments in HgTe and InAs/GaSb quantum wells. 
We start by calculating the characteristic donor density $n_0$ and the density of puddles $n_p$ for HgTe quantum wells.
Considering a well of thickness $d_{QW}=7.0\,\text{nm}$, one has~\cite{konig08} $E_{g}=10\,\text{meV}$ and $v=5.5\times10^{5}\,\text{m/s}$. 
 With a dielectric constant~\cite{konig13} $\kappa=12.7$, we find 
$\alpha=e^2/\kappa v=0.3$. Ignoring the logarithm in Eq.~(\ref{eq:DonorDensityCharacteristic}), we get the characteristic donor density $n_{0}=3.2\times10^{10}/\text{cm}^{2}$.
The doping density in Ref.~\onlinecite{konig07}, $n_{d}\sim10^{11}/\text{cm}^{2}$, was somewhat higher than the above value.
If the sample of Ref.~\onlinecite{konig07} would be in the bulk-metallic phase, we would expect a high  bulk conductivity,~\cite{fu12} $\sigma > e^2/h$.
However, this is not what was observed:
the measured conductance of a long sample with dimensions $L\times W=20\times13\,\textrm{\ensuremath{\mu}m}$
was $G=0.3e^{2}/h$, giving an upper bound for bulk conductivity,
$\sigma \leq GL/W=0.45e^{2}/h$, and indicating an insulating bulk.
Therefore it seems that the samples of Ref.~\onlinecite{konig07} are near the transition to symplectic metal phase,~\cite{fu12} but still on the topological-insulating side of it, as the nonlocal transport~\cite{roth09} and direct measurements of the current distribution in the sample~\cite{Nowack12} indicate.

Our analysis of Subsection \ref{sub:Potential-fluctuations-in}, which assumed $n_d/n_0 \ll 1$, is strictly speaking not valid near the transition. However, in estimating the puddle density, we can still use Eq.~(\ref{eq:DensityOfPuddles}) by extrapolating to the crossover density, $n_d/n_0 \to 1$. This leads to $n_p \sim 1/\ell_{g}a_{B}$. 
 Using~\cite{konig07} $\ell_{g}=130\,\text{nm}$, and $a_{B}=2v\hbar/\alpha E_{g}=230\,\text{nm}$, we find~\footnote{\label{fn:lgvsab}
 Note that $a_B > \ell_g$, contrary to the assumptions of Eq.~(\ref{eq:DensityOfPuddles}). The smaller $\ell_g$ is, the more effective is the screening of the potential fluctuations that create the puddles. In the limiting case of $\ell_g$ shorter than the radius $r_0\sim\alpha a_B$ of ``deep'' states with energies close to the center of the gap, the exponent $n_0/2n_d$ in Eq.~(\ref{eq:DensityOfPuddles}) is multiplied by an additional small factor $\sim (r_0/\ell_g)^2$. We believe, however, that we are only slightly overestimating $n_p$ by using here Eq.~(\ref{eq:DensityOfPuddles}), as the heterostructures of Ref.~\onlinecite{konig07} are in an intermediate regime, $a_B\approx 2\ell_g\approx 3 r_0$.}  
 $n_p \approx 3.3\times10^{9}/\text{cm}^{2}$. Finally, using the wave function penetration depth $\Lambda=v/E_{g}=36\,\text{nm}$, we obtain the number of near-edge puddles per unit length,  $n_{p}\Lambda\approx1.1/\mu \textrm{m}$.
 Therefore it is unlikely that multiple puddles would exist near the edge in short samples of length $L=1\,\mu\text{m}$. 
 This allows us to explain the approximate quantization of conductance $G$ observed in samples with $L = 1\,\mu\text{m}$. 
We may associate the deviation of $G$ from the universal value $G_0=e^2/h$ with electron backscattering off a single, or at most a few, puddles. 
Furthermore, the observed~\cite{konig07} rapid fluctuation in the dependence of $G$ on gate voltage $V_g$ may be related to the sweeping of the chemical potential over the discrete electronic levels of a puddle.

In short samples ($L=1\textrm{\ensuremath{\mu}m}$ for the parameters of Ref.~\onlinecite{konig07}) the
theory of backscattering off a single puddle developed in Subsection~\ref{sub:QualitSingle} becomes relevant. 
We showed that at low temperatures ($T\ll\delta$) the puddle-induced correction $\Delta G$ to the edge conductance displays strong dependence on gate voltage $V_{g}$ (see Fig.~\ref{FigKondoComparison}a). 
This dependence comes from charging the puddle with electrons one by one. 
The gate voltage increment $\Delta V_{g}$ corresponding to the addition of an electron is $\Delta V_{g} \sim e/C_g$ where $C_{g}$ is the gate-puddle capacitance.
Replacing $C_g$ by the puddle total capacitance $C$, we have $\Delta V_{g} \sim E_C/e$, where $E_C=e^2/2C$ is the puddle charging energy.
From Eq.~(\ref{eq:PuddleParams}) we have $E_C \sim \alpha^2 E_g$ so that $\Delta V_{g} \sim 1\textrm{mV}$.
This is a lower estimate for $\Delta V_{g}$ since, by replacing $C_g$ with $C$, we ignored screening by other puddles and by the helical edges.
Indeed, from Ref.~\onlinecite{konigThesis} we estimate a higher value, $\Delta V_{g} \sim 50 \textrm{mV}$.
More importantly, our theory predicts the temperature dependence of the helical edge resistance $R$.
The increase $\Delta R$ of the resistance $R=R_0 +\Delta R$ above the single-edge universal value $R_0 = h/e^2$ is related to $\Delta G$ evaluated in Subsection~\ref{sub:QualitSingle}, $\Delta R=\Delta G/G_{0}^{2}$. Note that maxima in backscattering correspond to maxima in the resistance.
At $T\ll \delta$, the maxima and average value of $\Delta R$ depend weakly on temperature over a broad temperature interval [see Eqs.~(\ref{eq:QualitPeakHighTWithEC}), (\ref{eq:TypicalDeltaGPeakHighTWithEC}), (\ref{eq:DeltaGavWithECTggGamma})]. 
This is in contrast with the minima of $\Delta R$ which obey a stronger ($T^4$ power-law) dependence,
see Eqs.~(\ref{eq:QualitDeltaGevenLowT}) and (\ref{eq:TypicalDeltaG1ValleyLowTWithECCmb}).
 The amplitude of fluctuations, $\Delta R^{\text{max}}-\Delta R^{\text{min}}$,
decays with increasing temperature. The temperature scale above which
fluctuations become suppressed is the puddle charging energy, which typically is of the same order of magnitude as level spacing, $E_{C}\sim\delta$, see Eq.~(\ref{eq:PuddleParams}). 
 While there is no detailed measurements of temperature dependence of resistance  in short samples, it has been established that the resistance fluctuations decrease with temperature.~\cite{konig07,konigThesis,molenkamp14unpublished} 
 However, the suppression of fluctuations with increased temperature is not unique to the above mechanism due to puddles.
For example, coherent backscattering off correlated spins may produce
such a feature.~\cite{cheianov13,altshuler13} Considering spinful
puddles close to the edge, we can estimate the temperature
scale $T_{SG}$, at which the puddle spins become correlated.~\cite{cheianov13}
Using $n_{p}\Lambda=1.1/\mu m$ and $\rho J\sim 1$, we get $T_{SG}=v(\rho J)^{2}n_{p}\Lambda/4\pi\sim0.3\,\text{K}$, which is clearly an over-estimate, since we expect $\rho J\lesssim 1$.
Note also that the derivation of $T_{SG}$ assumes the presence of at least two  puddles near the edge, which may not be true in short samples.

Strong evidence for puddles was given in Ref.~\onlinecite{konig13}, where the edge conductance was shown to be very sensitive to local gating. However, we cannot fully explain the single instance where the authors were able to completely suppress ($\Delta G\approx e^2/h$) conduction of an edge state. 
In our theory, an accidental large ($\delta\ll T$) puddle near the edge would give a conductance correction $\Delta G=e^2/2h$, see Eq.~(\ref{eq:QualitDeltaGHighTSaturated}). 

In long samples ($L > n_p^{-1/2}$) multiple puddles may lie near the edge, leading to resistance scaling linearly with sample length, $R=\varrho L$.
Experiments with  HgTe samples of $L\geq10\,\textrm{\ensuremath{\mu}m}$ show a strongly reduced edge conductance,~\cite{konig07,Nowack12} or resistive
behavior.~\cite{gusev13} In long samples the current carried by edge modes may start to leak into the bulk.~\cite{vayrynen13} 
 This happens in samples longer than the leakage length $L^{*}=1/\sigma\varrho$, where $\sigma$ is the bulk conductivity and $\varrho$ the edge resistivity. 
In Ref.~\onlinecite{Nowack12} the authors showed that the current was carried by the edges in the topological state, despite the inferred high edge resistance $R_{\text{edge}}=16h/e^{2}$  at $T\approx3\,\text{K}$.
Using the estimated bulk resistance $R_{\text{bulk}}=58h/e^{2}$, and the sample dimensions $L\times W=50\times30\,\textrm{\ensuremath{\mu}m}^{2}$, we obtain $\sigma \leq L/WR_{\text{bulk}}\sim0.029e^{2}/h$, while the edge resistivity is $\varrho_{\text{exp}}\sim0.3(h/e^{2})/\textrm{\ensuremath{\mu}m}$.
This leads to leakage length $L^{*}\geq 110\,\mu\text{m}>L$, in agreement
with low leakage found in Ref.~\onlinecite{Nowack12}.
Finally, we can estimate the theoretical value for puddle-induced resistivity $\varrho_{\text{thy}}$, by calculating the pre-factor in Eq.~(\ref{eq:varrhoEcEqdelta}), $\varrho_{\text{thy}} \sim (h/e^2) n_{p}\Lambda$.
Taking~\cite{Nowack12} $\ell_{g}=84\,\text{nm}$ and using our previously calculated values 
 $a_{B}=230\,\text{nm}$ and $\Lambda=36\,\text{nm}$, we find $n_{p}\sim1/\ell_{g}a_{B}=5\times10^{9}\,\text{cm}^{-2}$.
This leads to $\varrho_{\text{thy}} \sim1.9 (h/e^{2})/\mu\textrm{m}$ and $R_{\text{thy, edge}}\sim 95h/e^{2}$. Our estimate $R_{\text{thy, edge}}$, which is about 6 times higher then the actual  $R_{\text{edge}}$, is reasonably close to the measured value, considering the probable overestimation~\cite{Note7} of $n_p$.

The temperature dependence of long edge resistance has been studied in HgTe,~\cite{Nowack12,gusev13}  as well as in Si-doped topological InAs/GaSb quantum wells.~\cite{spanton14,du13} 
In the latter system Ref.~\onlinecite{du13} finds a temperature-independent resistance over three decades of temperatures, up to $T\approx  4.2\,\textrm{K}$.
This is somewhat at odds with our predicted dependence at low temperatures, $\varrho \propto {1/\ln^{2}(\delta/T)}$, see Eq.~(\ref{eq:varrhoEcEqdelta}).
Measurements at higher temperatures,~\cite{spanton14}  $T \gtrsim 5\,\textrm{K}$, in the same system show very little $T$-dependence in edge resistance.
In HgTe, the resistance at $T \gtrsim 5\,\textrm{K}$ has been measured, and its observed $T$-dependence seems to indicate an onset of activated bulk conduction.~\cite{Nowack12,gusev13}  
Using the parameters of Ref.~\onlinecite{konig07}, we find $\delta \sim 10\, \mathrm{K}$, which makes it possible that $T\gtrsim \delta$ in these experiments.
Thus, to definitely support or rule out the existence of puddles,  experiments at lower temperatures, $T\ll \delta$, are needed. In the next, concluding section, we summarize  the predictions for the low-temperature resistance induced by puddles.

\section{Conclusions\label{sec:Conclusions}}

We have shown in the previous section how several key measurement results are explained qualitatively by the puddle physics. The explained features include the high resistance of the helical edges in long samples and the fluctuations of the excess resistance in short ones.
However, we are not aware of detailed low-temperature data which would allow quantitative comparison with the theory developed in this paper.
In this section, we will summarize the predictions for the resistance at low temperatures $T \ll \delta$.

We will start by considering a single puddle with relatively large charging energy, $E_C \gg  \delta$. Such a puddle may be found in short samples ($L=1\,\mu \text{m}$, using the parameters of Ref.~\onlinecite{konig07}), or perhaps even formed intentionally.~\cite{konig13} Either way, the resulting puddle-induced edge resistance $\Delta R$ at low temperatures ($T \ll \delta$) will display very distinct dependence on the temperature and gate voltage. (Here $\Delta R=R-R_0$ is the deviation of edge resistance from the universal value $R_0=h/e^2$.)
Plotted against gate voltage, the resistance $\Delta R$ displays narrow peaks separating valleys, see Fig.~\ref{FigKondoComparison}. 
The resistance $\Delta R$ strongly depends  on the parity of electron number in the puddle, and therefore is very different in neighboring valleys. When the puddle occupation is even, the puddle-induced resistance is small and $R$ decays fast, $\Delta R \propto T^4$, towards the universal value, see Eq.~(\ref{eq:QualitDeltaGevenLowT}). 
When the occupation is odd, however, the decay is slow, $\Delta R \propto \ln^2 T$, see Eq.~(\ref{eq:QualitDeltaGoddRgd}). This slow decay persists over a broad interval of temperatures, down to $T\sim \delta \exp(-E_C/\Gamma)$. (Here $\Gamma$ is the width of electron level in a puddle; $\Gamma$ reaches its maximum, $\Gamma \sim \delta$, when the puddle is very close to the edge.) 
Only at very low temperatures, below $T\sim \delta \exp(-\pi E_C/2\Gamma)$,  do we have $\Delta R \propto T^4$ throughout an odd valley, see Eqs.~(\ref{eq:QualitDeltaGevenLowT}) and (\ref{eq:QualitDeltaGoddLowT}).
In the peaks between the valleys, $\Delta R$ is approximately independent of temperature at $T \gg \Gamma$ [see Eq.~(\ref{eq:QualitPeakHighTWithEC})], and has the aforementioned power-law-dependence at lower temperatures, see Eq.~(\ref{eq:QualitPeakSubleadingNoEC}).
Combining the above description of peaks and valleys, we finally note that the amplitude of fluctuations in 
$\Delta R$ decreases with increasing temperature, while the gate-voltage-averaged $\Delta R$ grows.

As we argued in Subsection~\ref{sub:Potential-fluctuations-in}, in a typical doping-induced puddle we have $\delta \sim E_C$.
The signatures illustrated by Fig.~\ref{FigKondoComparison} remain valid. 
We must note, also, that if the puddle is very close to the edge, $\Gamma \sim \delta$, then the slow $T$-dependence, $\Delta R \propto \ln^2 T$, in the odd-occupied valleys may not persist over a broad temperature interval.

On a long edge, there are always puddles far from the edge.
Such remote puddles with an odd electron number dominate the low-temperature resistivity of a long edge.
Taking $E_C \sim \delta$, as is appropriate for doping-induced puddles, we find resistivity $\varrho \propto 1/\ln^2 (\delta/T)$, see Eq.~(\ref{eq:QualitvarrhoEcEqdelta}). 
This result is valid at temperatures below the average level spacing, $T\ll \delta$.
 The logarithmic temperature-dependence of $\varrho$ is one of the main testable predictions of this paper.

\begin{acknowledgments}
We thank Katja Nowack for discussions. 
This work was supported by NSF DMR Grant No. 1206612, U.S.-Israel Binational Science Foundation (Grant 2010366), the Simons Foundation, and the Bikura (FIRST) program of the Israel Academy of Science. LG is grateful to the Aspen Center for Physics, supported by NSF Grant No. PHYS-106629, for hospitality which helped to complete this work.
\end{acknowledgments}

\appendix

\section{Tunneling through an extended edge-dot contact\label{Adx:Tunneling-through-an}}

In the main text of this work we assumed a point-contact model for
the tunneling between the helical edge and the quantum dot. That led
us to momentum-independent, diagonal (in the Kramers index) tunneling
matrix elements in the Hamiltonian (\ref{eq:FreeHamiltonian}). The
goal of this section is to demonstrate that our main conclusions
remain valid for an extended edge-dot junction. First, we explicitly
show below how TR symmetry enforces reflectionless edge transport, at any electron energy, if there is no interaction inside the dot. Next, in Subsection~\ref{par:Presence-of-interaction}, we include
interaction, and show that in the low-temperature limit the
backscattering correction to the conductance remains universal, $\Delta G \propto T^{4}$. Finally, we compare the conductance corrections from quantum dots with extended and point-like contacts. We find that dots  with an extended contact yield a smaller correction to the conductance than those coupled to the edge by a point-contact.

We start by introducing the most general tunneling Hamiltonian,
\begin{flalign}
H_{\text{tun}}= & \sum_{kn}\sum_{\alpha\beta=L,R}t_{k\alpha n\beta}d_{k\alpha}^{\dagger}c_{n\beta}+h.c.\label{eq:GenericTunnelingHamiltonian}
\end{flalign}
The matrix elements of $H_{\text{tun}}$ satisfy the relation dictated by TR symmetry,
\begin{equation}
t_{k\alpha n\beta}=\alpha\beta t_{k\overline{\alpha}n\overline{\beta}}^{*}\,.\label{eq:TRtrfmationofHtun}
\end{equation}
The tunneling matrix elements $t_{k\alpha n\beta}$ form a $2\times2$
matrix in the Kramers indices $\alpha,\,\beta$. We can diagonalize
this matrix and make its elements real in the diagonal basis by performing a unitary transformation of the dot states. Introducing phases $\theta_{k\alpha n\beta} \in [0,2\pi)$, we can write $t_{k\alpha n\beta}=|t_{k\alpha n\beta}|e^{i\theta_{k\alpha n\beta}}$. Then the diagonalizing unitary transformation reads
\begin{align}
 & \left(\begin{array}{c}
c_{nL}\\
c_{nR}
\end{array}\right)\rightarrow\frac{1}{\sqrt{|t_{kLnR}|^{2}+|t_{kLnL}|^{2}}}\nonumber \\
 & \times\left(\begin{array}{cc}
e^{-i\theta_{kLnL}}|t_{kLnL}| & -e^{i\theta_{kLnR}}|t_{kLnR}|\\
e^{-i\theta_{kLnR}}|t_{kLnR}| & e^{i\theta_{kLnL}}|t_{kLnL}|
\end{array}\right)\left(\begin{array}{c}
c_{nL}\\
c_{nR}
\end{array}\right)\,.\label{eq:UnitaryDotTsfm}
\end{align}
This rotation generally depends on $k$. If $t_{k\alpha n\beta}$
is independent of $k$ (tunneling through point-contact), as assumed
in the main text, the rotation puts the Hamiltonian in the simple
form of Eq.~(\ref{eq:FreeHamiltonian})
\begin{flalign}
H_{\text{p-c-tun}}= & \sum_{n\gamma}t_{n}\psi_{\gamma}^{\dagger}(0)c_{n\gamma}+h.c.,\label{eq:PCTunnelingHamiltonian}
\end{flalign}
with $t_{n}=\sqrt{|t_{LnR}|^{2}+|t_{LnL}|^{2}}$.

In this form, the two complementary components of the Kramers doublets
for any level are coupled to the respective channels $L$, $R$; there are
no dot states which couple simultaneously to $L$ and $R$. Clearly, the Hamiltonian
(\ref{eq:PCTunnelingHamiltonian}) does not lead to backscattering.
In the general case of Eq.~(\ref{eq:GenericTunnelingHamiltonian})
the unitary transformation (\ref{eq:UnitaryDotTsfm}) does not lead
to such decoupling. Nevertheless, TR symmetry forbids elastic backscattering, as we will demonstrate next.

The rate $r_{i\rightarrow f}$ of transition from an initial state $|i\rangle$ to a final state $|f\rangle$ is
\begin{equation}
r_{i\rightarrow f}=2\pi\left|\langle f|\mathcal{T}(E_{i})|i\rangle\right|^{2}w_{i}\delta(E_{i}-E_{f})\,,
\end{equation}
where $w_{i}$ is the thermal probability factor of state $|i\rangle$.
The $T$-matrix is defined as 
\begin{equation}
\mathcal{T}(E)=H_{\text{tun}}+H_{\text{tun}}(E-H_{0})^{-1}\mathcal{T}(E)\,,\label{eq:TmatrixDefinition2nd}
\end{equation}
where $H_{0}$ is the unperturbed Hamiltonian, and $H_{\text{tun}}$, Eq.~(\ref{eq:GenericTunnelingHamiltonian}), is the pertubation. 
Without interaction, the unperturbed Hamiltonian in Eq.~(\ref{eq:TmatrixDefinition2nd})
is 
\begin{align*}
H_{0}= & \sum_{q\gamma}E_{q}d_{q\gamma}^{\dagger}d_{q\gamma}+\sum_{n\gamma}\varepsilon_{n}c_{n\gamma}^{\dagger}c_{n\gamma}\,.
\end{align*}
Consider first backscattering of a single electron. We will denote
the initial and final states as single-particle excitations over some
common vacuum state, $|i\rangle=|qL\rangle$ and $|f\rangle=|q'R\rangle$.
From the TR symmetry of the Hamiltonian $H_{0}$ and the perturbation $H_{\text{tun}}$, it follows that also the $T$-matrix is TR-symmetric.
Therefore we have ${\langle q'R|\mathcal{T}(E)|qL\rangle}={-\langle qR|\mathcal{T}(E)|q'L\rangle}$.
Since energy is conserved, $E_{q}=E_{q'}$, the amplitude vanishes,
\begin{align}
\langle q'R|\mathcal{T}(E)|qL\rangle= & 0\,.\label{eq:1particleTunnelingTRSNoEc}
\end{align}
Similarly, the backscattering amplitude of non-interacting many-particle states vanishes.

\subsection{Backscattering in the constant-interaction model \label{par:Presence-of-interaction}}
In this section we will show that even in the constant-interaction model there is backscattering if the edge-dot tunneling contact is extended.
We will also compare the resulting conductance correction, $\Delta G^{\text{ext}}$, to the one due to a quantum dot point-contact-coupled to the edge. 
It is found that generally a point-like edge-dot junction gives rise to a larger correction to edge conductance.

The simplest interacting model for the dot is the Anderson-type Hamiltonian of a single level,
\begin{equation}
H_{\text{dot}}=\sum_{\gamma=L,R}\varepsilon_{0}c_{\gamma}^{\dagger}c_{\gamma}+Vn_{L}n_{R}\,; \quad n_\gamma = c_{\gamma}^{\dagger}c_{\gamma}\,,
\end{equation}
where $V\gtrsim\varepsilon_{0}$ and $\varepsilon_{0}>0$ is above
the Fermi level so that the level is empty in its ground state. We will consider low temperatures $T\ll\varepsilon_{0}$ at which transport through
the level is dominated by virtual processes. Using the $T$-matrix
approach, we will calculate the backscattering current $\Delta I$,
\begin{equation}
\Delta I=2\pi e\sum_{if} \Delta N_{i \to f}\left|\langle f|T^{(4)}(E_{i})|i\rangle\right|^{2}w_{i}\delta(E_{i}-E_{f})\,,
\label{eq:DeltaIApndx1}
\end{equation}
where the sum is over initial and final states ($|i\rangle$ and $|f\rangle$), $\Delta N_{i \to f}$ is the number of backscattered particles, 
and $T^{(4)}$ is the $T$-matrix fourth order in tunneling, Eq.~(\ref{eq:TmatrixDefinition2nd}). 
Assuming that the level is empty in the initial and final states, we
have two contributions to the $T$-matrix, 
\begin{equation}
T^{(4)}(E_{i})=T_{-+-+}^{(4)}(E_{i})+T_{--++}^{(4)}(E_{i})\,,
\end{equation}
where
\begin{align}
 & T_{-+-+}^{(4)}(E_{i})\nonumber \\
 & =H_{\text{tun}}^{-}\frac{1}{E_{i}-H}H_{\text{tun}}^{+}\frac{1}{E_{i}-H}H_{\text{tun}}^{-}\frac{1}{E_{i}-H}H_{\text{tun}}^{+}\,,\\
 & T_{--++}^{(4)}(E_{i})\nonumber \\ \label{eq:outoutinin}
 & =H_{\text{tun}}^{-}\frac{1}{E_{i}-H}H_{\text{tun}}^{-}\frac{1}{E_{i}-H}H_{\text{tun}}^{+}\frac{1}{E_{i}-H}H_{\text{tun}}^{+}\,,\\ 
 & H_{\text{tun}}^{-}=(H_{\text{tun}}^{+})^{\dagger}=\sum_{kn}\sum_{\alpha\beta=L,R}t_{k\alpha n\beta}d_{k\alpha}^{\dagger}c_{n\beta}\,.
\end{align}
The term $T_{-+-+}^{(4)}$ corresponds to a process where at most one
virtual electron occupies the level at a given moment. The second
term $T_{--++}^{(4)}$ has a virtual intermediate state where the level is doubly
occupied and therefore the second denominator in  Eq.~(\ref{eq:outoutinin}) is of order $V$.
In the non-interacting limit $V\rightarrow0$ these two processes cancel each other and the amplitude for backscattering vanishes. 
Considering a final state $\langle f|=\langle i|d_{q_{1}L}^{\dagger}d_{q_{2}L}^{\dagger}d_{q_{3}L}d_{q_{4}R}$
(we assume unequal momenta and conserved total energy), the amplitudes
for these two 
processes are
\begin{align*}
 & \langle f|T_{-+-+}^{(4)}(E_{i})|i\rangle=\langle i|n_{q_{1}L}n_{q_{2}L}(1-n_{q_{3}L})(1-n_{q_{4}R})|i\rangle\\
 & \times\gamma_{f}\frac{2\varepsilon_{0}-E_{q_{1}}-E_{q_{2}}}{\prod_{i=1}^{4}(\varepsilon_{0}-E_{q_{i}})}\,,\\
 & \langle f|T_{--++}^{(4)}(E_{i})|i\rangle=-\langle i|n_{q_{1}L}n_{q_{2}L}(1-n_{q_{3}L})(1-n_{q_{4}R})|i\rangle\\
 & \times\gamma_{f}\frac{(2\varepsilon_{0}-E_{q_{1}}-E_{q_{2}})^{2}}{(2\varepsilon_{0}+V-E_{q_{1}}-E_{q_{2}})\prod_{i=1}^{4}(\varepsilon_{0}-E_{q_{i}})}\,,
\end{align*}
where 
\[
\gamma_{f}=\sum_{\gamma=L,R}\left(t_{q_{1}L\gamma}^{*}t_{q_{2}L\overline{\gamma}}^{*}-t_{q_{1}L\overline{\gamma}}^{*}t_{q_{2}L\gamma}^{*}\right)t_{q_{3}L\overline{\gamma}}t_{q_{4}R\gamma}\,.
\]
(Note that $\gamma_{f}$ vanishes if the tunneling amplitude $t_{q\alpha \beta}$ is independent of momentum.)
Accounting only for the above 1-particle backscattering process, correction to ideal conductance from Eq.~(\ref{eq:DeltaIApndx1}) becomes, 
\begin{align}
 & \Delta G_{1}^{\text{ext}}/G_{0} =\frac{4\pi}{T}\sum_{q_{i}}\frac{|\gamma_{f}|^{2}V^{2}}{(2\varepsilon_{0}+V-E_{q_{1}}-E_{q_{2}})^{2}}\nonumber \\
 & \times \frac{(2\varepsilon_{0}-E_{q_{1}}-E_{q_{2}})^{2}}{\prod_{i=1}^{4}(\varepsilon_{0}-E_{q_{i}})^{2} \cosh \frac{E_{q_i}}{2T}} \delta(E_{q_{1}}+E_{q_{2}}-E_{q_{3}}-E_{q_{4}})\,.
\end{align}

At low energies we can expand the tunneling amplitudes $t_{q\alpha\beta}$ in energy. In particular,
\[
\gamma_{f}=u^{-1}(E_{q_{1}}-E_{q_{2}})\sum_{\gamma,\alpha}\alpha t_{k_{F}\alpha\overline{\gamma}}t_{k_{F}\overline{\alpha}\gamma}t_{k_{F}L\overline{\gamma}}^{*}\left[\partial_{q}t_{qL\gamma}^{*}\right]_{q=k_{F}}\,.
\]
The conductance correction at low temperatures becomes (we drop from $\Delta G_{1}^{\text{ext}}$ the subscript indicating backscattering of one particle)
\begin{align}
 & \Delta G^{\text{ext}}/G_{0} =\pi(2\pi)^{4}\frac{16}{15}\frac{T^{4}V^{2}}{(2\varepsilon_{0}+V)^{2}\varepsilon_{0}^{6}}\nonumber \\
 & \times\left|\rho^{2}\sum_{\gamma,\alpha}\alpha t_{k_{F}\alpha\overline{\gamma}}t_{k_{F}\overline{\alpha}\gamma}t_{k_{F}L\overline{\gamma}}^{*}u^{-1}\left[\partial_{q}t_{qL\gamma}^{*}\right]_{q=k_{F}}\right|^{2}\,,
 \label{eq:G1ExtendedTunneling}
\end{align}
in agreement with the general $\Delta G \propto T^4$ result (see Section \ref{sec:Qualitative}). 

The derivative $\partial_{q}t_{qL\gamma}^{*}$ in Eq.~(\ref{eq:G1ExtendedTunneling}) introduces a length scale $l_t$, which can be interpreted as the size of the tunneling region.
We will next find out how large $l_t$ has to be, in order to yield $\Delta G^{\text{ext}}$ comparable to that given by a point-contact-coupled quantum dot. 

To get a rough estimate, we write $\partial_{q}t \sim l_tt$, and  introduce the level width $\Gamma \sim \rho t^2$. 
We also denote $\varepsilon_0 \equiv E_C$, to ease the comparison to a Coulomb blockaded dot, discussed in Subsection~\ref{sub:QualitSingle}. 
In the limit of large charging energy, the virtual double-occupied dot  states are rare, and we have $V \gtrsim E_C$.
Then Eq.~(\ref{eq:G1ExtendedTunneling}) becomes 
\begin{align}
 &\Delta G^{\text{ext}}/G_{0} \sim \frac{T^{4} \Gamma^4 (l_t/u)^2}{E_{C}^{6}} \,.
 \label{eq:G1ExtendedTunnelingAverage}
\end{align}
The conductance correction $\Delta G^{\text{p-c}}$ due to a dot identical-to-above, but with a point-like contact to the edge, is given in Eq.~(\ref{eq:QualitDeltaGevenLowT}).
From Eq.~(\ref{eq:G1ExtendedTunnelingAverage}), comparing with the point-contact result Eq.~(\ref{eq:QualitDeltaGevenLowT}), we see that
 $\Delta G^{\text{ext}} \gtrsim \Delta G^{\text{p-c}}$ only if the  tunneling region is large enough, $l_t \gtrsim u E_T^{-1} E_C/ \delta$, where $E_T^{-1}=1/g\delta$ is the dot Thouless time. 
 The Thouless time is the diffusion time of electron across the dot, and therefore always larger than the ballistic crossing time $w/u$, $w$ being the dot linear size. 
 Since also $E_C /\delta \gtrsim 1$, we find $l_t \gtrsim w$, \emph{i.e.}, junction has to be wider than the dot linear size.
 This is a justification for using a point-contact coupling between the edge and the dot in the main text.

\section{Averaging over disorder\label{sec:Averaging-over-disorder}}

In this section, we show how to obtain some of the averages used in the main text. 
In particular, in Subsection~\ref{sub:AverageU}, we calculate the average matrix element of the non-universal interaction $U$. We show that for screened Coulomb interaction $\langle U^2 \rangle \sim \delta^2 / g^2$.
In Subsections \ref{sec:AveragesOverLevels} and \ref{sec:AveragesOverLevels2}, we show how to obtain the RMT averages over the dot energy
levels cited in the main text, in Sections \ref{sec:BSinAbsenceOfEC} and \ref{sub:Even-valley-conductance}, respectively. 
The RMT averages used in Subsection \ref{sub:Odd-valleys-of} are given later, in  Appendix \ref{sec:Components-of-the}.

\subsection{Average of the non-universal part of interaction $U$ \label{sub:AverageU}}

In this subsection we give the typical value of the (squared) backscattering matrix elements of the interaction $U$. 
The backscattering matrix elements are given by the off-diagonal, non-universal part of the intra-dot interaction. 
In this subsection, we denote this off-diagonal part by $U$. 
(The same convention is followed in Section~\ref{sec:CoulombBlockade}, where the special notation $E_C$ is reserved for the diagonal part.)
Note that in the symplectic ensemble~\footnote{We use the symplectic ensemble since a generic single-particle Hamiltonian of the dot does not conserve
any component of the electron spin (i.e., the spin scattering rate is
larger than the level spacing, $\tau_{so}\delta\ll1$).} all the backscattering matrix elements are on average equal. Using the standard diagrammatic approach,~\cite{aleiner02} we find, 
\begin{equation}
\frac{\langle U^{2}\rangle}{\delta^{2}}=\frac{\nu_{0}^{2}}{8\pi^{2}A^{2}}\left(2U(0)-\left\langle U(k+k')\right\rangle _{FS}\right)^{2}\sum_{\gamma_{n}\neq0}\frac{\delta^{2}}{\gamma_{n}^{2}}\,.\label{eq:AverageInteractionMatrixelement}
\end{equation}
Here $\langle \dots \rangle $ denotes average over dot levels,  $\nu_{0}=1/\delta$ is the dot
density of levels, $A$ is the area of
the dot, $\langle\cdots\rangle_{FS}$ denotes average over the Fermi
surface, and $U(k)$ is the Fourier-transform of the interaction potential
$U(x-y)$. The sum is over the eigenvalues $\gamma_{n}$ of the diffusion
operator with Neumann boundary conditions. 
The lowest non-zero eigenvalue $\gamma_{1}=E_{T}$ is the inverse time
to diffuse through the system (Thouless energy), and gives the dimensionless
conductance of the dot, $g=E_{T}/\delta$. Introducing a shape-dependent
coefficient $c_{\gamma}=\sum_{\gamma_{n}\neq0}(\gamma_{1}/\gamma_{n})^{2}$
we can write $\sum_{\gamma_{n}\neq0}(\delta/\gamma_{n})^{2}=c_{\gamma}/g^{2}$
in Eq.~(\ref{eq:AverageInteractionMatrixelement}). For the special
but realistic case of screened Coulomb interaction the zero-momentum
component $U(0)=A/2\nu_{0}$ dominates, $\langle U(k)\rangle_{FS}\ll U(0)$,
and we find 
\begin{equation}
\langle U^{2}\rangle=c_{\gamma}\delta^{2}/8\pi^{2}g^{2}\,.\label{eq:UaverageCoulomb}
\end{equation}

\subsection{Averages over dot energy levels in Section \ref{sec:BSinAbsenceOfEC} \label{sec:AveragesOverLevels}}

In this subsection we show how to average over the level sums in Eqs.~(\ref{eq:G1-particlevalleyNocharging}), (\ref{eq:DeltaG1PeakHighTNoEC}), and (\ref{eq:DeltaG1PeakLowTNoEC}), to obtain Eqs.~(\ref{eq:TypicalDeltaG1PeakLowTNoEC})
and (\ref{eq:TypicalDeltaG1ValleyLowTNoEC}) of the main text. 
Our main tool in averaging over dot energy levels is the $n$-point correlation function defined as 
\begin{equation}
R_{n}(x_{1},x_{2},\dots,x_{n})=\langle\sum_{\{m_{i}\}}\,'\delta(x_{1}-\frac{\varepsilon_{m_{1}}}{\delta})\dots\delta(x_{n}-\frac{\varepsilon_{m_{n}}}{\delta})\rangle\,,
\label{eq:RnDef}
\end{equation}
where the brackets denote average over disorder, and the primed sum
is taken over all unequal levels. Therefore $R_{n}$ vanishes if any
of the coordinates coincide, which is a manifestation of level repulsion.
The correlation functions $R_{n}$ are known from RMT.\cite{mehta04} For us, the most important correlation function is
the 2-point function $R_{2}$. Its asymptotic behavior for matching
arguments is $R_{2}(x,y)\sim|x-y|^{\beta}$ when $|x-y|\ll1$, while
distant levels are uncorrelated, $R_{2}(x,y)\sim1$ as $|x-y|\gtrsim1$.
Here $\beta=1,2,4$ for Gaussian orthogonal, unitary, and symplectic ensembles, respectively. 
Due to the strong spin orbit coupling in the topological insulators, we use the symplectic ensemble throughout.~\cite{Note8} 
The higher order correlation functions behave similarly.

The numbers $r_{1}$ and $r_{2}$ in Eqs.~(\ref{eq:TypicalDeltaG1PeakLowTNoEC})
and (\ref{eq:TypicalDeltaG1ValleyLowTNoEC}) are given by 
\begin{align}
r_{1}= & \delta^{2}\left\langle\sum_{n,n'\neq1}\frac{1}{\varepsilon_{n}\varepsilon_{n'}}\right\rangle \,,\\
r_{2}= & \delta^{10}\left\langle \sum_{m_{1}\neq m_{2}}\sum_{n_{1}\neq n_{2}}\sum_{m_{3,4},n_{3,4}}\right.\nonumber \\*
 & \times\left.\frac{1}{\varepsilon_{n_{1}}\varepsilon_{m_{1}}\varepsilon_{n_{2}}^{2}\varepsilon_{m_{2}}^{2}\varepsilon_{n_{3}}\varepsilon_{m_{3}}\varepsilon_{n_{4}}\varepsilon_{m_{4}}}\right\rangle .
\end{align}
 These averages can be calculated using the correlation functions
$R_{n}$. Note that in Eqs.~(\ref{eq:TypicalDeltaG1PeakLowTNoEC}) and (\ref{eq:TypicalDeltaG1ValleyLowTNoEC})
we have the extra condition that level $\varepsilon_{1}$ is
excluded from the sum, \emph{i.e.}, none of the levels summed over is at
the Fermi energy. For example, we then get 
\begin{align}
r_{1}=&\int_{-\infty}^{\infty}dx\frac{1}{x^{2}}R_{2}(x,0)+\int_{-\infty}^{\infty}dx\int_{-\infty}^{\infty}dy\frac{1}{xy}R_{3}(x,y,0) \nonumber \\
\approx & \, 30 \,.
\end{align}
We leave out evaluation of the level sum $r_2$, since it requires, for example, integration of the 6-point function $R_6$. 
In principle, $r_2$ can be evaluated similarly to $r_1$.

\subsection{Averages over dot energy levels in Subsection \ref{sub:Even-valley-conductance} \label{sec:AveragesOverLevels2}}

In this section we derive from Eq.~(\ref{eq:DeltaG1EvenValley}) the disorder-averaged conductance correction, Eq.~(\ref{eq:TypicalDeltaG1ValleyLowTWithECCmb}), due to 1-particle backscattering in the even Coulomb blockade valley. 
The averages over dot levels are evaluated in similar fashion as in the previous subsection \ref{sec:AveragesOverLevels}.

 The main contribution to the sum over levels in Eq.~(\ref{eq:DeltaG1EvenValley}) comes from all levels
unequal. Dispensing with numerical pre-factors, we get from Eq.~(\ref{eq:DeltaG1EvenValley})
\begin{equation}
\left\langle \Delta G\right\rangle \sim\frac{\Gamma^{4}}{g^{2}\delta^{4}}T^{4}\left[\frac{1}{E_{-}}+\frac{1}{E_{+}}\right]^{4} \mathcal{I}[\text{min}(E_{+},E_{-})]
\label{eq:DeltaGavAppendixB2Intermediate}
\end{equation}
where the dimensionless factor is \begin{widetext}
\begin{align*}
\mathcal{I}(E)= & g^{2}\delta^{4}\left\langle \left|\sum_{n_{i}}(U_{n_{1}Ln_{2}L;n_{3}Ln_{4}R}-U_{n_{1}Ln_{2}L;n_{3}Rn_{4}L})\right.\right.\\
\times & \left\{ \frac{\Theta(\varepsilon_{n_{1}})\Theta(\varepsilon_{n_{2}})\Theta(-\varepsilon_{n_{3}})\Theta(-\varepsilon_{n_{4}})+\Theta(-\varepsilon_{n_{1}})\Theta(-\varepsilon_{n_{2}})\Theta(\varepsilon_{n_{3}})\Theta(\varepsilon_{n_{4}})}{(\varepsilon_{n_{1}}+\varepsilon_{n_{2}}-\varepsilon_{n_{3}}-\varepsilon_{n_{4}})}\frac{2\varepsilon_{n_{3}}-(\varepsilon_{n_{1}}+\varepsilon_{n_{2}})}{(\varepsilon_{n_{3}}-\varepsilon_{n_{1}})^{2}(\varepsilon_{n_{3}}-\varepsilon_{n_{2}})^{2}}(\varepsilon_{n_{1}}-\varepsilon_{n_{2}})\right.\\
 & \left.\left.\left.+\Theta(-\varepsilon_{n_{1}})\Theta(\varepsilon_{n_{2}})\Theta(\varepsilon_{n_{3}})\Theta(-\varepsilon_{n_{4}})\frac{(\varepsilon_{n_{1}}-\varepsilon_{n_{2}}-\varepsilon_{n_{3}}+\varepsilon_{n_{4}})}{(\varepsilon_{n_{4}}-\varepsilon_{n_{2}})^{2}(\varepsilon_{n_{3}}-\varepsilon_{n_{1}})^{2}}\right\} \right|^{2}\right\rangle \,,
\end{align*}
The high-energy cutoff $E$ on the level sum, $E>|\varepsilon_{n}-\varepsilon_{m}|$, makes $\mathcal{I}(E)$ dependent on $E$. 
The main contribution to $\mathcal{I}$ comes from unequal levels, \emph{i.e.}, unequal $n_{i}$ in the sum. Since we are not interested in the pre-factors, we will only demonstrate that $\mathcal{I}(E)\sim\ln E/\delta$. 
For example, the last term gives
\begin{align}
\mathcal{I}(E)\sim & \dots +  \int_{1}^{E/\delta}dx_{1}dx_{2}dx_{3}dx_{4}R_{2}(x_{1},x_{4})R_{2}(x_{2},x_{3})\frac{(x_{1}+x_{2}+x_{3}+x_{4})^{2}}{(x_{4}+x_{2})^{4}(x_{3}+x_{1})^{4}}\,,
\end{align}
\end{widetext}where $x_i=\varepsilon_i/\delta$, and we changed variables $x_{1,4}\rightarrow-x_{1,4}$.
To leading order in $\delta/E$, we can replace the two-point correlator [Eq.~(\ref{eq:RnDef})] by one, $R_{2}\rightarrow1$.
The remaining integral gives to leading order 
\begin{align}
\mathcal{I}(E)\sim &\dots + \int_{1}^{E/\delta}dx_{1}dx_{2}dx_{3}dx_{4}\frac{1}{(x_{4}+x_{2})^{2}(x_{3}+x_{1})^{4}}\nonumber \\
\sim &\ln\frac{E}{\delta}\,.
\end{align}
Inserting this into Eq.~(\ref{eq:DeltaGavAppendixB2Intermediate})  leads to Eq.~(\ref{eq:TypicalDeltaG1ValleyLowTWithECCmb}) of the
main text. 

\section{An exactly-solvable toy-model with arbitrary interaction strength\label{Adx:AppendixToymodel}}

In this section we derive the substitution, Eq.~(\ref{eq:ToymodelReplacement}), used in Subsection \ref{sub:HighTempWithoutEC}. To account for the
intra-dot interaction $U$ exactly, we use the $T$-matrix formalism.
This amounts to replacing in the cross section, in Eq.~(\ref{eq:S}), the
interaction matrix elements by those of the
$T$-matrix, $T=U(1-K_{0}U)^{-1}$. Here $K_{0}$ is the two-particle
Green function of non-interacting electrons. At high enough temperatures where direct tunneling
dominates (which is the case when Eq.~(\ref{eq:two-level}) is valid),
we consider only diagonal parts of the single-particle Green functions,
see Eq.~(\ref{eq:GreenFnNarrowDiag}). Then the two-particle Green function
takes the form 
\begin{align*}
 & \langle n_{1}\gamma_{1},n_{2}\gamma_{2}|K_{0}(E)|n_{3}\gamma_{3},n_{4}\gamma_{4}\rangle\\
= & \frac{\delta_{\gamma_{1}\gamma_{3}}\delta_{n_{1}n_{3}}\delta_{\gamma_{2}\gamma_{4}}\delta_{n_{2}n_{4}}-\delta_{\gamma_{1}\gamma_{4}}\delta_{n_{1}n_{4}}\delta_{\gamma_{2}\gamma_{3}}\delta_{n_{2}n_{3}}}{E-\varepsilon_{n_{1}}-\varepsilon_{n_{2}}+i\Gamma_{n_{1}}+i\Gamma_{n_{2}}}\,.
\end{align*}

Let us consider a toy-model of only two dot levels, $1$ and $2$, and an interaction $U$ that allows only 2-particle backscattering, $\gamma\gamma\rightarrow\overline{\gamma}\overline{\gamma}$.
We want to calculate the matrix element $T_{1L2L1R2R}$. In the basis
$\{|1\gamma_{1},2\gamma_{2}\rangle\}_{\gamma_{1},\gamma_{2}=L,R}$
(we can neglect doubly occupied levels since they are always annihilated
by $U$) the matrix for $K_{0}(E)$ is proportional to unit matrix
(and for our purposes can be treated as a scalar), while 
\begin{equation}
U=\left(\begin{array}{cccc}
0 & 0 & 0 & U_{1L2L1R2R}\\
0 & 0 & 0 & 0\\
0 & 0 & 0 & 0\\
U_{1L2L1R2R}^{*} & 0 & 0 & 0
\end{array}\right)\,.
\end{equation}
The inverse $(1-K_{0}U)^{-1}$ is easily calculated and the element of the $T$-matrix corresponding to two-particle backscattering becomes, 
\begin{flalign}
\langle1L,2L|T(E)|1R,2R\rangle= & \frac{U_{1L2L1R2R}}{1-K_{0}(E)^{2}|U_{1L2L1R2R}|^{2}}\,.
\label{eq:TmatrixElementToyModel}
\end{flalign}
In resonant tunneling, Eq.~(\ref{eq:DeltaGDirectTunneling}), we have
$E\approx\varepsilon_{1}+\varepsilon_{2}$ and thus $K_{0}(E)\approx-i/(\Gamma_{1}+\Gamma_{2})$.
Inserting this in Eq.~(\ref{eq:TmatrixElementToyModel}) leads to Eq.~(\ref{eq:ToymodelReplacement}). 

\section{Derivation of Eq.~(\ref{eq:DeltaG1EvenValley})\label{sec:Detailed-derivation-of}}

In this section, we discuss the conductance correction in the even valley of a Coulomb blockaded quantum dot at low temperatures. We present the detailed derivation of $\Delta G$, Eq.~(\ref{eq:DeltaG1EvenValley}), due to backscattering of one particle, and show how inclusion of two-particle backscattering modifies the result. We also give the relative importance of different terms in the 5th order $T$-matrix, see Table \ref{table-Tmatrix}.

As explained in Subsection \ref{sub:Even-valley-conductance} of the
main text, in the even valley the main contribution to the $T$-matrix
comes from the 12 terms with small denominators $E_{\pm}E_{\pm}\delta^{2}$
(independent signs). These terms are the elements in rows 1, 3, and 6 in Table~\ref{table-Tmatrix}.
 In particular, the third elements in each of the rows 1 and 6, and third and fourth elements of the row 3, were used in the example in the main text, Eq.~(\ref{eq:ExampleTmatrixTerm}).
The transition amplitude between states $|i\rangle$ and ${\langle f|=\langle i|d_{k_{4}\gamma_{4}}d_{k_{3}\gamma_{3}}d_{k_{1}\gamma_{1}}^{\dagger}d_{k_{2}\gamma_{2}}^{\dagger}}$
induced by the 12 terms from rows 1, 3, and 6, of Table~\ref{table-Tmatrix} is 
\begin{flalign}
\langle f|\mathcal{T}(E_{i})|i\rangle= & \langle i|n_{k_{1}\gamma_{1}}n_{k_{2}\gamma_{2}}(1-n_{k_{3}\gamma_{3}})(1-n_{k_{4}\gamma_{4}})|i\rangle\nonumber \\
\times & \left[\frac{1}{E_{-}}+\frac{1}{E_{+}}\right]^{2}M(k_{1}\gamma_{1},k_{2}\gamma_{2};k_{3}\gamma_{3},k_{4}\gamma_{4})\,,
\label{eq:TmatrixEvenValFull}
\end{flalign}
where $n_{k\gamma}=d_{k\gamma}^{\dagger}d_{k\gamma}$ and 
\begin{flalign}
 & M(k_{1}\gamma_{1},k_{2}\gamma_{2};k_{3}\gamma_{3},k_{4}\gamma_{4})\nonumber \\
 & =-\sum_{n_{1}n_{2}n_{3}n_{4}}U_{\{n_{i}\gamma_{i}\}}t_{n_{1}}t_{n_{2}}t_{n_{3}}t_{n_{4}}\nonumber \\
 & \left[\frac{\Theta_{-n_{1}}\Theta_{-n_{2}}\Theta_{n_{3}}\Theta_{n_{4}}}{\Delta E_{1}+\Delta E_{2}-\Delta E_{3}-\Delta E_{4}}\frac{1}{\Delta E_{1}-\Delta E_{3}}\right.\nonumber \\
 & +\frac{\Theta_{-n_{1}}\Theta_{n_{2}}\Theta_{n_{3}}\Theta_{-n_{4}}}{\Delta E_{4}-\Delta E_{2}}\frac{1}{\Delta E_{1}-\Delta E_{3}}\nonumber \\
 & +\frac{\Theta_{n_{1}}\Theta_{n_{2}}\Theta_{-n_{3}}\Theta_{-n_{4}}}{\Delta E_{1}+\Delta E_{2}-\Delta E_{3}-\Delta E_{4}}\frac{1}{\Delta E_{1}-\Delta E_{3}}\nonumber \\
 & \left.+(1\leftrightarrow2)+(3\leftrightarrow4)\vphantom{\left|U_{n_{1}Ln_{2}Rn_{3}Rn_{4}R}\right|^{2}}\right]\,,\label{eq:MCoulombBlockade}
\end{flalign}
where $\Delta E_{i}=\varepsilon_{n_{i}}-E_{k_{i}}\ll E_{\pm}$. 
Since the dot has an even number of electrons in its ground state, and we are at low temperatures $T\ll\delta$, we used a zero-temperature distribution function for the dot states: $\Theta_{\pm n}=\Theta(\pm\varepsilon_{n}\mp\varepsilon_{1})$,  level 1 being the lowest unoccupied level. 
We can now evaluate the contribution $I_{\gamma_{1}\gamma_{2}\rightarrow\gamma_{3}\gamma_{4}}$, due to two electrons in Kramers states $\gamma_{1}\gamma_{2}$ scattering into states $\gamma_{3}\gamma_{4}$, to the total backscattering current.
The partial current  $I_{\gamma_{1}\gamma_{2}\rightarrow\gamma_{3}\gamma_{4}}$ is a sum (over the initial and final states) of the transition rates $r_{i\rightarrow f}$ from state $|i\rangle$ to the final state
 $\langle f|=\langle i|d_{k_{4}\gamma_{4}}d_{k_{3}\gamma_{3}}d_{k_{1}\gamma_{1}}^{\dagger}d_{k_{2}\gamma_{2}}^{\dagger}$. 
Equation~(\ref{eq:RateTmatrixGeneral}) relates the rate $r_{i\rightarrow f}$ to the amplitude in Eq.~(\ref{eq:TmatrixEvenValFull}). 
Summing over the initial states $|i\rangle$ and the final state momenta $\{k_{i}\}$, we find 
\begin{widetext}
\begin{align}
I_{\gamma_{1}\gamma_{2}\rightarrow\gamma_{3}\gamma_{4}}= & 2\pi e\left[\frac{1}{E_{-}}+\frac{1}{E_{+}}\right]^{4}\sum_{k_{1},k_{2},k_{3},k_{4}}|M(k_{1}\gamma_{1},k_{2}\gamma_{2};k_{3}\gamma_{3},k_{4}\gamma_{4})|^{2}\nonumber \\
\times & \delta(E_{k_{1}}+E_{k_{2}}-E_{k_{3}}-E_{k_{4}})f_{k_{1}\gamma_{1}}f_{k_{2}\gamma_{2}}(1-f_{k_{3}\gamma_{3}})(1-f_{k_{4}\gamma_{4}})\,.
\end{align}
\end{widetext}The distribution functions ${f_{k\gamma}=1/[e^{(E_{k}+\gamma eV/2)/T}+1]}$
are the result of summing the factors ${\langle i|n_{k_{i}\gamma_{i}}|i\rangle}$
over initial states with weight $w_{i}$. The correction to the ideal
current is then 
\begin{align}
 & \Delta I=e\sum_{\gamma_{1},\gamma_{2},\gamma_{3},\gamma_{4}}\Delta N_{\gamma_{1}\gamma_{2};\gamma_{3}\gamma_{4}}2\pi\left[\frac{1}{E_{-}}+\frac{1}{E_{+}}\right]^{4}\sum_{k_{1},k_{2},k_{3},k_{4}}\nonumber \\*
 & \times|M(k_{1}\gamma_{1},k_{2}\gamma_{2};k_{3}\gamma_{3},k_{4}\gamma_{4})|^{2}\delta(E_{k_{1}}+E_{k_{2}}-E_{k_{3}}-E_{k_{4}})\nonumber \\*
 & \times\big[f_{k_{1}\gamma_{1}}f_{k_{2}\gamma_{2}}(1-f_{k_{3}\gamma_{3}})(1-f_{k_{4}\gamma_{4}})\nonumber \\*
 & -f_{k_{3}\gamma_{3}}f_{k_{4}\gamma_{4}}(1-f_{k_{1}\gamma_{1}})(1-f_{k_{2}\gamma_{2}})\big]\,,
 \label{eq:CurrentRevisited}
\end{align}
where $\Delta N_{\gamma_{1}\gamma_{2};\gamma_{3}\gamma_{4}}=(\gamma_{3}+\gamma_{4}-\gamma_{1}-\gamma_{2})/2$ counts
the number of backscattered particles. [Naturally, Eq.~(\ref{eq:CurrentRevisited}) consists of contributions arising from backscattering of both one and two particles.] 
The above Eq.~(\ref{eq:CurrentRevisited}) is a generalization of Eq.~(\ref{eq:CurrentGeneral}) to the case of a Coulomb blockaded quantum dot in the even valley.

Let us next consider the contribution to $\Delta G$ due to 1-particle
backscattering. From Eq.~(\ref{eq:MCoulombBlockade}) one sees that
because $U$ is symmetric under time-reversal, $|M|^{2}$ is the same
for all combinations $\{\gamma_{i}\}$ that backscatter one particle.
At low temperatures $T\ll\delta$ we can expand the denominators of $M$ for small $E_{k_{i}}$ which brings out a factor $E_{k_{1}}-E_{k_{2}}\sim T$, leading to $\Delta G_{1}\propto T^{4}$ as in the case without charging interaction, Eq.~(\ref{eq:G1-particlevalleyNocharging}); see also the amplitude  Eq.~(\ref{eq:AmplitudeSymmetrized}) and its derivation. 
Linearizing the Fermi functions $f_{k_{i}\gamma_{i}}$ in Eq.~(\ref{eq:CurrentRevisited}) in bias voltage, leads to Eq.~(\ref{eq:DeltaG1EvenValley}) given in Subsection \ref{sub:Even-valley-conductance}. 

The conductance correction $\Delta G_{2}$ due to 2-particle backscattering is obtained similarly from Eq.~(\ref{eq:CurrentRevisited}). We find \begin{widetext}
\begin{flalign}
\Delta G_{2}/G_{0}= & 4\frac{64\pi^{3}}{35}T^{6}\left[\frac{1}{E_{-}}+\frac{1}{E_{+}}\right]^{4}\left|\sum_{n_{1}n_{2}n_{3}n_{4}}U_{n_{1}Ln_{2}L;n_{3}Rn_{4}R}\sqrt{\Gamma_{n_{1}}\Gamma_{n_{2}}\Gamma_{n_{3}}\Gamma_{n_{4}}}\right.\nonumber \\
 & \times\left\{ \frac{\Theta_{n_{1}}\Theta_{n_{2}}\Theta_{-n_{3}}\Theta_{-n_{4}}+\Theta_{-n_{1}}\Theta_{-n_{2}}\Theta_{n_{3}}\Theta_{n_{4}}}{(\varepsilon_{n_{1}}+\varepsilon_{n_{2}}-\varepsilon_{n_{3}}-\varepsilon_{n_{4}})(\varepsilon_{n_{1}}-\varepsilon_{n_{3}})^{3}}\right.\nonumber \\
 & \left.\left.+\Theta_{-n_{1}}\Theta_{n_{2}}\Theta_{n_{3}}\Theta_{-n_{4}}\frac{(\varepsilon_{n_{4}}-\varepsilon_{n_{2}})(\varepsilon_{n_{1}}-\varepsilon_{n_{3}})-(\varepsilon_{n_{4}}-\varepsilon_{n_{2}}+\varepsilon_{n_{1}}-\varepsilon_{n_{3}})^{2}}{(\varepsilon_{n_{1}}-\varepsilon_{n_{3}})^{3}(\varepsilon_{n_{2}}-\varepsilon_{n_{4}})^{3}}\right\} \right|^{2}\,.
 \label{eq:DeltaG2EvenValley}
\end{flalign}
\end{widetext}
From Eq.~(\ref{eq:DeltaG2EvenValley}), comparing to Eq.~(\ref{eq:DeltaG1EvenValley}), we see that $\Delta G_{2} \ll \Delta G_{1}$ at low temperatures.

\subsection{2-particle contribution near the peak\label{sub:2-particle-near-Pk}}

In this section we show how 2-particle backscattering modifies the
ideal conductance in the even valley near the peak, where $\varepsilon_{1}+E_{+} \ll \delta$ and level 1 is the lowest unoccupied level. The 1-particle contribution was given in Eqs.~(\ref{eq:DeltaG1NearPkLowT}) and Eq.~(\ref{eq:DeltaG1NearPkHighT}), calculated in Subsection \ref{sub:Cond-near-the-Pk-Even}.

For backscattering of two particles, we set $\gamma_{1,2}=\overline{\gamma}_{3,4}$ in the amplitude, Eq.~(\ref{eq:TmatrixElementEvenValley}).  
Summing the rates, Eq.~(\ref{eq:RateTmatrixGeneral}), leads to 
\begin{flalign}
 & \Delta G_{2}/G_{0}\nonumber \\
 & =\frac{\Theta(E_{+}+\varepsilon_{1})}{32\pi^{3}}\int dE_{1}dE_{2}dE_{3}dE_{4}\delta(E_{1}+E_{2}-E_{3}-E_{4})\nonumber \\
 & \times \frac{\Gamma_{1}^{2}}{T} (E_{1}-E_{2})^{2}(E_{3}-E_{4})^{2}\prod_{i=1}^{4}\frac{\text{sech}\frac{\beta E_{i}}{2}}{(\varepsilon_{1}+E_{+}-E_{i})^{2}}\nonumber \\
 & \times\left|\sum_{n,m}\frac{\sqrt{\Gamma_{n}\Gamma_{m}}}{\varepsilon_{n}-\varepsilon_{m}}\left( \frac{\Theta_{-m}}{\varepsilon_{1}-\varepsilon_{m}}-\frac{\Theta_{-n}}{\varepsilon_{1}-\varepsilon_{n}} \right) U_{1LnL;1RmR}\right|^{2}\,.\label{eq:DeltaG2pkWithEc}
\end{flalign}
The integrals are performed the same way as in Subsection \ref{sub:Cond-near-the-Pk-Even}.
In the low-temperature limit, $T\ll \varepsilon_{1}+E_{+}$, we find  
\begin{flalign}
 & \Delta G_{2}/G_{0}=\frac{4\pi^{3}}{35}\frac{\Theta(E_{+}+\varepsilon_{1})T^{6}\Gamma_{1}^{2}}{(\varepsilon_{1}+E_{+})^{8}}\nonumber \\
 & \times\left|\sum_{n,m}\frac{\sqrt{\Gamma_{n}\Gamma_{m}}}{\varepsilon_{n}-\varepsilon_{m}}\left( \frac{\Theta_{-m}}{\varepsilon_{1}-\varepsilon_{m}}-\frac{\Theta_{-n}}{\varepsilon_{1}-\varepsilon_{n}} \right) \right.\nonumber \\
 & \times\left.U_{1LnL;mR1R}\vphantom{\left|U_{n_{1}Ln_{2}Rn_{3}Rn_{4}R}\right|^{2}}\right|^{2}\,,\quad\varepsilon_{1}+E_{+}\gg T\,,\Gamma_1\,.\label{eq:DeltaG2NearPkLowT}
\end{flalign}
Compared to the 1-particle backscattering contribution, Eq.~(\ref{eq:DeltaG1NearPkLowT}),
$\Delta G_{2}$ is smaller by a factor $T^{2}/(\varepsilon_{1}+E_{+})^{2}$.

At high temperatures, $T\gg \varepsilon_{1}+E_{+}$,  we get from Eq.~(\ref{eq:DeltaG2pkWithEc}), 
\begin{flalign}
 & \Delta G_{2}/G_{0}=\frac{1}{2\pi}\Theta(\varepsilon_{1}+E_{+})\nonumber \\
 & \left|\sum_{n,m}\frac{\sqrt{\Gamma_{n}\Gamma_{m}}}{\varepsilon_{n}-\varepsilon_{m}}\left(\frac{\Theta_{-m}}{\varepsilon_{1}-\varepsilon_{m}}-\frac{\Theta_{-n}}{\varepsilon_{1}-\varepsilon_{n}}\right)U_{1LnL;1RmR}\right|^{2}\,,\nonumber \\*
 & \vphantom{\sum_{n,m}\frac{\sqrt{\Gamma_{n}\Gamma_{m}}}{\varepsilon_{n}-\varepsilon_{m}}\left(\frac{\Theta(-\varepsilon_{m})}{\varepsilon_{1}-\varepsilon_{m}}-\frac{\Theta(-\varepsilon_{n})}{\varepsilon_{1}-\varepsilon_{n}}\right)}\quad T\gg\varepsilon_{1}+E_{+}\,,\Gamma_1\,.\label{eq:DeltaG2NearPkHighT}
\end{flalign}
Comparing Eq.~(\ref{eq:DeltaG2NearPkHighT}) to $\Delta G_1$ in Eq.~(\ref{eq:DeltaG1NearPkHighT}), we see that at high temperatures the 1- and 2-particle contributions to $\Delta G$ are of the same order of magnitude.

\section{Components of the exchange tensor
 and averages in Subsection \ref{sub:Odd-valleys-of}
\label{sec:Components-of-the}}

In this section we give the components of the exchange tensor $J_{ij}$, introduced in Eq.~(\ref{eq:JtensorRotated}) of Subsection~\ref{sub:Odd-valleys-of}, in terms of microscopic dot parameters. 
We will also calculate the disorder-averaged odd-valley conductance correction $\Delta G$ in Subsection~\ref{sub:AverageOverDeltaJ}.
This requires averaging over the proper exchange couplings $J_{ij}$.

The components $J_{ij}$ can be obtained by using the microscopic dot Hamiltonian [given in Eqs.~(\ref{eq:HdotWithEc})--(\ref{eq:PerturbationCoulombBlockade})] in Eq.~(\ref{eq:HeffWithProjectors}) and matching terms with Eq.~(\ref{eq:Heffective}).

For the isotropic part $J_{0}$ we find 
\begin{flalign}
 & J_{0}=2t_{1}^{2}(\frac{1}{E_{-}-\varepsilon_{1}}+\frac{1}{E_{+}+\varepsilon_{1}})\nonumber \\
 & -\sum_{\mu\nu}\sum_{n,m<1}\frac{t_{1}^{2}U_{m\mu n\nu;m\mu n\nu}}{(E_{-}-\varepsilon_{1})^{2}}-\sum_{\mu\nu}\sum_{n,m\leq1}\frac{t_{1}^{2}U_{m\mu n\nu;m\mu n\nu}}{(E_{+}+\varepsilon_{1})^{2}}\nonumber \\
 & -2\sum_{\mu\nu}\sum_{n<1}\sum_{m<1}\frac{t_{1}t_{n}U_{1\nu m\mu;n\nu m\mu}}{(E_{-}-\varepsilon_{1})(\varepsilon_{n}-E_{-})}\nonumber \\
 & -2\sum_{\mu\nu}\sum_{n>1}\sum_{m<1}\frac{t_{1}t_{n}U_{1\nu m\mu;n\nu m\mu}}{(E_{+}+\varepsilon_{1})(\varepsilon_{n}+E_{+})}\nonumber \\
 & -2\sum_{\mu\nu}\sum_{m<1}\sum_{n>1}\frac{t_{1}t_{n}U_{1\nu m\mu;n\nu m\mu}}{(E_{-}-\varepsilon_{1})(\varepsilon_{n}-\varepsilon_{1})}\nonumber \\
 & -2\sum_{\mu\nu}\sum_{m<1}\sum_{n<1}\frac{t_{1}t_{n}U_{1\nu m\mu;n\nu m\mu}}{(E_{+}+\varepsilon_{1})(\varepsilon_{n}-\varepsilon_{1})}\,.\label{eq:J0Full}
\end{flalign}
The first term in $J_{0}$ is independent of interaction $U$ and
corresponds to the exchange coupling of the usual Anderson Hamiltonian.
The components $\delta J_{ij}$ {[}see Eq.~(\ref{eq:JtensorRotated}){]}
are given in Table \ref{table-Jcomponents}.  

\subsection{Averages in Subsection \ref{sub:Odd-valleys-of} \label{sub:AverageOverDeltaJ}}

In this subsection we obtain the disorder-averaged conductance correction in the odd Coulomb blockade valley, Eqs.~(\ref{eq:TypicalDeltaGOddValley0})--(\ref{eq:TypicalDeltaGOddValley2}) of Subsection \ref{sub:The-correction-to}.

In Eq.~(\ref{eq:DeltaGOddValleyAboveTKNonRGd}) we expressed $\Delta G$
in terms of the bare exchange tensor components $\delta J_{ij}$. 
The conductance correction takes the simple form
\begin{equation}
\Delta G=-\frac{\pi\rho^{2}}{4}\left[16|J_{++}|^{2}+|J_{z+}|^{2}\right]\,,\label{eq:DeltaGOddValSimpleForm}
\end{equation}
 if we write $\delta J_{xx}-\delta J_{yy}=4\text{Re}J_{++}$,
$\delta J_{yx}=-4\text{Im}J_{++}$, $\delta J_{zx}=2\text{Re}J_{z+}$, and
$\delta J_{zy}=-2\text{Im}J_{z+}$. From Table~\ref{table-Jcomponents} we find  \begin{widetext}\begin{subequations}
\begin{flalign}
J_{++}= & \sum_{n,m}t_{m}t_{n}U_{nL1L;mR1R}\nonumber \\
\times & \left\{ \frac{\Theta_{-m}\Theta_{-n}}{(\varepsilon_{m}-E_{-})(\varepsilon_{n}-E_{-})}+\frac{\Theta_{m}\Theta_{n}}{(\varepsilon_{m}+E_{+})(\varepsilon_{n}+E_{+})}-2\frac{\Theta_{m}\Theta_{-n}}{(\varepsilon_{n}-\varepsilon_{m})}\left[\frac{1}{(\varepsilon_{n}-E_{-})}+\frac{1}{(\varepsilon_{m}+E_{+})}\right]\right\} \,,
\end{flalign}
and
\begin{flalign}
J_{z+}= & 2\sum_{n,m}t_{n}\left\{ t_{m}(U_{nL1L;mR1L}+U_{nL1L;mL1R})\left[\frac{\Theta_{-m}\Theta_{-n}}{(\varepsilon_{m}-E_{-})(\varepsilon_{n}-E_{-})}+\frac{\Theta_{m}\Theta_{n}}{(\varepsilon_{m}+E_{+})(\varepsilon_{n}+E_{+})}\right]\right.\nonumber \\
+ & t_{m}(U_{mR1R;nL1R}+U_{nR1R;mL1R})\frac{\Theta_{m}\Theta_{-n}}{(\varepsilon_{n}-\varepsilon_{m})}\left[\frac{1}{(\varepsilon_{n}-E_{-})}+\frac{1}{(\varepsilon_{m}+E_{+})}\right]\nonumber \\
- & \left.t_{1}\sum_{\mu}U_{nLm\mu;1Rm\mu}\frac{\Theta_{-m}}{(\varepsilon_{n}-\varepsilon_{1})}\left[\frac{\Theta_{-n}}{(\varepsilon_{n}-E_{-})}+\frac{\Theta_{n}}{(\varepsilon_{n}+E_{+})}\right]\right\} \,,
\end{flalign}
\end{subequations}with $\Theta_{n}=\Theta(\varepsilon_{n}-\varepsilon_{1})$.

Averaging the terms $|J_{++}|^{2}$ and $|J_{z+}|^{2}$ in Eq.~(\ref{eq:DeltaGOddValSimpleForm}) over disorder is straightforward and follows Subsection \ref{sec:AveragesOverLevels}. However, due to the large number of different terms we will settle with an order of magnitude estimate.

Let us first consider the behavior near the peak, and take $E_{+}\ll\delta$.
Then one easily finds 
\begin{equation}
\left\langle |J_{++}|^{2}\right\rangle \sim\left\langle |J_{z+}|^{2}\right\rangle \sim\frac{1}{g^{2}\delta^{2}}\frac{\Gamma^{2}}{\rho^{2}}\,,
\end{equation}
which, upon inserting to Eq.~(\ref{eq:DeltaGOddValSimpleForm}), leads to Eq.~(\ref{eq:TypicalDeltaGOddValley0}). 

Next we take $E_{\pm}\gg\delta$. Then
\begin{equation}
 16\left\langle |J_{++}|^{2}\right\rangle +\left\langle |J_{z+}|^{2}\right\rangle =  \frac{\delta^{2}}{g^{2}}\frac{\Gamma^{2}}{\rho^{2}}\left\{ c_{1}(F_{4}^{+}+F_{4}^{-})+c_{2}(F_{2}^{++}+F_{2}^{--})+c_{3}F_{2}^{+-}+c_{4}(F_{1}^{-}+F_{1}^{+}+2F_{1b})+c_{5}(F_{0}^{+}+F_{0}^{-})\right\} \,,
\label{eq:AverageJsAppendix}
\end{equation}
where $c_{1,2,3,4,5}$ are numerical coefficients, and where we have introduced
the averages ($\kappa=\pm1$)\begin{subequations} 
\begin{flalign}
F_{1b}= & \left\langle \sum_{n,m}\frac{1}{(\varepsilon_{m}-\varepsilon_{n})^{2}}\frac{\Theta_{-n}}{(\varepsilon_{n}-E_{-})}\frac{\Theta_{m}}{(\varepsilon_{m}+E_{+})}\right\rangle \,,\\
F_{1}^{\kappa}= & \left\langle \sum_{n,m}\frac{1}{(\varepsilon_{m}-\varepsilon_{n})^{2}}\frac{\Theta_{\kappa n}\Theta_{\overline{\kappa}m}}{[\varepsilon_{n}+\kappa E_{\kappa}]^{2}}\right\rangle \,;\quad
F_{0}^{\kappa}= \left\langle \sum_{n}\frac{\Theta_{\kappa n}}{(\varepsilon_{n}-\varepsilon_{1})^{2}[\varepsilon_{n}+\kappa E_{\kappa}]^{2}}\right\rangle \,,\\
F_{2}^{\kappa\kappa'}= & \left\langle \sum_{n\neq m}\frac{\Theta_{\kappa m}\Theta_{\kappa'n}}{(\varepsilon_{m}+\kappa E_{\kappa})^{2}(\varepsilon_{n}+\kappa'E_{\kappa'})^{2}}\right\rangle \,;\quad
F_{4}^{\kappa}= \left\langle \sum_{n}\frac{\Theta_{\kappa n}}{(\varepsilon_{n}+\kappa E_{\kappa})^{4}}\right\rangle \,.
\end{flalign}
\end{subequations}
 As long as $|E_{+}-E_{-}|\sim E_{C}$, the largest
sums are $F_{1}^{\pm}$ and $F_{1b}$. To leading order in $E_{\pm}/\delta$,

\begin{flalign}
F_{1}^{\pm}= & \frac{1}{\delta^{2}E_{\pm}^{2}}\ln\frac{E_{\pm}}{\delta}\,;\quad 
F_{1b}= -\frac{1}{2}\frac{\ln\frac{E_{+}E_{-}}{\delta^{2}}}{\delta^{2}E_{+}E_{-}}-\frac{1}{4}\frac{(E_{+}-E_{-})\ln\frac{E_{-}}{E_{+}}}{\delta^{2}E_{+}E_{-}E_{C}}\,.
\end{flalign}
\end{widetext}
Keeping only the terms $F_{1}^{\pm}$ and $F_{1b}$ in Eq.~(\ref{eq:AverageJsAppendix}) leads to Eq.~(\ref{eq:TypicalDeltaGOddValley1})
of the main text. Closer to the bottom of the valley, $|E_{+}-E_{-}|\ll E_{C}$,
the sum of the logarithmic terms in $F_{1}^{-}+F_{1}^{+}+2F_{1b}$ in Eq.~(\ref{eq:AverageJsAppendix}) above becomes subleading. The main contribution to Eq.~(\ref{eq:AverageJsAppendix}), and $\Delta G$, comes from the terms of the order of $1/\delta^{2}E_{C}^{2}$ in the sums $F_{1}^{\kappa}$, $F_{1b}$, $F_{0}^{\kappa}$, $F_{2}^{\kappa\kappa'}$. This leads to Eq.~(\ref{eq:TypicalDeltaGOddValley2}) of the main text.

\section{Renormalization group for the anisotropic Kondo model\label{sec:RG-for-anisotropic}}

We saw in Subsection \ref{sub:The-correction-to} of the main text that
in the odd valley the correction to the conductance, $\Delta G$, arises due to anisotropy in the exchange coupling. 
In this section we study, using perturbative renormalization group,  how the weakly anisotropic Kondo model, Eq.~(\ref{eq:Heffective}), flows to the isotropic fixed point at low energies.~\cite{anderson70} 
We find that the renormalization leads to a logarithmic suppression of $\Delta G$ as one approaches low temperatures $T\rightarrow T_{K}$, see Eq.~(\ref{eq:DeltaGOddValleyAboveTK}).

In Subsection \ref{sub:Odd-valleys-of} we derived the Hamiltonian for
the effective interaction between the helical edge and a quantum dot
with spin, valid at temperatures $T\ll D={\min(E_{\pm},\delta)}$,
\begin{equation}
H=\sum_{k}\sum_{\gamma}E_{k}d_{k\gamma}^{\dagger}d_{k\gamma}+\sum_{i,j=x,y,z}S_{i}J_{ij}(D)s_{j}\,.
\label{eq:HRGAppendix}
\end{equation}
The first term in Eq.~(\ref{eq:HRGAppendix}) corresponds to the free itinerant electrons of the helical edge.
The sum over momenta is restricted to a band of energies $|E_{k}-E_{F}|\ll D$, and we set $E_{F}=0$.
The edge spin density at the point contact is denoted $\mathbf{s}$, while $\mathbf{S}$ is the spin-1/2 operator of the dot. We have explicitly indicated the cutoff-dependence of the exchange couplings $J_{ij}$.

Next, we will follow Ref. \onlinecite{anderson70} and eliminate the  high-energy
states in a small strip of width $\Delta D\ll D$ near the edge of
the band, thus reducing the cutoff energy $D$ to $D-\Delta D$. The
resulting low-energy Hamiltonian contains information about the rare
virtual transitions between the low- and high-energy sectors. Considering
only the exchange part of this low-energy Hamiltonian, we can relate
its coupling constants $J_{ij}(D-\Delta D)$ to the couplings $J_{ij}(D)$
of the original Hamiltonian. 
 To second order in the couplings this relation is 
\begin{align}
& J_{ij}(D-\Delta D) - J_{ij}(D) \nonumber\\
&=\frac{1}{2}\frac{\rho\Delta D}{D}\sum_{klmn}\varepsilon_{ikm}\varepsilon_{jln}J_{kl}(D)J_{mn}(D)\,.
\end{align}
Here $\varepsilon_{ijk}$ is the Levi-Civita symbol. Let us assume
that $J_{ij}(D)$ is almost isotropic, $J_{ij}=J_{0}\delta_{ij}+\delta J_{ij}$,
where the anisotropic part $\delta J_{ij}$ is traceless and symmetric
(trace would add to the isotropic part $J_{0}$, and antisymmetric part corresponds to
a rotation of $\mathbf{S}$). Taking $\delta J_{ij}$ to be small,
we can write the renormalization group equations to linear order in
$\delta J_{ij}$, \begin{subequations}
\begin{flalign}
\frac{dJ_{0}(D)}{d\ln D}= & -\rho J_{0}(D)^{2}\,,\\
\frac{d\delta J_{ij}(D)}{d\ln D}= & \rho J_{0}(D)\delta J_{ij}(D)\,.
\end{flalign}
\end{subequations}Solving the equation for $J_{0}$ gives the known
result that $J_{0}$ flows to strong coupling at low energies, 
\begin{equation}
\rho J_{0}(D)=\frac{1}{\ln\frac{D}{T_{K}}}\,.
\end{equation}
The Kondo temperature $T_{K}=De^{-\frac{1}{\rho J_{0}(D)}}$ is invariant under reduction of the cutoff $D$. 
The anisotropic perturbation flows towards zero upon reduction of bandwidth from $D$ to $D'$, 
\begin{equation}
\delta J_{ij}(D')=\frac{\ln\frac{D'}{T_{K}}}{\ln\frac{D}{T_{K}}}\delta J_{ij}(D)\,.
\label{eq:DeltaJRGd}
\end{equation}
Setting $D'=T<D$, Eq.~(\ref{eq:DeltaJRGd}) gives the logarithmic suppression of backscattering used in Eq.~(\ref{eq:DeltaGOddValleyAboveTK}).

\section{Single-dot conductance correction at temperatures $\delta\ll T\ll E_{C}$ \label{sec:Cotunneling}}

In this section we consider the correction to the conductance at intermediate temperatures, $\delta\ll T\ll E_{C}$, alluded to in Subsection~\ref{sub:HighTCB} of the main text. 

We will first discuss qualitatively the derivation of $\Delta G$. For that, we follow the part of Section~\ref{sec:Qualitative} devoted to the electron backscattering by the exchange interaction. The spin of the dot remains finite (albeit not equal to $0$ or $1/2$) even at $T\gg\delta$. That allows us to relate the backscattered current to the time-averaged derivative of the $z$-component of the total spin,
\begin{equation}
\Delta I = -e\overline{\langle\frac{d}{dt}\left(\hat{S}_z+\hat{S}_z^{\rm edge}\right)\rangle}\,.
\label{eq:DeltaI}
\end{equation}
The number of electrons $N$ in a quantum dot is well-defined at $T\ll E_C$. However, its spin $S$ is not determined by the electron number, if $T\gg\delta$. That makes the spin transfer from the electrons of the edge to the electrons of the dot effective in the backscattering at any value of $N$. Assuming the dot is ``deep'' in the Coulomb blockade regime, electrons can not tunnel into it. The backscattering then is governed by an exchange Hamiltonian of a form similar to Eq.~(\ref{eq:Kondo1}),
\begin{equation}
H_{\rm ex}\!=\!\!\! \sum_{i,j;n,m} \! J_{ij}^{nm}S_{i}^{nm}s_{j}(x_0)\,,\,\,\,\,
\mathbf{S}^{nm}=\frac{1}{2}\!\!\! \sum_{n,m;\alpha,\beta} \! c_{n\alpha}^{\dagger}\boldsymbol{\sigma}_{\alpha\beta}c_{m\beta}\,.
\label{eq:KondoG}
\end{equation}
Unlike Eq.~(\ref{eq:Kondo1}), the form of Eq.~(\ref{eq:KondoG}) includes off-diagonal in the orbital index matrix elements of the total spin of the dot. That makes Eq.~(\ref{eq:KondoG}) applicable in the energy band $|\varepsilon|\ll E_C$ and allows for an electron transition between the dot levels in the course of the edge electron scattering. At $T\gg\delta$ it is sufficient for us to keep only the largest, isotropic part of exchange [cf. Eq.~(\ref{eq:QualitJ0})],
\begin{equation}
J^{nm}= 2t_{n}t_{m}\left(\frac{1}{E_{-}}+\frac{1}{E_{+}}\right)\,.
\label{eq:QualitJG}
\end{equation}
(Here $t_n$ and $t_m$ are the tunneling matrix elements for the dot levels $n$ and $m$; unlike in Eq.~(\ref{eq:QualitJ0}), we keep in the denominators of Eq.~(\ref{eq:QualitJG}) only the charging energy, as we are not aiming here at the description of the Coulomb blockade peaks in $\Delta G$). Indeed, at $T\gg\delta$, contrary to the low-temperature limit, the spin transferred to the dot in a backscattering event may relax due to the intra-dot scattering processes, see Subsection~\ref{sub:Kinetic-equation}. 

The isotropic exchange Hamiltonian, Eqs.~(\ref{eq:KondoG}) and (\ref{eq:QualitJG}), commutes with the total spin, therefore $\Delta I\sim-\overline{\langle i[\hat{U},\hat{S}_z]\rangle}$. Here $\hat{U}$ is the intra-dot interaction Hamiltonian which causes the relaxation of the dot spin, $\langle i[\hat{U},\hat{S}_z]\rangle\sim -\langle\hat{S}_z\rangle/\tau_{e-e}(T)$. In this way, we relate $\Delta I$ to the spin kinetics of the dot,
\begin{equation}
\Delta I\sim \tau_{e-e}^{-1}(T)\langle\hat{S}_z\rangle\,.
\label{eq:DeltaI2}
\end{equation}
Here $\tau_{e-e}^{-1}(T)$ is the characteristic rate of the intra-dot electron relaxation at energy $E\sim T$, cf. Eq.~(\ref{eq:CollisionTermAveraged}).

In the steady state, $\langle\hat{S}_z\rangle$ is determined by the balance between the above-mentioned intra-dot spin relaxation and transfer of the spin from the edge electrons due to the exchange interaction~(\ref{eq:KondoG}). In the absence of any intra-dot relaxation, the equilibrium with the edge would be reached at $\langle\hat{S}_z\rangle= eV/2\delta$, assuming $T\gg\delta$ (this relation is derived similarly to the one for the $T\ll\delta$ limit considered in Section~\ref{sec:Qualitative}). At $\langle\hat{S}_z\rangle\neq eV/2\delta$, the rate of the spin transfer can be estimated as $(eV/2\delta-\langle\hat{S}_z\rangle)/\tau_{\rm cot}(T)$ with $\tau_{\rm cot}^{-1}(T)\sim (\rho J)^2T^2/\delta$ being the analogue of the inelastic co-tunneling rate.~\cite{glazman05} The rate balance equation for $\langle\hat{S}_z\rangle$ thus reads
\begin{equation}
-\tau_{e-e}^{-1}(T)\langle\hat{S}_z\rangle+\tau_{\rm cot}^{-1}(T)(eV/2\delta-\langle\hat{S}_z\rangle)=0\,.
\label{eq:balanceG}
\end{equation}
Solving it and substituting the result in Eq.~(\ref{eq:DeltaI2}), we find for $\Delta G=\Delta I/V$
\begin{equation}
\Delta G\sim G_0 \frac{1}{\delta} \frac{\tau_{e-e}^{-1}(T)\tau^{-1}_{\rm cot}(T)}{\tau_{e-e}^{-1}(T)+\tau^{-1}_{\rm cot}(T)}\,.
\label{eq:DeltaGG}
\end{equation}
The conductance correction scales with temperature as $T^2$, with the coefficient depending on the ratio between the typical values of $J^2$ and $U^2$.

Now we proceed with a more detailed derivation of $\Delta G$. 
Since $T\gg\delta$, it is convenient to use the average distribution function
of the dot electrons, $p_{\gamma}(E)=\langle\sum_{n}\delta(E-\varepsilon_{n})p_{n\gamma}\rangle/\nu_{0}$,
introduced in Subsection~\ref{sub:Kinetic-equation}. (Here $\nu_{0}=\delta^{-1}$
is the electron density of states in the dot.) In the absence of bias voltage, $p_\gamma$ is independent of $\gamma=L,\,R$ and  equal to a Fermi function. Hereafter, we refer to it as the  equilibrium distribution.

The correction $\Delta I$ to the ideal current is obtained from the number of backscattered, say, left-moving
edge electrons per unit time. Electrons of the helical edge cannot tunnel into the dot at $T\ll E_C$, and their scattering off the dot is mostly inelastic at $T\gg\delta$ (inelastic co-tunneling~\cite{glazman05}). An inelastic scattering process leaves behind a particle-hole excitation in the dot.  Accounting for the inelastic backscattering, we have 
\begin{align}
\Delta I= & e\nu_{0}\int dE\left(\frac{d}{dt}p_{L}(E)\right)_{\text{cot}} \nonumber \\
= & \frac{1}{2}e\nu_{0} \int dE \sum_{\gamma=\pm=L,R}\gamma \left(\frac{d}{dt}p_{\gamma}(E)\right)_{\text{cot}}\,.\label{eq:DeltaIKinEqWithCot1}
\end{align}
The second equality here follows from the fact that the co-tunneling process
preserves the total number of electrons in the dot. 
The time-derivative $(dp_\gamma/dt)_\text{cot}$ is the the counter-part of the second term in Eq.~(\ref{eq:balanceG}) and contributes to the rate equation for the dot distribution function, as  shown below.

The full rate equation for the distribution function $p_{\gamma}$ consists of contributions from both co-tunneling and relaxation inside the dot, 
\begin{align}
\frac{dp_{\gamma}(E)}{dt}= & \left(\frac{dp_{\gamma}(E)}{dt}\right)_{\text{cot}}+\left(\frac{dp_{\gamma}(E)}{dt}\right)_{e-e}\,.\label{eq:rateEqWithCot}
\end{align}
(We neglect the contribution from direct tunneling, which is suppressed
at temperatures $T\ll E_{C}$.) The rate of intra-dot relaxation $\tau_{e-e}^{-1}$
was evaluated in Eq.~(\ref{eq:ScatteringrateScrndClmb}), see Subsection~\ref{sub:Kinetic-equation}.
Considering a small deviation of $p_\gamma$ from equilibrium distribution (\emph{i.e.}, a Fermi function), we have to linear order in the bias voltage $eV\ll T$,
\begin{equation}
\sum_{\gamma}\gamma\left(\frac{d}{dt}p_{\gamma}(E)\right)_{e-e}=  -\tau_{e-e}^{-1}(E)\sum_{\gamma}\gamma p_{\gamma}(E)\,.\label{eq:dpdtRelLinearized}
\end{equation}
This is obtained from Eq.~(\ref{eq:CollisionTermAveraged}) in the limit of strong spin-orbit coupling (so that the integration kernel there cancels away). It is valid at $T\gg g^{1/2}\delta$, since we replaced averages of products by the products of averages when performing the averaging over dot levels, see Subsection~\ref{sub:Kinetic-equation}. 

The backscattering current $\Delta I$ can be written in terms of the right-hand-side of Eq.~(\ref{eq:dpdtRelLinearized}) as we show next. 
In the steady state, we have $dp_\gamma/dt=0$ in Eq.~(\ref{eq:rateEqWithCot}). 
Using Eqs.~(\ref{eq:rateEqWithCot}) and (\ref{eq:dpdtRelLinearized}), we can then write  $\Delta I$ in the form
\begin{equation}
\Delta I= \frac{1}{2}e\nu_{0} \int dE \tau_{e-e}^{-1}(E)\sum_{\gamma}\gamma p_{\gamma}(E) \,.\label{eq:DeltaIKinEqWithCot2}
\end{equation}
Identifying $\langle\hat{S}_z\rangle =\frac{1}{2} \nu_0 \int dE \sum_\gamma \gamma p_\gamma (E)  $, we see that  Eq.~(\ref{eq:DeltaIKinEqWithCot2}) above is the analogue of  Eq.~(\ref{eq:DeltaI2}) in the introduction to this section.
To  express $\Delta I$ in terms of known quantities, we will next solve for $\sum_\gamma \gamma p_\gamma$ by using the steady-state version of Eq.~(\ref{eq:rateEqWithCot}). For that, we need to first write the equation analogous to Eq.~(\ref{eq:dpdtRelLinearized}) for the co-tunneling term in Eq.~(\ref{eq:rateEqWithCot}).

The co-tunneling term in Eq.~(\ref{eq:rateEqWithCot}) is evaluated
using second order perturbation theory in tunneling or, equivalently, by using the Hamiltonian $H_{\rm ex}$ of Eq.~(\ref{eq:KondoG}) with $J_{ij}^{nm}$ given by Eq.~(\ref{eq:QualitJG}). We find \begin{widetext}
\begin{align}
\left(\frac{d}{dt}p_{\gamma}(E)\right)_{\text{cot}} & =\frac{1}{2\pi}\Gamma^{2} \left(\frac{1}{E_{+}}+\frac{1}{E_{-}}\right)^{2} \sum_{\gamma'}\nu_0 \int dE'dE_{k}dE_{k'}\delta(E_{k'}+E'-E_{k}-E)\nonumber \\
\times & \left\lbrace [1-f_{\gamma'}(E_{k})][1-p_{\gamma}(E)]f_{\gamma}(E_{k'})p_{\gamma'}(E')-f_{\gamma'}(E_{k})p_{\gamma}(E)[1-f_{\gamma}(E_{k'})][1-p_{\gamma'}(E')]\right\rbrace \,.\label{eq:rateCotunneling}
\end{align}
\end{widetext}
 The relation of Eq.~(\ref{eq:rateCotunneling}) to the Hamiltonian (\ref{eq:KondoG}) is evident since averaging over different dot levels there gives $\langle \rho^2 J^{nm}J^{mn} \rangle = \pi^{-2} \Gamma^2 {(\frac{1}{E_{+}}+\frac{1}{E_{-}})^{2}}$. 
 Likewise, Eq.~(\ref{eq:rateCotunneling}) is valid at temperatures less than the charging energy, ${T\ll E_{\pm}}$. 
The distribution function of the edge electrons in Eq.~(\ref{eq:rateCotunneling}) is $f_{\gamma}(E)=1/[e^{(E+\gamma eV/2)/T}+1]$.

To find the co-tunneling version of Eq.~(\ref{eq:dpdtRelLinearized}), we again consider a small deviation of $p_\gamma$ from equilibrium. 
To the first order in bias voltage, we find from Eq.~(\ref{eq:rateCotunneling}) 
\begin{align}
&\sum_{\gamma}\gamma\left(\frac{d}{dt}p_{\gamma}(E)\right)_{\text{cot}}= \nonumber \\  &-\tau_{\text{cot}}^{-1}(E)\left[\sum_{\gamma}\gamma p_{\gamma}(E) - \frac{eV}{4T}\frac{1}{\cosh^{2}\frac{E}{2T}}\right] \,,\label{eq:dpdtCotLinearized}
\end{align}
where we have introduced the inelastic co-tunneling rate $\tau_{\text{cot}}^{-1}(E)$, 
\begin{equation}
\tau_{\text{cot}}^{-1}(E)=\frac{1}{2\pi}\frac{\Gamma^{2}}{\delta}\left(\frac{1}{E_{+}}+\frac{1}{E_{-}}\right)^{2}(\pi^{2}T^{2}+E^{2})\,.
\end{equation}

In the steady state, Eqs.~(\ref{eq:rateEqWithCot}), (\ref{eq:dpdtRelLinearized}),
and (\ref{eq:dpdtCotLinearized}) yield the rate balance equation in the form
\begin{align}
&-\tau_{e-e}^{-1}(E)\sum_{\gamma}\gamma p_{\gamma}(E) \nonumber \\
&+ \tau_{\text{cot}}^{-1}(E)\left( \frac{eV}{4T}\frac{1}{\cosh^{2}\frac{E}{2T}}-\sum_{\gamma}\gamma p_{\gamma}(E)\right)=0 \,,
\label{eq:balanceG2}
\end{align}
equivalent to  Eq.~(\ref{eq:balanceG}). We find from Eq.~(\ref{eq:balanceG2}) 
\begin{equation}
\sum_{\gamma}\gamma p_{\gamma}(E)=\frac{eV}{4T}\frac{\tau_{\text{cot}}^{-1}(E)}{\tau_{\text{cot}}^{-1}(E)+\tau_{e-e}^{-1}(E)} \frac{1}{\cosh^{2}\frac{E}{2T}}\,.
\label{eq:deltapSolution}
\end{equation}
Inserting Eq.~(\ref{eq:deltapSolution}) into the expression for $\Delta I$, Eq.~(\ref{eq:DeltaIKinEqWithCot2}), we find the correction to the conductance,
\begin{equation}
\Delta G=\frac{e^{2}}{8T}\nu_{0}\int dE\frac{\tau_{\text{cot}}^{-1}(E)\tau_{e-e}^{-1}(E)}{\tau_{\text{cot}}^{-1}(E)+\tau_{e-e}^{-1}(E)}\frac{1}{\cosh^{2}\frac{E}{2T}}\,.\label{eq:DeltaGKinEqWithCot}
\end{equation}
Comparing to Eq.~(\ref{eq:GkinEqSmalltausoEqualtunnelingrates}), we see that the result of Subsection~\ref{sub:Kinetic-equation}
remains qualitatively valid here. However, the tunneling rate $\Gamma$ there 
gets replaced by the co-tunneling rate $\tau_{\text{cot}}^{-1}(E)$
here since, in presence of charging energy, direct tunneling (described by $\Gamma$) is strongly suppressed.

Evaluating the integral in Eq.~(\ref{eq:DeltaGKinEqWithCot}) in
the limiting cases of fast ($\tau_{\text{cot}}^{-1}\gg\tau_{e-e}^{-1}$) and slow ($\tau_{\text{cot}}^{-1}\ll\tau_{e-e}^{-1}$)  co-tunneling, leads to 
\begin{equation}
\Delta G/G_{0}=\frac{2\pi}{3}\frac{T^{2}}{\delta^{2}}\begin{cases}
c_{\gamma}/g^{2}\,, & g\gg E_{C}/\Gamma\,,\\
\dfrac{\Gamma^{2}}{2}\left(\dfrac{1}{E_{+}}+\dfrac{1}{E_{-}}\right)^{2}\,, & g\ll E_{C}/\Gamma\,,
\end{cases}\label{eq:KineticConductanceFinalWithCot}
\end{equation}
valid in the temperature interval $g^{1/2}\delta\ll T\ll E_{\pm}$.
Here the first line corresponds to the case $\tau_{\text{cot}}^{-1}\gg\tau_{e-e}^{-1}$,
where the slow intra-dot relaxation is the bottleneck for backscattering,
and $\Delta G\sim\nu_{0}\tau_{e-e}^{-1}$. In the second line the
bottleneck is shifted to the co-tunneling process, and $\Delta G\sim\nu_{0}\tau_{\text{cot}}^{-1}$.
Extrapolating towards low-temperatures, $T\gtrsim \delta$, the latter result  can be written as $\Delta G \sim (\rho J_0)^2$ where $J_0$ is the (disorder-averaged) isotropic part of exchange in  Eq.~(\ref{eq:Heffective}) [see also Eq.~(\ref{eq:Kondo1})]. 
This is similar in form to the $T\ll\delta$ result in the odd valley, $\Delta G \sim (\rho \delta J)^2$, see Eqs.~(\ref{eq:DeltaGOddValleyAboveTKNonRGd}) and (\ref{eq:TypicalDeltaGOddValleyApprx}). 
In the former case, however, the anisotropic part of exchange, $\delta J$, is not necessary for backscattering, since the dot spin relaxes fast in the intra-dot scattering processes. 

\bibliographystyle{apsrev4-1}
\bibliography{HelEdgeReferences}

\begin{thebibliography}{53}%
\makeatletter
\providecommand \@ifxundefined [1]{%
 \@ifx{#1\undefined}
}%
\providecommand \@ifnum [1]{%
 \ifnum #1\expandafter \@firstoftwo
 \else \expandafter \@secondoftwo
 \fi
}%
\providecommand \@ifx [1]{%
 \ifx #1\expandafter \@firstoftwo
 \else \expandafter \@secondoftwo
 \fi
}%
\providecommand \natexlab [1]{#1}%
\providecommand \enquote  [1]{``#1''}%
\providecommand \bibnamefont  [1]{#1}%
\providecommand \bibfnamefont [1]{#1}%
\providecommand \citenamefont [1]{#1}%
\providecommand \href@noop [0]{\@secondoftwo}%
\providecommand \href [0]{\begingroup \@sanitize@url \@href}%
\providecommand \@href[1]{\@@startlink{#1}\@@href}%
\providecommand \@@href[1]{\endgroup#1\@@endlink}%
\providecommand \@sanitize@url [0]{\catcode `\\12\catcode `\$12\catcode
  `\&12\catcode `\#12\catcode `\^12\catcode `\_12\catcode `\%12\relax}%
\providecommand \@@startlink[1]{}%
\providecommand \@@endlink[0]{}%
\providecommand \url  [0]{\begingroup\@sanitize@url \@url }%
\providecommand \@url [1]{\endgroup\@href {#1}{\urlprefix }}%
\providecommand \urlprefix  [0]{URL }%
\providecommand \Eprint [0]{\href }%
\providecommand \doibase [0]{http://dx.doi.org/}%
\providecommand \selectlanguage [0]{\@gobble}%
\providecommand \bibinfo  [0]{\@secondoftwo}%
\providecommand \bibfield  [0]{\@secondoftwo}%
\providecommand \translation [1]{[#1]}%
\providecommand \BibitemOpen [0]{}%
\providecommand \bibitemStop [0]{}%
\providecommand \bibitemNoStop [0]{.\EOS\space}%
\providecommand \EOS [0]{\spacefactor3000\relax}%
\providecommand \BibitemShut  [1]{\csname bibitem#1\endcsname}%
\let\auto@bib@innerbib\@empty
\bibitem [{\citenamefont {Kane}\ and\ \citenamefont
  {Mele}(2005{\natexlab{a}})}]{kane05a}%
  \BibitemOpen
  \bibfield  {author} {\bibinfo {author} {\bibfnamefont {C.~L.}\ \bibnamefont
  {Kane}}\ and\ \bibinfo {author} {\bibfnamefont {E.~J.}\ \bibnamefont
  {Mele}},\ }\href {\doibase 10.1103/PhysRevLett.95.146802} {\bibfield
  {journal} {\bibinfo  {journal} {Phys. Rev. Lett.}\ }\textbf {\bibinfo
  {volume} {95}},\ \bibinfo {pages} {146802} (\bibinfo {year}
  {2005}{\natexlab{a}})}\BibitemShut {NoStop}%
\bibitem [{\citenamefont {Kane}\ and\ \citenamefont
  {Mele}(2005{\natexlab{b}})}]{kane05b}%
  \BibitemOpen
  \bibfield  {author} {\bibinfo {author} {\bibfnamefont {C.~L.}\ \bibnamefont
  {Kane}}\ and\ \bibinfo {author} {\bibfnamefont {E.~J.}\ \bibnamefont
  {Mele}},\ }\href {\doibase 10.1103/PhysRevLett.95.226801} {\bibfield
  {journal} {\bibinfo  {journal} {Phys. Rev. Lett.}\ }\textbf {\bibinfo
  {volume} {95}},\ \bibinfo {pages} {226801} (\bibinfo {year}
  {2005}{\natexlab{b}})}\BibitemShut {NoStop}%
\bibitem [{\citenamefont {Bernevig}\ \emph {et~al.}(2006)\citenamefont
  {Bernevig}, \citenamefont {Hughes},\ and\ \citenamefont
  {Zhang}}]{bernevig06}%
  \BibitemOpen
  \bibfield  {author} {\bibinfo {author} {\bibfnamefont {B.~A.}\ \bibnamefont
  {Bernevig}}, \bibinfo {author} {\bibfnamefont {T.~L.}\ \bibnamefont
  {Hughes}}, \ and\ \bibinfo {author} {\bibfnamefont {S.-C.}\ \bibnamefont
  {Zhang}},\ }\href {\doibase 10.1126/science.1133734} {\bibfield  {journal}
  {\bibinfo  {journal} {Science}\ }\textbf {\bibinfo {volume} {314}},\ \bibinfo
  {pages} {1757} (\bibinfo {year} {2006})}\BibitemShut {NoStop}%
\bibitem [{\citenamefont {Xu}\ and\ \citenamefont {Moore}(2006)}]{xu06}%
  \BibitemOpen
  \bibfield  {author} {\bibinfo {author} {\bibfnamefont {C.}~\bibnamefont
  {Xu}}\ and\ \bibinfo {author} {\bibfnamefont {J.~E.}\ \bibnamefont {Moore}},\
  }\href {\doibase 10.1103/PhysRevB.73.045322} {\bibfield  {journal} {\bibinfo
  {journal} {Phys. Rev. B}\ }\textbf {\bibinfo {volume} {73}},\ \bibinfo
  {pages} {045322} (\bibinfo {year} {2006})}\BibitemShut {NoStop}%
\bibitem [{\citenamefont {Wu}\ \emph {et~al.}(2006)\citenamefont {Wu},
  \citenamefont {Bernevig},\ and\ \citenamefont {Zhang}}]{wu06}%
  \BibitemOpen
  \bibfield  {author} {\bibinfo {author} {\bibfnamefont {C.}~\bibnamefont
  {Wu}}, \bibinfo {author} {\bibfnamefont {B.~A.}\ \bibnamefont {Bernevig}}, \
  and\ \bibinfo {author} {\bibfnamefont {S.-C.}\ \bibnamefont {Zhang}},\ }\href
  {\doibase 10.1103/PhysRevLett.96.106401} {\bibfield  {journal} {\bibinfo
  {journal} {Phys. Rev. Lett.}\ }\textbf {\bibinfo {volume} {96}},\ \bibinfo
  {pages} {106401} (\bibinfo {year} {2006})}\BibitemShut {NoStop}%
\bibitem [{\citenamefont {Budich}\ \emph {et~al.}(2012)\citenamefont {Budich},
  \citenamefont {Dolcini}, \citenamefont {Recher},\ and\ \citenamefont
  {Trauzettel}}]{budich12}%
  \BibitemOpen
  \bibfield  {author} {\bibinfo {author} {\bibfnamefont {J.~C.}\ \bibnamefont
  {Budich}}, \bibinfo {author} {\bibfnamefont {F.}~\bibnamefont {Dolcini}},
  \bibinfo {author} {\bibfnamefont {P.}~\bibnamefont {Recher}}, \ and\ \bibinfo
  {author} {\bibfnamefont {B.}~\bibnamefont {Trauzettel}},\ }\href {\doibase
  10.1103/PhysRevLett.108.086602} {\bibfield  {journal} {\bibinfo  {journal}
  {Phys. Rev. Lett.}\ }\textbf {\bibinfo {volume} {108}},\ \bibinfo {pages}
  {086602} (\bibinfo {year} {2012})}\BibitemShut {NoStop}%
\bibitem [{\citenamefont {Schmidt}\ \emph {et~al.}(2012)\citenamefont
  {Schmidt}, \citenamefont {Rachel}, \citenamefont {von Oppen},\ and\
  \citenamefont {Glazman}}]{schmidt12}%
  \BibitemOpen
  \bibfield  {author} {\bibinfo {author} {\bibfnamefont {T.~L.}\ \bibnamefont
  {Schmidt}}, \bibinfo {author} {\bibfnamefont {S.}~\bibnamefont {Rachel}},
  \bibinfo {author} {\bibfnamefont {F.}~\bibnamefont {von Oppen}}, \ and\
  \bibinfo {author} {\bibfnamefont {L.~I.}\ \bibnamefont {Glazman}},\ }\href
  {\doibase 10.1103/PhysRevLett.108.156402} {\bibfield  {journal} {\bibinfo
  {journal} {Phys. Rev. Lett.}\ }\textbf {\bibinfo {volume} {108}},\ \bibinfo
  {pages} {156402} (\bibinfo {year} {2012})}\BibitemShut {NoStop}%
\bibitem [{\citenamefont {Lezmy}\ \emph {et~al.}(2012)\citenamefont {Lezmy},
  \citenamefont {Oreg},\ and\ \citenamefont {Berkooz}}]{lezmy12}%
  \BibitemOpen
  \bibfield  {author} {\bibinfo {author} {\bibfnamefont {N.}~\bibnamefont
  {Lezmy}}, \bibinfo {author} {\bibfnamefont {Y.}~\bibnamefont {Oreg}}, \ and\
  \bibinfo {author} {\bibfnamefont {M.}~\bibnamefont {Berkooz}},\ }\href
  {\doibase 10.1103/PhysRevB.85.235304} {\bibfield  {journal} {\bibinfo
  {journal} {Phys. Rev. B}\ }\textbf {\bibinfo {volume} {85}},\ \bibinfo
  {pages} {235304} (\bibinfo {year} {2012})}\BibitemShut {NoStop}%
\bibitem [{\citenamefont {{Kainaris}}\ \emph {et~al.}(2014)\citenamefont
  {{Kainaris}}, \citenamefont {{Gornyi}}, \citenamefont {{Carr}},\ and\
  \citenamefont {{Mirlin}}}]{kainaris14}%
  \BibitemOpen
  \bibfield  {author} {\bibinfo {author} {\bibfnamefont {N.}~\bibnamefont
  {{Kainaris}}}, \bibinfo {author} {\bibfnamefont {I.~V.}\ \bibnamefont
  {{Gornyi}}}, \bibinfo {author} {\bibfnamefont {S.~T.}\ \bibnamefont
  {{Carr}}}, \ and\ \bibinfo {author} {\bibfnamefont {A.~D.}\ \bibnamefont
  {{Mirlin}}},\ }\href@noop {} {\bibfield  {journal} {\bibinfo  {journal}
  {ArXiv e-prints}\ } (\bibinfo {year} {2014})},\ \Eprint
  {http://arxiv.org/abs/1404.3129} {arXiv:1404.3129 [cond-mat.mes-hall]}
  \BibitemShut {NoStop}%
\bibitem [{\citenamefont {Qi}\ and\ \citenamefont {Zhang}(2011)}]{qi_review}%
  \BibitemOpen
  \bibfield  {author} {\bibinfo {author} {\bibfnamefont {X.-L.}\ \bibnamefont
  {Qi}}\ and\ \bibinfo {author} {\bibfnamefont {S.-C.}\ \bibnamefont {Zhang}},\
  }\href {\doibase 10.1103/RevModPhys.83.1057} {\bibfield  {journal} {\bibinfo
  {journal} {Rev. Mod. Phys.}\ }\textbf {\bibinfo {volume} {83}},\ \bibinfo
  {pages} {1057} (\bibinfo {year} {2011})}\BibitemShut {NoStop}%
\bibitem [{\citenamefont {K{\"o}nig}\ \emph {et~al.}(2007)\citenamefont
  {K{\"o}nig}, \citenamefont {Wiedmann}, \citenamefont {Br{\"u}ne},
  \citenamefont {Roth}, \citenamefont {Buhmann}, \citenamefont {Molenkamp},
  \citenamefont {Qi},\ and\ \citenamefont {Zhang}}]{konig07}%
  \BibitemOpen
  \bibfield  {author} {\bibinfo {author} {\bibfnamefont {M.}~\bibnamefont
  {K{\"o}nig}}, \bibinfo {author} {\bibfnamefont {S.}~\bibnamefont {Wiedmann}},
  \bibinfo {author} {\bibfnamefont {C.}~\bibnamefont {Br{\"u}ne}}, \bibinfo
  {author} {\bibfnamefont {A.}~\bibnamefont {Roth}}, \bibinfo {author}
  {\bibfnamefont {H.}~\bibnamefont {Buhmann}}, \bibinfo {author} {\bibfnamefont
  {L.~W.}\ \bibnamefont {Molenkamp}}, \bibinfo {author} {\bibfnamefont {X.-L.}\
  \bibnamefont {Qi}}, \ and\ \bibinfo {author} {\bibfnamefont {S.-C.}\
  \bibnamefont {Zhang}},\ }\href {\doibase 10.1126/science.1148047} {\bibfield
  {journal} {\bibinfo  {journal} {Science}\ }\textbf {\bibinfo {volume}
  {318}},\ \bibinfo {pages} {766} (\bibinfo {year} {2007})}\BibitemShut
  {NoStop}%
\bibitem [{\citenamefont {Gusev}\ \emph {et~al.}(2014)\citenamefont {Gusev},
  \citenamefont {Kvon}, \citenamefont {Olshanetsky}, \citenamefont {Levin},
  \citenamefont {Krupko}, \citenamefont {Portal}, \citenamefont {Mikhailov},\
  and\ \citenamefont {Dvoretsky}}]{gusev13}%
  \BibitemOpen
  \bibfield  {author} {\bibinfo {author} {\bibfnamefont {G.~M.}\ \bibnamefont
  {Gusev}}, \bibinfo {author} {\bibfnamefont {Z.~D.}\ \bibnamefont {Kvon}},
  \bibinfo {author} {\bibfnamefont {E.~B.}\ \bibnamefont {Olshanetsky}},
  \bibinfo {author} {\bibfnamefont {A.~D.}\ \bibnamefont {Levin}}, \bibinfo
  {author} {\bibfnamefont {Y.}~\bibnamefont {Krupko}}, \bibinfo {author}
  {\bibfnamefont {J.~C.}\ \bibnamefont {Portal}}, \bibinfo {author}
  {\bibfnamefont {N.~N.}\ \bibnamefont {Mikhailov}}, \ and\ \bibinfo {author}
  {\bibfnamefont {S.~A.}\ \bibnamefont {Dvoretsky}},\ }\href {\doibase
  10.1103/PhysRevB.89.125305} {\bibfield  {journal} {\bibinfo  {journal} {Phys.
  Rev. B}\ }\textbf {\bibinfo {volume} {89}},\ \bibinfo {pages} {125305}
  (\bibinfo {year} {2014})}\BibitemShut {NoStop}%
\bibitem [{\citenamefont {{Du}}\ \emph {et~al.}(2013)\citenamefont {{Du}},
  \citenamefont {{Knez}}, \citenamefont {{Sullivan}},\ and\ \citenamefont
  {{Du}}}]{du13}%
  \BibitemOpen
  \bibfield  {author} {\bibinfo {author} {\bibfnamefont {L.}~\bibnamefont
  {{Du}}}, \bibinfo {author} {\bibfnamefont {I.}~\bibnamefont {{Knez}}},
  \bibinfo {author} {\bibfnamefont {G.}~\bibnamefont {{Sullivan}}}, \ and\
  \bibinfo {author} {\bibfnamefont {R.-R.}\ \bibnamefont {{Du}}},\ }\href@noop
  {} {\bibfield  {journal} {\bibinfo  {journal} {ArXiv e-prints}\ } (\bibinfo
  {year} {2013})},\ \Eprint {http://arxiv.org/abs/1306.1925} {arXiv:1306.1925
  [cond-mat.mes-hall]} \BibitemShut {NoStop}%
\bibitem [{\citenamefont {{Spanton}}\ \emph {et~al.}(2014)\citenamefont
  {{Spanton}}, \citenamefont {{Nowack}}, \citenamefont {{Du}}, \citenamefont
  {{Du}},\ and\ \citenamefont {{Moler}}}]{spanton14}%
  \BibitemOpen
  \bibfield  {author} {\bibinfo {author} {\bibfnamefont {E.~M.}\ \bibnamefont
  {{Spanton}}}, \bibinfo {author} {\bibfnamefont {K.~C.}\ \bibnamefont
  {{Nowack}}}, \bibinfo {author} {\bibfnamefont {L.}~\bibnamefont {{Du}}},
  \bibinfo {author} {\bibfnamefont {R.-R.}\ \bibnamefont {{Du}}}, \ and\
  \bibinfo {author} {\bibfnamefont {K.~A.}\ \bibnamefont {{Moler}}},\
  }\href@noop {} {\bibfield  {journal} {\bibinfo  {journal} {ArXiv e-prints}\ }
  (\bibinfo {year} {2014})},\ \Eprint {http://arxiv.org/abs/1401.1531}
  {arXiv:1401.1531 [cond-mat.mes-hall]} \BibitemShut {NoStop}%
\bibitem [{\citenamefont {Liu}\ \emph {et~al.}(2008)\citenamefont {Liu},
  \citenamefont {Hughes}, \citenamefont {Qi}, \citenamefont {Wang},\ and\
  \citenamefont {Zhang}}]{liu08}%
  \BibitemOpen
  \bibfield  {author} {\bibinfo {author} {\bibfnamefont {C.}~\bibnamefont
  {Liu}}, \bibinfo {author} {\bibfnamefont {T.~L.}\ \bibnamefont {Hughes}},
  \bibinfo {author} {\bibfnamefont {X.-L.}\ \bibnamefont {Qi}}, \bibinfo
  {author} {\bibfnamefont {K.}~\bibnamefont {Wang}}, \ and\ \bibinfo {author}
  {\bibfnamefont {S.-C.}\ \bibnamefont {Zhang}},\ }\href {\doibase
  10.1103/PhysRevLett.100.236601} {\bibfield  {journal} {\bibinfo  {journal}
  {Phys. Rev. Lett.}\ }\textbf {\bibinfo {volume} {100}},\ \bibinfo {pages}
  {236601} (\bibinfo {year} {2008})}\BibitemShut {NoStop}%
\bibitem [{\citenamefont {V\"ayrynen}\ \emph {et~al.}(2013)\citenamefont
  {V\"ayrynen}, \citenamefont {Goldstein},\ and\ \citenamefont
  {Glazman}}]{vayrynen13}%
  \BibitemOpen
  \bibfield  {author} {\bibinfo {author} {\bibfnamefont {J.~I.}\ \bibnamefont
  {V\"ayrynen}}, \bibinfo {author} {\bibfnamefont {M.}~\bibnamefont
  {Goldstein}}, \ and\ \bibinfo {author} {\bibfnamefont {L.~I.}\ \bibnamefont
  {Glazman}},\ }\href {\doibase 10.1103/PhysRevLett.110.216402} {\bibfield
  {journal} {\bibinfo  {journal} {Phys. Rev. Lett.}\ }\textbf {\bibinfo
  {volume} {110}},\ \bibinfo {pages} {216402} (\bibinfo {year}
  {2013})}\BibitemShut {NoStop}%
\bibitem [{\citenamefont {Kouwenhoven}\ \emph {et~al.}(1997)\citenamefont
  {Kouwenhoven}, \citenamefont {Marcus}, \citenamefont {McEuen}, \citenamefont
  {Tarucha}, \citenamefont {Westervelt},\ and\ \citenamefont
  {Wingreen}}]{Kouwenhoven97}%
  \BibitemOpen
  \bibfield  {author} {\bibinfo {author} {\bibfnamefont {L.~P.}\ \bibnamefont
  {Kouwenhoven}}, \bibinfo {author} {\bibfnamefont {C.~M.}\ \bibnamefont
  {Marcus}}, \bibinfo {author} {\bibfnamefont {P.~L.}\ \bibnamefont {McEuen}},
  \bibinfo {author} {\bibfnamefont {S.}~\bibnamefont {Tarucha}}, \bibinfo
  {author} {\bibfnamefont {R.~M.}\ \bibnamefont {Westervelt}}, \ and\ \bibinfo
  {author} {\bibfnamefont {N.~S.}\ \bibnamefont {Wingreen}},\ }\href@noop {}
  {\bibfield  {journal} {\bibinfo  {journal} {Proceedings of the NATO Advanced
  Study Institute on Mesoscopic Electron Transport}\ }\textbf {\bibinfo
  {volume} {E345}},\ \bibinfo {pages} {105} (\bibinfo {year}
  {1997})}\BibitemShut {NoStop}%
\bibitem [{\citenamefont {Glazman}\ and\ \citenamefont
  {Pustilnik}(2005)}]{glazman05}%
  \BibitemOpen
  \bibfield  {author} {\bibinfo {author} {\bibfnamefont {L.~I.}\ \bibnamefont
  {Glazman}}\ and\ \bibinfo {author} {\bibfnamefont {M.}~\bibnamefont
  {Pustilnik}},\ }in\ \href {\doibase 10.1016/S0924-8099(05)80050-2} {\emph
  {\bibinfo {booktitle} {Nanophysics: Coherence and Transport École d'été de
  Physique des Houches Session LXXXI}}},\ \bibinfo {series} {Les Houches},
  Vol.~\bibinfo {volume} {81},\ \bibinfo {editor} {edited by\ \bibinfo {editor}
  {\bibfnamefont {S.~G. G.~M.}\ \bibnamefont {H.~Bouchiat}, \bibfnamefont
  {Y.~Gefen}}\ and\ \bibinfo {editor} {\bibfnamefont {J.}~\bibnamefont
  {Dalibard}}}\ (\bibinfo  {publisher} {Elsevier},\ \bibinfo {year} {2005})\
  pp.\ \bibinfo {pages} {427 -- 478}\BibitemShut {NoStop}%
\bibitem [{\citenamefont {Roth}\ \emph {et~al.}(2009)\citenamefont {Roth},
  \citenamefont {Br{\"u}ne}, \citenamefont {Buhmann}, \citenamefont
  {Molenkamp}, \citenamefont {Maciejko}, \citenamefont {Qi},\ and\
  \citenamefont {Zhang}}]{roth09}%
  \BibitemOpen
  \bibfield  {author} {\bibinfo {author} {\bibfnamefont {A.}~\bibnamefont
  {Roth}}, \bibinfo {author} {\bibfnamefont {C.}~\bibnamefont {Br{\"u}ne}},
  \bibinfo {author} {\bibfnamefont {H.}~\bibnamefont {Buhmann}}, \bibinfo
  {author} {\bibfnamefont {L.~W.}\ \bibnamefont {Molenkamp}}, \bibinfo {author}
  {\bibfnamefont {J.}~\bibnamefont {Maciejko}}, \bibinfo {author}
  {\bibfnamefont {X.-L.}\ \bibnamefont {Qi}}, \ and\ \bibinfo {author}
  {\bibfnamefont {S.-C.}\ \bibnamefont {Zhang}},\ }\href {\doibase
  10.1126/science.1174736} {\bibfield  {journal} {\bibinfo  {journal}
  {Science}\ }\textbf {\bibinfo {volume} {325}},\ \bibinfo {pages} {294}
  (\bibinfo {year} {2009})}\BibitemShut {NoStop}%
\bibitem [{\citenamefont {Nozi{\`e}res}\ and\ \citenamefont
  {Pines}(1999)}]{nozieresBook}%
  \BibitemOpen
  \bibfield  {author} {\bibinfo {author} {\bibfnamefont {P.}~\bibnamefont
  {Nozi{\`e}res}}\ and\ \bibinfo {author} {\bibfnamefont {D.}~\bibnamefont
  {Pines}},\ }\href@noop {} {\emph {\bibinfo {title} {{The theory of quantum
  liquids}}}},\ Advanced book classics\ (\bibinfo  {publisher} {Perseus},\
  \bibinfo {address} {Cambridge, MA},\ \bibinfo {year} {1999})\BibitemShut
  {NoStop}%
\bibitem [{\citenamefont {Aleiner}\ \emph {et~al.}(2002)\citenamefont
  {Aleiner}, \citenamefont {Brouwer},\ and\ \citenamefont
  {Glazman}}]{aleiner02}%
  \BibitemOpen
  \bibfield  {author} {\bibinfo {author} {\bibfnamefont {I.~L.}\ \bibnamefont
  {Aleiner}}, \bibinfo {author} {\bibfnamefont {P.~W.}\ \bibnamefont
  {Brouwer}}, \ and\ \bibinfo {author} {\bibfnamefont {L.~I.}\ \bibnamefont
  {Glazman}},\ }\href@noop {} {\bibfield  {journal} {\bibinfo  {journal}
  {Physics Reports}\ }\textbf {\bibinfo {volume} {358}},\ \bibinfo {pages}
  {309} (\bibinfo {year} {2002})}\BibitemShut {NoStop}%
\bibitem [{\citenamefont {Anderson}(1966)}]{anderson66}%
  \BibitemOpen
  \bibfield  {author} {\bibinfo {author} {\bibfnamefont {P.~W.}\ \bibnamefont
  {Anderson}},\ }\href {\doibase 10.1103/PhysRevLett.17.95} {\bibfield
  {journal} {\bibinfo  {journal} {Phys. Rev. Lett.}\ }\textbf {\bibinfo
  {volume} {17}},\ \bibinfo {pages} {95} (\bibinfo {year} {1966})}\BibitemShut
  {NoStop}%
\bibitem [{Note1()}]{Note1}%
  \BibitemOpen
  \bibinfo {note} {This limitation is lifted if the tunneling region is
  extended, see Appendix \ref {Adx:Tunneling-through-an}}\BibitemShut {NoStop}%
\bibitem [{\citenamefont {Anderson}(1970)}]{anderson70}%
  \BibitemOpen
  \bibfield  {author} {\bibinfo {author} {\bibfnamefont {P.}~\bibnamefont
  {Anderson}},\ }\href@noop {} {\bibfield  {journal} {\bibinfo  {journal}
  {Journal of Physics C: Solid State Physics}\ }\textbf {\bibinfo {volume}
  {3}},\ \bibinfo {pages} {2436} (\bibinfo {year} {1970})}\BibitemShut
  {NoStop}%
\bibitem [{\citenamefont {Tanaka}\ \emph {et~al.}(2011)\citenamefont {Tanaka},
  \citenamefont {Furusaki},\ and\ \citenamefont {Matveev}}]{tanaka11}%
  \BibitemOpen
  \bibfield  {author} {\bibinfo {author} {\bibfnamefont {Y.}~\bibnamefont
  {Tanaka}}, \bibinfo {author} {\bibfnamefont {A.}~\bibnamefont {Furusaki}}, \
  and\ \bibinfo {author} {\bibfnamefont {K.~A.}\ \bibnamefont {Matveev}},\
  }\href {\doibase 10.1103/PhysRevLett.106.236402} {\bibfield  {journal}
  {\bibinfo  {journal} {Phys. Rev. Lett.}\ }\textbf {\bibinfo {volume} {106}},\
  \bibinfo {pages} {236402} (\bibinfo {year} {2011})}\BibitemShut {NoStop}%
\bibitem [{\citenamefont {Landau}\ and\ \citenamefont
  {Lifshitz}(1980)}]{LLvol5}%
  \BibitemOpen
  \bibfield  {author} {\bibinfo {author} {\bibfnamefont {L.}~\bibnamefont
  {Landau}}\ and\ \bibinfo {author} {\bibfnamefont {E.}~\bibnamefont
  {Lifshitz}},\ }\href@noop {} {\emph {\bibinfo {title} {Statistical
  Physics}}},\ \bibinfo {edition} {3rd}\ ed.,\ Vol.~\bibinfo {volume} {5}\
  (\bibinfo  {publisher} {Butterworth-Heinemann},\ \bibinfo {year}
  {1980})\BibitemShut {NoStop}%
\bibitem [{Note2()}]{Note2}%
  \BibitemOpen
  \bibinfo {note} {Using the equilibrium distribution $\protect \mathaccentV
  {hat}05E\rho $ is compatible with having different chemical potentials of the
  two fermion species due to the presence of the integral of motion $\protect
  \mathaccentV {hat}05ES_z^{\protect \rm tot}$}\BibitemShut {NoStop}%
\bibitem [{\citenamefont {Bloch}(1946)}]{bloch46}%
  \BibitemOpen
  \bibfield  {author} {\bibinfo {author} {\bibfnamefont {F.}~\bibnamefont
  {Bloch}},\ }\href {\doibase 10.1103/PhysRev.70.460} {\bibfield  {journal}
  {\bibinfo  {journal} {Phys. Rev.}\ }\textbf {\bibinfo {volume} {70}},\
  \bibinfo {pages} {460} (\bibinfo {year} {1946})}\BibitemShut {NoStop}%
\bibitem [{\citenamefont {Korringa}(1950)}]{korringa50}%
  \BibitemOpen
  \bibfield  {author} {\bibinfo {author} {\bibfnamefont {J.}~\bibnamefont
  {Korringa}},\ }\href {\doibase
  http://dx.doi.org/10.1016/0031-8914(50)90105-4} {\bibfield  {journal}
  {\bibinfo  {journal} {Physica}\ }\textbf {\bibinfo {volume} {16}},\ \bibinfo
  {pages} {601 } (\bibinfo {year} {1950})}\BibitemShut {NoStop}%
\bibitem [{\citenamefont {van~der Wiel}\ \emph {et~al.}(2000)\citenamefont
  {van~der Wiel}, \citenamefont {Franceschi}, \citenamefont {Fujisawa},
  \citenamefont {Elzerman}, \citenamefont {Tarucha},\ and\ \citenamefont
  {Kouwenhoven}}]{vanderWiel00}%
  \BibitemOpen
  \bibfield  {author} {\bibinfo {author} {\bibfnamefont {W.~G.}\ \bibnamefont
  {van~der Wiel}}, \bibinfo {author} {\bibfnamefont {S.~D.}\ \bibnamefont
  {Franceschi}}, \bibinfo {author} {\bibfnamefont {T.}~\bibnamefont
  {Fujisawa}}, \bibinfo {author} {\bibfnamefont {J.~M.}\ \bibnamefont
  {Elzerman}}, \bibinfo {author} {\bibfnamefont {S.}~\bibnamefont {Tarucha}}, \
  and\ \bibinfo {author} {\bibfnamefont {L.~P.}\ \bibnamefont {Kouwenhoven}},\
  }\href {\doibase 10.1126/science.289.5487.2105} {\bibfield  {journal}
  {\bibinfo  {journal} {Science}\ }\textbf {\bibinfo {volume} {289}},\ \bibinfo
  {pages} {2105} (\bibinfo {year} {2000})}\BibitemShut {NoStop}%
\bibitem [{\citenamefont {{Gergel'}}\ and\ \citenamefont
  {{Suris}}(1978)}]{gergel78}%
  \BibitemOpen
  \bibfield  {author} {\bibinfo {author} {\bibfnamefont {V.~A.}\ \bibnamefont
  {{Gergel'}}}\ and\ \bibinfo {author} {\bibfnamefont {R.~A.}\ \bibnamefont
  {{Suris}}},\ }\href@noop {} {\bibfield  {journal} {\bibinfo  {journal} {Zh.
  Eksp. Teor. Fiz.}\ }\textbf {\bibinfo {volume} {75}},\ \bibinfo {pages} {191}
  (\bibinfo {year} {1978})},\ \bibinfo {note} {[Sov. Phys. JETP \textbf{48}, 95
  (1978)]}\BibitemShut {NoStop}%
\bibitem [{Note3()}]{Note3}%
  \BibitemOpen
  \bibinfo {note} {The lower limit is non-zero for a finite width sample where
  $d<W$. Here we however assume a wide enough sample, so that the lower limit
  can be taken smaller than any other scale\label {fn:samplewidth}}\BibitemShut
  {NoStop}%
\bibitem [{\citenamefont {Shklovskii}\ and\ \citenamefont
  {Efros}(1984)}]{shklovskii84}%
  \BibitemOpen
  \bibfield  {author} {\bibinfo {author} {\bibfnamefont {B.}~\bibnamefont
  {Shklovskii}}\ and\ \bibinfo {author} {\bibfnamefont {A.}~\bibnamefont
  {Efros}},\ }\href@noop {} {\emph {\bibinfo {title} {{Electronic properties of
  doped semiconductors}}}},\ Springer Series in Solid-State Sciences\ (\bibinfo
   {publisher} {Springer, Berlin},\ \bibinfo {year} {1984})\BibitemShut
  {NoStop}%
\bibitem [{Note8()}]{Note8}%
  \BibitemOpen
  \bibinfo {note} {We use the symplectic ensemble since a generic
  single-particle Hamiltonian of the dot does not conserve any component of the
  electron spin (i.e., the spin scattering rate is larger than the level
  spacing, $\tau _{so}\delta \ll 1$).}\BibitemShut {Stop}%
\bibitem [{\citenamefont {Sivan}\ \emph {et~al.}(1994)\citenamefont {Sivan},
  \citenamefont {Imry},\ and\ \citenamefont {Aronov}}]{sivan94}%
  \BibitemOpen
  \bibfield  {author} {\bibinfo {author} {\bibfnamefont {U.}~\bibnamefont
  {Sivan}}, \bibinfo {author} {\bibfnamefont {Y.}~\bibnamefont {Imry}}, \ and\
  \bibinfo {author} {\bibfnamefont {A.}~\bibnamefont {Aronov}},\ }\href@noop {}
  {\bibfield  {journal} {\bibinfo  {journal} {EPL (Europhysics Letters)}\
  }\textbf {\bibinfo {volume} {28}},\ \bibinfo {pages} {115} (\bibinfo {year}
  {1994})}\BibitemShut {NoStop}%
\bibitem [{\citenamefont {Altshuler}\ \emph {et~al.}(1997)\citenamefont
  {Altshuler}, \citenamefont {Gefen}, \citenamefont {Kamenev},\ and\
  \citenamefont {Levitov}}]{altshuler97}%
  \BibitemOpen
  \bibfield  {author} {\bibinfo {author} {\bibfnamefont {B.~L.}\ \bibnamefont
  {Altshuler}}, \bibinfo {author} {\bibfnamefont {Y.}~\bibnamefont {Gefen}},
  \bibinfo {author} {\bibfnamefont {A.}~\bibnamefont {Kamenev}}, \ and\
  \bibinfo {author} {\bibfnamefont {L.~S.}\ \bibnamefont {Levitov}},\ }\href
  {\doibase 10.1103/PhysRevLett.78.2803} {\bibfield  {journal} {\bibinfo
  {journal} {Phys. Rev. Lett.}\ }\textbf {\bibinfo {volume} {78}},\ \bibinfo
  {pages} {2803} (\bibinfo {year} {1997})}\BibitemShut {NoStop}%
\bibitem [{Note4()}]{Note4}%
  \BibitemOpen
  \bibinfo {note} {For example, for HgTe with $E_g \approx 20\protect \tmspace
  +\thinmuskip {.1667em} \protect \text {meV}$, $v \approx {5.5\times
  10^5\protect \tmspace +\thinmuskip {.1667em} \protect \text {m/s}}$, and
  $\alpha \approx 0.3$ one finds~\cite {gusev13} $a_B \approx 120\protect
  \tmspace +\thinmuskip {.1667em} \protect \text {nm}$.}\BibitemShut {Stop}%
\bibitem [{\citenamefont {Hewson}(1997)}]{hewson97}%
  \BibitemOpen
  \bibfield  {author} {\bibinfo {author} {\bibfnamefont {A.~C.}\ \bibnamefont
  {Hewson}},\ }\href@noop {} {\emph {\bibinfo {title} {The Kondo problem to
  heavy fermions}}}\ (\bibinfo  {publisher} {Cambridge University Press},\
  \bibinfo {year} {1997})\BibitemShut {NoStop}%
\bibitem [{\citenamefont {Nozieres}(1974)}]{nozieres74}%
  \BibitemOpen
  \bibfield  {author} {\bibinfo {author} {\bibfnamefont {P.}~\bibnamefont
  {Nozieres}},\ }\href@noop {} {\bibfield  {journal} {\bibinfo  {journal}
  {Journal of Low Temperature Physics}\ }\textbf {\bibinfo {volume} {17}},\
  \bibinfo {pages} {31} (\bibinfo {year} {1974})}\BibitemShut {NoStop}%
\bibitem [{Note5()}]{Note5}%
  \BibitemOpen
  \bibinfo {note} {This is the gate voltage at which the lowest energy
  excitation of the dot changes between one with charge and one with spin. The
  nature of these excitations defines the form of denominators in the
  perturbative result for $\delta J_{ij}$, see Eq.~(\ref {eq:Jyx}) and
  Table~\ref {table-Jcomponents} of Appendix \ref
  {sec:Components-of-the}.}\BibitemShut {Stop}%
\bibitem [{\citenamefont {{Knap}}\ \emph {et~al.}(2014)\citenamefont {{Knap}},
  \citenamefont {{Sau}}, \citenamefont {{Halperin}},\ and\ \citenamefont
  {{Demler}}}]{knap14}%
  \BibitemOpen
  \bibfield  {author} {\bibinfo {author} {\bibfnamefont {M.}~\bibnamefont
  {{Knap}}}, \bibinfo {author} {\bibfnamefont {J.~D.}\ \bibnamefont {{Sau}}},
  \bibinfo {author} {\bibfnamefont {B.~I.}\ \bibnamefont {{Halperin}}}, \ and\
  \bibinfo {author} {\bibfnamefont {E.}~\bibnamefont {{Demler}}},\ }\href@noop
  {} {\bibfield  {journal} {\bibinfo  {journal} {ArXiv e-prints}\ } (\bibinfo
  {year} {2014})},\ \Eprint {http://arxiv.org/abs/1405.0277} {arXiv:1405.0277
  [cond-mat.mes-hall]} \BibitemShut {NoStop}%
\bibitem [{\citenamefont {Fu}\ and\ \citenamefont {Kane}(2012)}]{fu12}%
  \BibitemOpen
  \bibfield  {author} {\bibinfo {author} {\bibfnamefont {L.}~\bibnamefont
  {Fu}}\ and\ \bibinfo {author} {\bibfnamefont {C.~L.}\ \bibnamefont {Kane}},\
  }\href {\doibase 10.1103/PhysRevLett.109.246605} {\bibfield  {journal}
  {\bibinfo  {journal} {Phys. Rev. Lett.}\ }\textbf {\bibinfo {volume} {109}},\
  \bibinfo {pages} {246605} (\bibinfo {year} {2012})}\BibitemShut {NoStop}%
\bibitem [{Note6()}]{Note6}%
  \BibitemOpen
  \bibinfo {note} {The second term in Eq.~(\ref
  {eq:TypicalDeltaG1PeakLowTNoEC}) was obtained from Eq.~(\ref
  {eq:DeltaGAvPkRarePuddle1}) by replacing $\Gamma _1 \to \Gamma $, which,
  however, we will not do here}\BibitemShut {NoStop}%
\bibitem [{\citenamefont {Haake}(2010)}]{haake01}%
  \BibitemOpen
  \bibfield  {author} {\bibinfo {author} {\bibfnamefont {F.}~\bibnamefont
  {Haake}},\ }\href@noop {} {\emph {\bibinfo {title} {Quantum signatures of
  chaos}}},\ Vol.~\bibinfo {volume} {54}\ (\bibinfo  {publisher} {Springer},\
  \bibinfo {year} {2010})\BibitemShut {NoStop}%
\bibitem [{\citenamefont {K{\"o}nig}\ \emph {et~al.}(2008)\citenamefont
  {K{\"o}nig}, \citenamefont {Buhmann}, \citenamefont {W.~Molenkamp},
  \citenamefont {Hughes}, \citenamefont {Liu}, \citenamefont {Qi},\ and\
  \citenamefont {Zhang}}]{konig08}%
  \BibitemOpen
  \bibfield  {author} {\bibinfo {author} {\bibfnamefont {M.}~\bibnamefont
  {K{\"o}nig}}, \bibinfo {author} {\bibfnamefont {H.}~\bibnamefont {Buhmann}},
  \bibinfo {author} {\bibfnamefont {L.}~\bibnamefont {W.~Molenkamp}}, \bibinfo
  {author} {\bibfnamefont {T.}~\bibnamefont {Hughes}}, \bibinfo {author}
  {\bibfnamefont {C.-X.}\ \bibnamefont {Liu}}, \bibinfo {author} {\bibfnamefont
  {X.-L.}\ \bibnamefont {Qi}}, \ and\ \bibinfo {author} {\bibfnamefont {S.-C.}\
  \bibnamefont {Zhang}},\ }\href {\doibase 10.1143/JPSJ.77.031007} {\bibfield
  {journal} {\bibinfo  {journal} {Journal of the Physical Society of Japan}\
  }\textbf {\bibinfo {volume} {77}},\ \bibinfo {pages} {031007} (\bibinfo
  {year} {2008})}\BibitemShut {NoStop}%
\bibitem [{\citenamefont {K\"onig}\ \emph {et~al.}(2013)\citenamefont
  {K\"onig}, \citenamefont {Baenninger}, \citenamefont {Garcia}, \citenamefont
  {Harjee}, \citenamefont {Pruitt}, \citenamefont {Ames}, \citenamefont
  {Leubner}, \citenamefont {Br\"une}, \citenamefont {Buhmann}, \citenamefont
  {Molenkamp},\ and\ \citenamefont {Goldhaber-Gordon}}]{konig13}%
  \BibitemOpen
  \bibfield  {author} {\bibinfo {author} {\bibfnamefont {M.}~\bibnamefont
  {K\"onig}}, \bibinfo {author} {\bibfnamefont {M.}~\bibnamefont {Baenninger}},
  \bibinfo {author} {\bibfnamefont {A.~G.~F.}\ \bibnamefont {Garcia}}, \bibinfo
  {author} {\bibfnamefont {N.}~\bibnamefont {Harjee}}, \bibinfo {author}
  {\bibfnamefont {B.~L.}\ \bibnamefont {Pruitt}}, \bibinfo {author}
  {\bibfnamefont {C.}~\bibnamefont {Ames}}, \bibinfo {author} {\bibfnamefont
  {P.}~\bibnamefont {Leubner}}, \bibinfo {author} {\bibfnamefont
  {C.}~\bibnamefont {Br\"une}}, \bibinfo {author} {\bibfnamefont
  {H.}~\bibnamefont {Buhmann}}, \bibinfo {author} {\bibfnamefont {L.~W.}\
  \bibnamefont {Molenkamp}}, \ and\ \bibinfo {author} {\bibfnamefont
  {D.}~\bibnamefont {Goldhaber-Gordon}},\ }\href {\doibase
  10.1103/PhysRevX.3.021003} {\bibfield  {journal} {\bibinfo  {journal} {Phys.
  Rev. X}\ }\textbf {\bibinfo {volume} {3}},\ \bibinfo {pages} {021003}
  (\bibinfo {year} {2013})}\BibitemShut {NoStop}%
\bibitem [{\citenamefont {Nowack}\ \emph {et~al.}(2013)\citenamefont {Nowack},
  \citenamefont {Spanton}, \citenamefont {Baenninger}, \citenamefont
  {K{\"o}nig}, \citenamefont {Kirtley}, \citenamefont {Kalisky}, \citenamefont
  {Ames}, \citenamefont {Leubner}, \citenamefont {Br{\"u}ne}, \citenamefont
  {Buhmann} \emph {et~al.}}]{Nowack12}%
  \BibitemOpen
  \bibfield  {author} {\bibinfo {author} {\bibfnamefont {K.~C.}\ \bibnamefont
  {Nowack}}, \bibinfo {author} {\bibfnamefont {E.~M.}\ \bibnamefont {Spanton}},
  \bibinfo {author} {\bibfnamefont {M.}~\bibnamefont {Baenninger}}, \bibinfo
  {author} {\bibfnamefont {M.}~\bibnamefont {K{\"o}nig}}, \bibinfo {author}
  {\bibfnamefont {J.~R.}\ \bibnamefont {Kirtley}}, \bibinfo {author}
  {\bibfnamefont {B.}~\bibnamefont {Kalisky}}, \bibinfo {author} {\bibfnamefont
  {C.}~\bibnamefont {Ames}}, \bibinfo {author} {\bibfnamefont {P.}~\bibnamefont
  {Leubner}}, \bibinfo {author} {\bibfnamefont {C.}~\bibnamefont {Br{\"u}ne}},
  \bibinfo {author} {\bibfnamefont {H.}~\bibnamefont {Buhmann}},  \emph
  {et~al.},\ }\href
  {http://www.nature.com/nmat/journal/v12/n9/full/nmat3682.html} {\bibfield
  {journal} {\bibinfo  {journal} {Nature materials}\ }\textbf {\bibinfo
  {volume} {12}},\ \bibinfo {pages} {787} (\bibinfo {year} {2013})}\BibitemShut
  {NoStop}%
\bibitem [{Note7()}]{Note7}%
  \BibitemOpen
  \bibinfo {note} {\label {fn:lgvsab} Note that $a_B > \ell _g$, contrary to
  the assumptions of Eq.~(\ref {eq:DensityOfPuddles}). The smaller $\ell _g$
  is, the more effective is the screening of the potential fluctuations that
  create the puddles. In the limiting case of $\ell _g$ shorter than the radius
  $r_0\sim \alpha a_B$ of ``deep'' states with energies close to the center of
  the gap, the exponent $n_0/2n_d$ in Eq.~(\ref {eq:DensityOfPuddles}) is
  multiplied by an additional small factor $\sim (r_0/\ell _g)^2$. We believe,
  however, that we are only slightly overestimating $n_p$ by using here
  Eq.~(\ref {eq:DensityOfPuddles}), as the heterostructures of Ref.~\protect
  \rev@citealpnum {konig07} are in an intermediate regime, $a_B\approx 2\ell
  _g\approx 3 r_0$.}\BibitemShut {Stop}%
\bibitem [{\citenamefont {K{\"o}nig}()}]{konigThesis}%
  \BibitemOpen
  \bibfield  {author} {\bibinfo {author} {\bibfnamefont {M.}~\bibnamefont
  {K{\"o}nig}},\ }\href@noop {} {Ph.D. thesis}\BibitemShut {NoStop}%
\bibitem [{\citenamefont {Eikenberg}\ \emph {et~al.}(2014)\citenamefont
  {Eikenberg} \emph {et~al.}}]{molenkamp14unpublished}%
  \BibitemOpen
  \bibfield  {author} {\bibinfo {author} {\bibfnamefont {N.}~\bibnamefont
  {Eikenberg}} \emph {et~al.},\ }\href@noop {} {} (\bibinfo {year} {2014}),\
  \bibinfo {note} {\textit{unpublished}}\BibitemShut {NoStop}%
\bibitem [{\citenamefont {Cheianov}\ and\ \citenamefont
  {Glazman}(2013)}]{cheianov13}%
  \BibitemOpen
  \bibfield  {author} {\bibinfo {author} {\bibfnamefont {V.}~\bibnamefont
  {Cheianov}}\ and\ \bibinfo {author} {\bibfnamefont {L.~I.}\ \bibnamefont
  {Glazman}},\ }\href {\doibase 10.1103/PhysRevLett.110.206803} {\bibfield
  {journal} {\bibinfo  {journal} {Phys. Rev. Lett.}\ }\textbf {\bibinfo
  {volume} {110}},\ \bibinfo {pages} {206803} (\bibinfo {year}
  {2013})}\BibitemShut {NoStop}%
\bibitem [{\citenamefont {Altshuler}\ \emph {et~al.}(2013)\citenamefont
  {Altshuler}, \citenamefont {Aleiner},\ and\ \citenamefont
  {Yudson}}]{altshuler13}%
  \BibitemOpen
  \bibfield  {author} {\bibinfo {author} {\bibfnamefont {B.~L.}\ \bibnamefont
  {Altshuler}}, \bibinfo {author} {\bibfnamefont {I.~L.}\ \bibnamefont
  {Aleiner}}, \ and\ \bibinfo {author} {\bibfnamefont {V.~I.}\ \bibnamefont
  {Yudson}},\ }\href {\doibase 10.1103/PhysRevLett.111.086401} {\bibfield
  {journal} {\bibinfo  {journal} {Phys. Rev. Lett.}\ }\textbf {\bibinfo
  {volume} {111}},\ \bibinfo {pages} {086401} (\bibinfo {year}
  {2013})}\BibitemShut {NoStop}%
\bibitem [{\citenamefont {{Mehta}}(2004)}]{mehta04}%
  \BibitemOpen
  \bibfield  {author} {\bibinfo {author} {\bibfnamefont {M.~L.}\ \bibnamefont
  {{Mehta}}},\ }\href@noop {} {\emph {\bibinfo {title} {{Random matrices}}}}\
  (\bibinfo  {publisher} {Elsevier/Academic Press},\ \bibinfo {year}
  {2004})\BibitemShut {NoStop}%
\end{thebibliography}%

\begin{table*}[p] \begin{tabular}{|c|c|} \hline  $\begin{array}{cc} J_{xx}= & 2\sum\limits_{n,m<1}\dfrac{t_{m}t_{n}\text{Re}(U_{nL1L;mR1R}+U_{nR1L;mL1R})}{(\varepsilon_{m}-E_{-})(\varepsilon_{n}-E_{-})}\\  & +2\sum\limits_{n,m>1}\dfrac{t_{n}t_{m}\text{Re}(U_{1LmR;1RnL}+U_{1RmR;1LnL})}{(\varepsilon_{m}+E_{+})(\varepsilon_{n}+E_{+})}\\  & +4\sum\limits_{\mu}\sum\limits_{n,m<1}\dfrac{t_{1}t_{n}\text{Re}U_{nLm\mu;1Lm\mu}}{(\varepsilon_{n}-\varepsilon_{1})(\varepsilon_{n}-E_{-})}\\  & +4\sum\limits_{\mu}\sum\limits_{n>1}\sum\limits_{m<1}\dfrac{t_{1}t_{n}\text{Re}U_{nLm\mu;1Lm\mu}}{(\varepsilon_{n}-\varepsilon_{1})(\varepsilon_{n}+E_{+})}\\  & -4\sum\limits_{n<1}\sum\limits_{m>1}\dfrac{t_{m}t_{n}\text{Re}(U_{nL1L;mR1R}+U_{nR1L;mL1R})}{(\varepsilon_{n}-\varepsilon_{m})(\varepsilon_{n}-E_{-})}\\  & -4\sum\limits_{n<1}\sum\limits_{m>1}\dfrac{t_{m}t_{n}\text{Re}(U_{nL1L;mR1R}+U_{nR1L;mL1R})}{(\varepsilon_{n}-\varepsilon_{m})(\varepsilon_{m}+E_{+})} \end{array}$ & $\begin{array}{cc} J_{yy}= & 2\sum\limits_{n,m<1}\dfrac{t_{m}t_{n}\text{Re}(U_{nR1L;mL1R}-U_{nL1L;mR1R})}{(\varepsilon_{m}-E_{-})(\varepsilon_{n}-E_{-})}\\  & +2\sum\limits_{n,m>1}\dfrac{t_{n}t_{m}\text{Re}(U_{1LmR;1RnL}-U_{1RmR;1LnL})}{(\varepsilon_{m}+E_{+})(\varepsilon_{n}+E_{+})}\\  & +4\sum\limits_{\mu}\sum\limits_{n,m<1}\dfrac{t_{1}t_{n}\text{Re}U_{nLm\mu;1Lm\mu}}{(\varepsilon_{n}-\varepsilon_{1})(\varepsilon_{n}-E_{-})}\\  & +4\sum\limits_{\mu}\sum\limits_{n>1}\sum\limits_{m<1}\dfrac{t_{n}t_{1}\text{Re}U_{nLm\mu;1Lm\mu}}{(\varepsilon_{n}-\varepsilon_{1})(\varepsilon_{n}+E_{+})}\\  & -4\sum\limits_{n<1}\sum\limits_{m>1}\dfrac{t_{m}t_{n}\text{Re}(U_{nR1L;mL1R}-U_{nL1L;mR1R})}{(\varepsilon_{n}-\varepsilon_{m})(\varepsilon_{n}-E_{-})}\\  & -4\sum\limits_{n<1}\sum\limits_{m>1}\dfrac{t_{m}t_{n}\text{Re}(U_{nR1L;mL1R}-U_{nL1L;mR1R})}{(\varepsilon_{n}-\varepsilon_{m})(\varepsilon_{m}+E_{+})} \end{array}$\tabularnewline \hline  $\begin{array}{cc} J_{zz}= & 2\sum\limits_{m\neq n<1}\sum\limits_{n<1}\dfrac{t_{m}t_{n}\text{Re}(U_{nL1L;mL1L}-U_{nR1L;mR1L})}{(\varepsilon_{m}-E_{-})(\varepsilon_{n}-E_{-})}\\  & +2\sum\limits_{m\neq n>1}\sum\limits_{n>1}\dfrac{t_{n}t_{m}\text{Re}(U_{mL1L;nL1L}-U_{mR1L;nR1L})}{(\varepsilon_{m}+E_{+})(\varepsilon_{n}+E_{+})}\\  & -4\sum\limits_{\mu}\sum\limits_{n,m<1}\dfrac{t_{1}t_{n}\text{Re}U_{nRm\mu;m\mu1R}}{(\varepsilon_{n}-\varepsilon_{1})(\varepsilon_{n}-E_{-})}\\  & -4\sum\limits_{\mu}\sum\limits_{n>1}\sum\limits_{m<1}\dfrac{t_{1}t_{n}\text{Re}U_{nRm\mu;m\mu1R}}{(\varepsilon_{n}-\varepsilon_{1})(\varepsilon_{n}+E_{+})}\\  & -4\sum\limits_{n<1}\sum\limits_{m>1}\dfrac{t_{m}t_{n}\text{Re}(U_{nL1L;mL1L}-U_{nL1R;mL1R})}{(\varepsilon_{n}-\varepsilon_{m})(\varepsilon_{n}-E_{-})}\\  & -4\sum\limits_{n<1}\sum\limits_{m>1}\dfrac{t_{m}t_{n}\text{Re}(U_{nL1L;mL1L}-U_{nL1R;mL1R})}{(\varepsilon_{n}-\varepsilon_{m})(\varepsilon_{m}+E_{+})} \end{array}$ & $\begin{array}{cc} J_{yx}= & -4\sum\limits_{n,m<1}\dfrac{t_{m}t_{n}\text{Im}U_{nL1L;mR1R}}{(\varepsilon_{m}-E_{-})(\varepsilon_{n}-E_{-})}\\  & -4\sum\limits_{n,m>1}\dfrac{t_{m}t_{n}\text{Im}U_{nL1L;mR1R}}{(\varepsilon_{m}+E_{+})(\varepsilon_{n}+E_{+})}\\  & +8\sum\limits_{n<1}\sum\limits_{m>1}\dfrac{t_{m}t_{n}\text{Im}U_{nL1L;mR1R}}{(\varepsilon_{n}-\varepsilon_{m})(\varepsilon_{n}-E_{-})}\\  & +8\sum\limits_{n<1}\sum\limits_{m>1}\dfrac{t_{m}t_{n}\text{Im}U_{nL1L;mR1R}}{(\varepsilon_{n}-\varepsilon_{m})(\varepsilon_{m}+E_{+})} \end{array}$\tabularnewline \hline  $\begin{array}{cc} J_{zx}= & 4\sum\limits_{n,m<1}\dfrac{t_{n}t_{m}\text{Re}(U_{nL1L;mR1L}+U_{nL1L;mL1R})}{(\varepsilon_{m}-E_{-})(\varepsilon_{n}-E_{-})}\\  & +4\sum\limits_{n,m>1}\dfrac{t_{n}t_{m}\text{Re}(U_{nL1L;mR1L}+U_{nL1L;mL1R})}{(\varepsilon_{m}+E_{+})(\varepsilon_{n}+E_{+})}\\  & -4\sum\limits_{\mu}\sum\limits_{n,m<1}\dfrac{t_{1}t_{n}\text{Re}U_{nLm\mu;1Rm\mu}}{(\varepsilon_{n}-\varepsilon_{1})(\varepsilon_{n}-E_{-})}\\  & -4\sum\limits_{\mu}\sum\limits_{n>1}\sum\limits_{m<1}\dfrac{t_{1}t_{n}\text{Re}U_{nLm\mu;1Rm\mu}}{(\varepsilon_{n}-\varepsilon_{1})(\varepsilon_{n}+E_{+})}\\  & +4\sum\limits_{n<1}\sum\limits_{m>1}\dfrac{t_{m}t_{n}\text{Re}(U_{mR1R;nL1R}+U_{mL1R;nR1R})}{(\varepsilon_{n}-\varepsilon_{m})(\varepsilon_{n}-E_{-})}\\  & +4\sum\limits_{n<1}\sum\limits_{m>1}\dfrac{t_{m}t_{n}\text{Re}(U_{mR1R;nL1R}+U_{mL1R;nR1R})}{(\varepsilon_{n}-\varepsilon_{m})(\varepsilon_{m}+E_{+})} \end{array}$ & $\begin{array}{cc} J_{zy}= & -4\sum\limits_{n,m<1}\dfrac{t_{m}t_{n}\text{Im}(U_{nL1L;mR1L}+U_{nL1L;mL1R})}{(\varepsilon_{m}-E_{-})(\varepsilon_{n}-E_{-})}\\  & -4\sum\limits_{n,m>1}\dfrac{t_{m}t_{n}\text{Im}(U_{nL1L;mR1L}+U_{nL1L;mL1R})}{(\varepsilon_{m}+E_{+})(\varepsilon_{n}+E_{+})}\\  & +4\sum\limits_{\mu}\sum\limits_{n,m<1}\dfrac{t_{1}t_{n}\text{Im}U_{nLm\mu;1Rm\mu}}{(\varepsilon_{n}-\varepsilon_{1})(\varepsilon_{n}-E_{-})}\\  & +4\sum\limits_{\mu}\sum\limits_{n>1}\sum\limits_{m<1}\dfrac{t_{1}t_{n}\text{Im}U_{nLm\mu;1Rm\mu}}{(\varepsilon_{n}-\varepsilon_{1})(\varepsilon_{n}+E_{+})}\\  & -4\sum\limits_{n<1}\sum\limits_{m>1}\dfrac{t_{m}t_{n}\text{Im}(U_{mR1R;nL1R}+U_{nR1R;mL1R})}{(\varepsilon_{n}-\varepsilon_{m})(\varepsilon_{n}-E_{-})}\\  & -4\sum\limits_{n<1}\sum\limits_{m>1}\dfrac{t_{m}t_{n}\text{Im}(U_{mR1R;nL1R}+U_{nR1R;mL1R})}{(\varepsilon_{n}-\varepsilon_{m})(\varepsilon_{m}+E_{+})} \end{array}$\tabularnewline \hline  \end{tabular}\caption{Components of the exchange tensor $\mathbf{J}$. The components are obtained by matching terms in Eqs.~(\ref{eq:HeffWithProjectors}) and (\ref{eq:Heffective}).} \label{table-Jcomponents} \end{table*}

\begin{table}[p]
\begin{tabular}{|c|c|}
\hline 
Term in $T$-matrix & Denominator\tabularnewline
\hline 
\hline 
$UH^{out}H^{in}H^{out}H^{in}$, $H^{out}H^{in}UH^{out}H^{in}$, $H^{out}H^{in}H^{out}H^{in}U$ & $E_{+}^{2}\delta^{2}$\tabularnewline
\hline 
$H^{out}UH^{in}H^{out}H^{in}$, $H^{out}H^{in}H^{out}UH^{in}$, & $E_{+}^{3}\delta$\tabularnewline
\hline 
$UH^{out}H^{in}H^{in}H^{out}$, $H^{out}H^{in}UH^{in}H^{out}$, $H^{out}H^{in}H^{in}H^{out}U$,  & $E_{-}E_{+}\delta^{2}$\tabularnewline
$H^{in}H^{out}H^{out}H^{in}U$, $H^{in}H^{out}UH^{out}H^{in}$, $UH^{in}H^{out}H^{out}H^{in}$ & \tabularnewline
\hline 
$H^{out}UH^{in}H^{in}H^{out}$, $H^{in}H^{out}H^{out}UH^{in}$ & $E_{-}E_{+}^{2}\delta$\tabularnewline
\hline 
 $UH^{out}H^{out}H^{in}H^{in}$, $H^{out}H^{out}H^{in}H^{in}U$ & $E_{C}E_{+}^{2}\delta$\tabularnewline
\hline 
$UH^{in}H^{out}H^{in}H^{out}$, $H^{in}H^{out}UH^{in}H^{out}$, $H^{in}H^{out}H^{in}H^{out}U$ & $E_{-}^{2}\delta^{2}$\tabularnewline
\hline 
$H^{in}UH^{out}H^{in}H^{out}$, $H^{in}H^{out}H^{in}UH^{out}$ & $E_{-}^{3}\delta$\tabularnewline
\hline 
$H^{out}H^{in}H^{in}UH^{out}$, $H^{in}UH^{out}H^{out}H^{in}$ & $E_{-}^{2}E_{+}\delta$\tabularnewline
\hline 
$H^{in}H^{in}H^{out}H^{out}U$, $UH^{in}H^{in}H^{out}H^{out}$ & $E_{C}E_{-}^{2}\delta$\tabularnewline
\hline 
$H^{out}H^{out}H^{in}UH^{in}$, $H^{out}UH^{out}H^{in}H^{in}$ & $E_{C}E_{+}^{3}$\tabularnewline
\hline 
$H^{in}UH^{in}H^{out}H^{out}$, $H^{in}H^{in}H^{out}UH^{out}$ & $E_{C}E_{-}^{3}$\tabularnewline
\hline 
$H^{out}H^{out}UH^{in}H^{in}$ & $E_{C}^{2}E_{+}^{2}$\tabularnewline
\hline 
$H^{in}H^{in}UH^{out}H^{out}$ & $E_{C}^{2}E_{-}^{2}$\tabularnewline
\hline 
\end{tabular}\caption{Relative importance of the 30 different terms in the 5th order $T$-matrix.
Here $H^{\text{out}}=(H^{\text{in}})^{\dagger}=\sum_{n,\gamma}t_{n}\psi_{\gamma}^{\dagger}(0)c_{n\gamma}$.
The virtual state with an excess or deficit of 2 particles has the
largest denominator $\sim E_{C}$.}
\label{table-Tmatrix}
\end{table}

\end{document}